\documentclass{article}
\usepackage[english]{babel}
\usepackage[latin1]{inputenc}
\usepackage{amsfonts}
\usepackage{amssymb,phystex}
\usepackage{graphicx}
\usepackage{slashed}
\usepackage{feynmp}
\usepackage{fancybox}
\usepackage{mathrsfs}
\usepackage{float}
\usepackage{dsfont}

\title{\Huge Properties of D-Mesons from QCD Sum Rules with an improved spectral function.}
\author{J\"org Pfannm\"oller \\
        Diplomathesis}
\date{31. January 2006}

\begin{document}
\newcommand{\braket}[2]{\langle#1|#2\rangle}
\newcommand{\mele}[1]{\langle #1 \rangle}
\maketitle
\vspace{5cm}
\huge
\begin{displaymath}
\parbox[c]{3cm}{
 \begin{fmffile}{cover01}
      \begin{fmfgraph*}(30,20)
        \fmfleft{i} 
        \fmfright{o}
        \fmf{dots}{i,v1} 
        \fmf{dots}{v2,o}
        \fmf{plain,left,tension=0.2,tag=1}{v1,v2}
        \fmf{plain,left,tension=0.2,tag=2}{v2,v1}
        \fmfdot{v1,v2}
        \fmfposition
        \fmfipath{p[]}
        \fmfiset{p1}{vpath1(__v1,__v2)}
        \fmfiset{p2}{vpath2(__v2,__v1)}
        \fmfi{gluon,left,tension=0.2}{point length(p1)/2 of p1 -- point length(p2)/2 of p2}
      \end{fmfgraph*}
     \end{fmffile}}\cdot\mathds{1}+
 \parbox[c]{3cm}{
 \begin{fmffile}{cover02}
      \begin{fmfgraph*}(30,20)
        \fmfleft{i} 
        \fmfright{o}
        \fmf{dots}{i,v1} 
        \fmf{dots}{v4,o}
        \fmf{plain,left,tension=0.0001,tag=1}{v1,v4}
        \fmf{plain,left,tension=0.0001,tag=2}{v4,v1}
	
	\fmf{phantom}{v1,v2}
	\fmf{phantom,tag=3}{v2,v3}
	\fmf{phantom}{v3,v4}
 
	\fmffixedx{0}{v2,v3}
	\fmffixedy{0.4cm}{v2,v3}
	\fmffixedy{0}{v1,v4}
	
        \fmfposition
        \fmfipath{p[]}
        \fmfiset{p1}{vpath1(__v1,__v4)}
        \fmfiset{p2}{vpath2(__v4,__v1)}
	 \fmfiset{p3}{vpath3(__v2,__v3)}
	\fmfi{gluon,right}{point 0 of p3 -- point length(p2)/2 of p2} 
       \fmfi{gluon,right}{point length(p3) of p3 -- point length(p1)/2 of p1} 
       \fmfiv{d.sh=cross,d.ang=0,d.siz=5thick}{point 0 of p3}
	 \fmfiv{d.sh=cross,d.ang=0,d.siz=5thick}{point length(p3) of p3} 
	 \fmfdot{v1}
	\fmfdot{v4}
      \end{fmfgraph*}
     \end{fmffile}}\cdot\mele{GG}   
    =\frac{1}{\pi}\int\limits_{0}^{\infty}\frac{\includegraphics[scale=0.25]{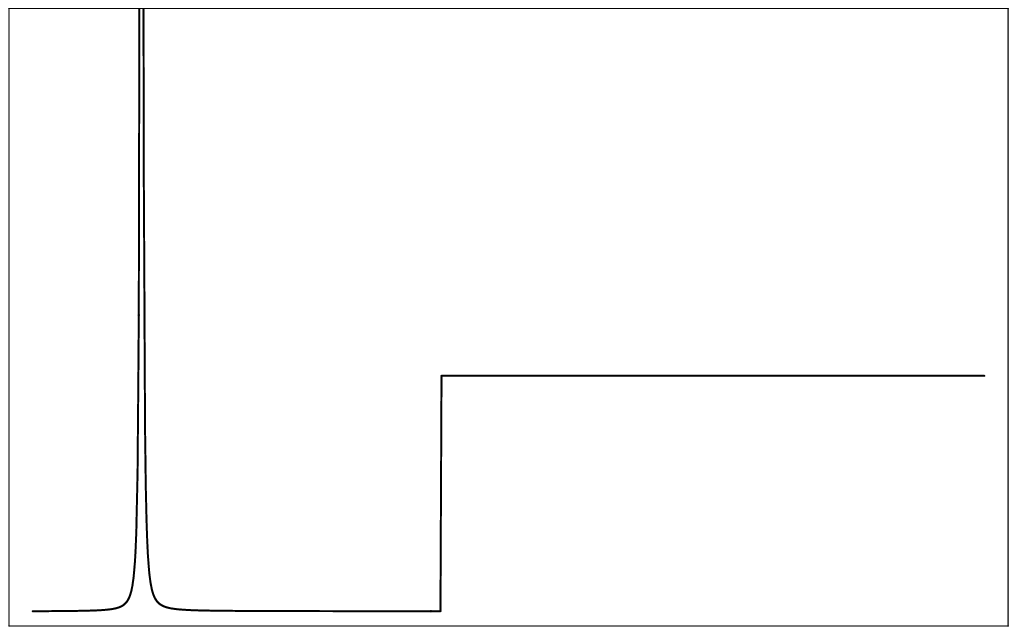}}{s-q^2}ds 
\end{displaymath}
\normalsize
\newpage

\tableofcontents
\newpage

\section{Introduction}
The purpose of this thesis is the computation of the observables for heavy-light mesons, mesons with a light and a heavy valence quark, in QCD. There are various methods available to accomplish this task. Hence, the first question is which method is chosen for the computations. The degrees of freedom which should be used give the first constraint on the methods in question. In this attempt quarks and gluons should be used, the elementary degrees of freedom of QCD. This choice reduces the number of methods which can be used. In particular QCD Sum Rules (QSR) are chosen to do the job.\\
Those few sentences raise several questions. There are two which are of special interests. How are the properties of resonances calculated in QCD the theory of strong interactions ? What are QCD Sum Rules ? The answers to these questions directly show how the calculations performed in this thesis work. Based on those calculations conclusions can be drawn.\\ 
The classical computations based on the QCD Lagrangian have been perturbative ones. They where based on a perturbative solution of the QCD Lagrangian where the solution is expanded in the coupling constant $\alpha_{s}$ and every order can be calculated via the Feynman diagram technique. Characteristic computations based on perturbation theory have been calculations of running coupling constants, running masses or of charges and deep inelastic scattering. Perturbation theory was also used to check whether QCD is the theory of strong interactions or not. \\
Unfortunately the properties of mesons can not be calculated via perturbation theory. The perturbative solution of the QCD Lagrangian is useful in the case where it can be truncated after a certain order. There are infinitely many orders of the solution in  $\alpha_{s}$. A calculation of every order would be impossible. Therefore the coupling constant $\alpha_{s}$ has to be small if the truncation should be valid. However, the coupling constant is momentum dependent. The coupling constant is small for momenta larger than $\Lambda_{QCD}$ and large for momenta smaller than $\Lambda_{QCD}$. Thus, the perturbative solution is expected to be valid for large momenta and invalid for small momenta. This behavior is called asymptotic freedom because the interaction tends to zero if the energy goes to infinity. Asymptotic freedom is the property that allows for all perturbative calculations in QCD.\\ 
In the time when the tests and measurements based on perturbation theory have been performed a new phenomenon was recognized. Free quarks or gluons have never been found. This was called the confinement of quarks. To our knowledge its impossible to produce free quarks or gluons. They are always confined in a bound state. Though it is widely believed that at high energies QCD matter melts and reveals quasi free quarks and gluons. This would be something like a liquid where the molecules are quarks and gluons. The confinement has up to today not been explained and is possibly not explainable by perturbative physics, but it is most probably  close connected to hadron physics.\\ 
The physics around us takes place in the domain of small momenta whereas the physics which takes place at high momenta could only be found in the early universe. There are artificial sources where the physics at high momentum scales can be found like accelerators. Hence, if the QCD physics at the energy scales at which we live should be described a non-perturbative solution of the QCD Lagrangian is needed. Unfortunately there is only one method which is supposed to be capable to solve the theory in every domain. It is lattice QCD, but this method needs a lot of computer power and is not very instructive about the physics which stands behind the calculations. Therefore, a less direct method is used, but a method with considerably more physical insight than lattice QCD.\\
The dream of everybody who works in this field would be to find a formula as it is given for the hydrogen atom. Here every bound state has a main quantum number $n$ which characterizes the mass of the bound state. Something like that is aimed for by modern researchers. The fine structure of the problem is completely neglected because the raw structure is already that difficult. One method to represent the states are spectral functions where every bound state is identified with a peak in a function of the energy. The properties of the bound states characterize the peaks. Thus, if the properties are known the peaks can be calculated or vice versa. The mathematical object which contains the spectral function of a certain bound states are correlators. This has been explored in the K/"al\'en-Lehmann spectral representation. In the domain with time-like momentum the correlator is given by the spectral function of the system. In a perturbative calculation bound states are not generated.\\
However, there is a possibility to estimate the spectral function taking non-perturbative effects of the QCD Lagrangian into account. One exploits the fact that the correlator in the space like domain can be calculated by the so called operator product expansion (OPE). Via a dispersion relation time-like and space-like momenta can be connected with the domain where the momentum squared is negative. A phenomenological ansatz for the spectral function is made and the properties of this ansatz are fitted in order to reproduce the OPE. In a certain way the spectral function is then calculated by the scheme. QCD enters the scheme through the OPE which is the quantity where theoretical calculations enter the game. The so called Wilson coefficients of the OPE can be calculated using the QCD Lagrangian. Hence, this scheme has two parts a phenomenological part which is the spectral function and a theoretical part which is the OPE. This scheme is called a QCD Sum Rule. It has two features that provide the non-perturbative information. Through the OPE non-perturbative information enters the calculation and the dispersion relation connects the spectral function with the non-perturbative information. \\
The thesis is structured in four parts:
\begin{itemize}
\item In the first part, sections \ref{reguandreno}, \ref{OPE}, \ref{wilsoncoefficient} and \ref{condensates}, the OPE and its calculation are introduced with a special glance on some subtle but important features of the OPE. 

\item In the second part, section \ref{dispersionrelations}, dispersions relations are discussed. 

\item The third part, sections \ref{sumrules}, \ref{charmonium} and \ref{borel}, introduce correlators, spectral functions, QCD Sum Rules and the Borel Sum Rules. 

\item The fourth part, section \ref{hl_systems}, QCD Sum Rule calculations of heavy-light systems are presented. In this part novel calculations are discussed. In addition previous QCD Sum Rule calculations are reviewed. In these examples the Wilson coefficients are calculated in QCD and in the heavy quark effective theory (HQET).
\end{itemize}

\section{Regularization and renormalization \label{reguandreno}}
In this section important terminologies and techniques in the field of regularization and renormalization are introduced. In the first part terminologies are discussed and in the second examples of some common techniques are presented. The results of the examples will be used in subsequent sections. 

\subsection{The terminology of superficial divergence}

Let D be the superficial degree of divergence: an amplitude with the degree D diverges like:
\begin{eqnarray}
\int^{\infty}_{0} k^{D-1} dk.
\end{eqnarray}
Some examples are given here:
\begin{eqnarray}
D=0~~~~~~~~~~~\int^{\infty}_{0} k^{-1} dk=\int^{\infty}_{0} \frac{1}{k}dk=\left(\ln[k]\right)_{0}^{\infty}
\end{eqnarray}
\begin{eqnarray}
D=1~~~~~~~~~~~\int^{\infty}_{0} k^{0} dk=\int^{\infty}_{0} 1 dk=\left(k\right)_{0}^{\infty}
\end{eqnarray}
\begin{eqnarray}
D=2~~~~~~~~~~~\int^{\infty}_{0} k^{1} dk=\int^{\infty}_{0} k dk=\left(\frac{k^2}{2}\right)_{0}^{\infty}.
\end{eqnarray}
This means: all amplitudes with $D\geq0$ are divergent, it is important to notice that also the amplitude with D=0 is divergent. Such amplitudes are logarithmically divergent. Examples for divergent diagrams are loop diagrams. A simple example occurs in the real scalar $\phi^4$ theory where the tadpole diagram is divergent:
\begin{eqnarray}
\parbox[c]{3cm}{
\begin{fmffile}{phi4tadpole01}
\begin{fmfgraph*}(23,20)
 \fmfleft{i} 
 \fmfright{o}
 \fmf{plain_arrow,label=$q$}{i,v}
 \fmf{plain_arrow,label=$k$,tension=0.8}{v,v}
 \fmf{plain_arrow,label=$q$}{v,o}
 \fmfdot{v}
\end{fmfgraph*}
\end{fmffile}}
\backsim\int d^{4}k\frac{i}{k^2-m^2} \label{tadpole01}.
\end{eqnarray}
In this case D=2. The external legs have been amputated in the explicit expression. The number of loop integrals in each theory is infinite but fortunately there is a class of quantum field theories in which the number of elementary diagrams which are divergent is limited. In this class the treatment of these divergences is simpler than in a class of theories with infinitely many divergent diagrams. An example for a diagram which is finite and contains a loop is\\
\begin{eqnarray}
\parbox[c]{4cm}{
\begin{fmffile}{phi46}
\begin{fmfgraph*}(20,15)
 \fmfleft{i1,i2,i3,i4} 
 \fmfright{o1,o2,o3,o4}
 
 \fmf{plain}{i1,v2}
 \fmf{plain}{i2,v1}
 \fmf{plain}{i3,v1}
 \fmf{plain}{i4,v3}
 
 \fmf{plain}{o1,v2}
 \fmf{plain}{o2,v4}
 \fmf{plain}{o3,v4}
 \fmf{plain}{o4,v3}
 
 \fmf{plain}{v1,v2,v4,v3,v1}
 \fmffixedx{0}{v2,v3}
 \fmffixedy{20}{v2,v3}
 \fmfdot{v1,v2,v3,v4}
 
 \fmflabel{$q_{1}$}{i4}
 \fmflabel{$q_{2}$}{i3}
 \fmflabel{$q_{3}$}{i2}
 \fmflabel{$q_{4}$}{i1}
 
 \fmflabel{$q'_{1}$}{o4}
 \fmflabel{$q'_{2}$}{o3}
 \fmflabel{$q'_{3}$}{o2}
 \fmflabel{$q'_{4}$}{o1}
\end{fmfgraph*}
\end{fmffile}}\backsim\int d^{4}k\left(\frac{i}{k^2-m^2}\right)^4.
\end{eqnarray}\\
Here D=-4, the integral is not divergent. The external legs are again amputated and the external momenta are set to 0.

\subsection{Regularization and renormalization of the tadpole diagram in $\lambda\phi^4$-theory}

\begin{figure}[htbp]
 \begin{center}
  \begin{picture}(0,0)%
\includegraphics{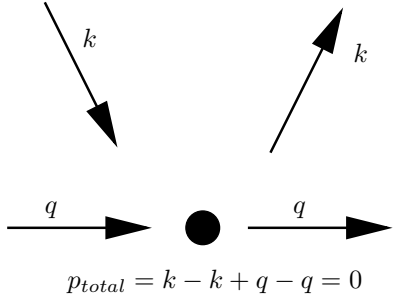}%
\end{picture}%
\setlength{\unitlength}{4144sp}%
\begingroup\makeatletter\ifx\SetFigFont\undefined%
\gdef\SetFigFont#1#2#3#4#5{%
  \reset@font\fontsize{#1}{#2pt}%
  \fontfamily{#3}\fontseries{#4}\fontshape{#5}%
  \selectfont}%
\fi\endgroup%
\begin{picture}(2384,1805)(879,-1844)
\put(1261,-1771){$p_{total}=k-k+q-q=0$}%
\put(1351,-331){$k$}%
\put(1126,-1321){$q$}%
\put(2611,-1321){$q$}%
\put(2971,-421){$k$}%
\end{picture}%
  \caption{Momentum conservation at the vertex.\label{tadpolemomentum}}
  \end{center}
  \end{figure}

The Lagrangian of the theory is given by
\begin{eqnarray}
      \mathscr{L}=\frac{1}{2}\left(\partial_{\mu}\phi\right)^2-\frac{1}{2}m^2\phi^2-\frac{\lambda}{4!}\
      \phi^4.
\end{eqnarray}
There is only one diagram to order $\lambda^1$ (see equation (\ref{tadpole01})). This diagram is the simplest divergent diagram that exists. Divergent diagrams can be regularized. These regularized expressions parameterize the divergence of the diagram. After the regularization it is possible to renormalize the diagram. Renormalization is a process in which the infinities are absorbed from observable quantities which have a relation to the amplitude, in order to eliminate them from the results. To achieve that a reference momentum is chosen at which the observable quantity is measured. This point is called the renormalization point. Taking the difference of the quantity given by the diagram at the renormalization point and another point results in an expression which is given by the value of the observable at the renormalization point and some momentum dependent function. At this stage the infinities cancel and the theory is defined at an arbitrary renormalization point.\\
The full analytic expression for the diagram (\ref{tadpole01}) is 
\begin{eqnarray}
\frac{i}{q^2-m^2}\left(-i\lambda\right)\int\frac{dk^4}{(2\pi)^4}\frac{i}{k^2-m^2}\frac{i}{q^2-m^2}=\frac{\left(-\lambda\right)}{(q^2-m^2)^2}\int\frac{dk^4}{(2\pi)^4}\frac{1}{q^2-m^2}\label{tadpolefull}.
\end{eqnarray}
The momentum conservation at the vertex in (\ref{tadpole01}) shows that the external momenta do not contribute to the internal momentum (see figure \ref{tadpolemomentum}).
The diagram is factorized in the contribution of two free propagators and the loop. In the following only the integral is kept. This integral which represents the loop is divergent, its superficial degree of divergence is D=1. There exist various methods to regularize and renormalize divergent diagrams. Every method has advantages and disadvantages therefore the three most common are applied to the tadpole diagram for a comparison and an overview.\\
An explicit integration of the loop integral is needed. One method to accomplish the integration is the Wick rotation. Here the integration in the four dimensional Minkowski space is transformed to an integration in a four dimensional Euclidean space. The only thing employed is the residue theorem.\\
 \begin{figure}[htbp]
 \begin{center}
  \begin{picture}(0,0)%
\includegraphics{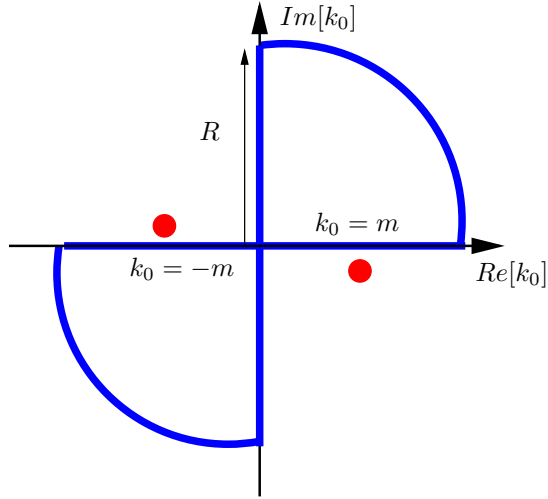}%
\end{picture}%
\setlength{\unitlength}{4144sp}%
\begingroup\makeatletter\ifx\SetFigFont\undefined%
\gdef\SetFigFont#1#2#3#4#5{%
  \reset@font\fontsize{#1}{#2pt}%
  \fontfamily{#3}\fontseries{#4}\fontshape{#5}%
  \selectfont}%
\fi\endgroup%
\begin{picture}(3044,3044)(2199,-4283)
\put(5011,-2971){$Re[k_{0}]$}%
\put(3841,-1441){$Im[k_{0}]$}%
\put(2941,-2941){$k_{0}=-m$}%
\put(4051,-2671){$k_{0}=m$}%
\put(3361,-2131){$R$}%
\end{picture}%

  \caption{Contour of integration in the zero component of (\ref{tadpolefull}). The red dots are mass poles shifted by $\epsilon$ away from the real axis. If the limit $R\rightarrow \infty$ is taken the integration over the bows vanish.\label{wickrotation}}
  \end{center}
  \end{figure}
The integration over the contour in figure \ref{wickrotation} is zero. In the limit $R\rightarrow \infty$ only the integration along the real and the imaginary axis contribute. Therefore the integrations must be equal except of their signs. This enables the replacement of the integral along the real, $I_{1}$, axis with the integral along the imaginary axes, $I_{2}$. The integral $I_{2}$ is not affected from the poles sketched in figure (\ref{wickrotation}) and furthermore the replacement transforms the integral from Minkowski to Euclidean space. This transformation is further investigated in section \ref{wilsoncoefficient}. The whole equation can also be merged in a graphic picture. The integration contour of $I_{1}$ can be rotated into the integration contour of $I_{2}$ without crossing the mass poles. Hence the name Wick rotation is explained as a simple rule for the procedure just outlined.\\
After the Wick rotation the integral is given by the following expression:
\begin{eqnarray}
\int\frac{dk^4}{(2\pi)^4}\frac{1}{k^2-m^2}=\frac{-i}{8\pi^2}\int_{0}^{\infty} 
 dk_{E}\frac{k_{E}^3}{k^2_{E}+m^2}=\frac{-i}{16\pi^2}\left(k_{E}^2-m^2\ln[k_{E}^2+m^2]\right)_{0}^{\infty}.
\end{eqnarray}
The pole in the integrand has vanished and the integration runs in Euclidean space, this is symbolized by the index E. 
The transformation law is simply $l^0=il^{0}_{E}$, $\vec{l}=\vec{l}_{E}$, the minus sign in front of the integration was explained above.\\ 
Thus the integral is computed and the comparison of the renormalization methods can be started.
\begin{enumerate}
\item The cutoff method:\\ \\
      This method is the most simple one, application of the Feynman rules results in an integration from 0 to
      $\infty$. This integral is divergent and has therefore to be regularized. Regularization means a parameterization of the divergence or in other words a method to make the integral finite with a rule that tells how the original integral can be derived from the regularized one. Thus one way to achieve that is by replacing the upper limit by $\Lambda$. The result is:
      \begin{eqnarray}
      \int\frac{dk^4}{(2\pi)^4}\frac{1}{k^2-m^2}=\frac{-i}{16\pi^2}\left(k_{E}^2-m^2\ln[k_{E}^2+m^2]\right)_{0}^{\Lambda}\nonumber\\
      =\frac{-i}{16\pi^2}\left(\Lambda^2-m^2\ln\left[\frac{\Lambda^2}{m^2}+1\right]\right)\underbrace{\approx}_{for 
      \Lambda\gg m}\frac{-i}{16\pi^2}\left(\Lambda^2-m^2\ln\left[\frac{\Lambda^2}{m^2}\right]\right)=-if\left(\Lambda\right) \label{power}.
      \end{eqnarray}
      The original divergent result is retained in the limit $\Lambda\rightarrow\infty$. This is a parameterization of 
      the tadpoles divergence. Through this procedure a parameterization of the divergence is obtained, the integral is regularized. Without the cutoff the 
      integral is infinite.\\
      Renormalization is a method to get rid of the unphysical parameter $\Lambda$, which is infinite and therefore unphysical. The question is which effect this divergence has. As we know from measurements the propagation of a particle from a to b is 
      not infinitely likely. This gives us the strong hint that the divergence should be absorbed into something inside the 
      amplitude. If we neglect all other contributions to the propagator except of the tadpole we can form a geometrical 
      series, that looks like:
      \begin{eqnarray}
      D(p^2)=i\int d^4x~e^{ip^{\mu}x_{\mu}}\bra{0}T\{\phi(0)\phi(x)\}\ket{0}=\nonumber\\
      \frac{i}{p^2-m^2}\left[1-if\left(\Lambda\right)\frac{1}{p^2-m^2}+\left(-if\left(\Lambda\right)\frac{1}{p^2-m^2}\right)^2+...
      \right.=\frac{i}{p^2-m^2}\frac{1}{1-\frac{-if\left(\Lambda\right)}{p^2-m^2}}=\nonumber\\=\frac{i}{p^2-m^2+if\left(\Lambda
      \right)}=\frac{i}{p^2-m^2_{R}} \label{selfenergy}.
      \end{eqnarray}
      This shows that the mass can absorb the divergence which we parameterized by the cutoff. The combination 
      $m^2-if\left(\Lambda\right)$ is called the renormalized mass $m_{R}$. The massparameter that we called m is the naked mass and has to be divergent in order to absorb the divergent  $-if\left(\Lambda\right)$. 
      If we assume the naked mass to be divergent, this subtraction of divergences eliminates the problem of infinities. Thus renormalization of the diagram is achieved.\\
      The renormalization just performed has to be treated with care. The loop integral is independent of the external 
      momentum $p^{\mu}$. This means the renormalization of the mass by means of the tadpole is a shift of the mass it does not 
      produce a momentum dependent mass, the renormalize mass stays constant. The first momentum dependent loop emerges in second order 
      $\lambda$. Therefore no physical measurable effect exists that proves whether the mass is shifted by the tadpole or not.\\
      The consequence of this analysis is a simplification of our calculations! All loop diagrams in which the loops do not 
      depend on the external momentum p can be dropped from the perturbation expansion of the two point function. They produce no 
      momentum dependence of the mass but the mass is a parameter which enters the theory out of measurements. This means it is fixed if all contributions beyond first order $\lambda$ can be neglected.
\item The Pauli-Villars method:\\ \\
      Another method to regularize the tadpole diagram is the introduction of regulators. Where the regulators are just copies of the original particel's propagator with a very big "mass". The starting point is the loop integral
      \begin{eqnarray}
      \int\frac{dk^4}{(2\pi)^4}\frac{1}{k^2-m^2}-D_{1}\int\frac{dk^4}{(2\pi)^4}\frac{1}{k^2-M_{1}^2}-D_{2}\int\frac{dk^4}{(2\pi)^4}\frac{1}{k^2-M_{2}^2}\label{paulivillars}
      \end{eqnarray}
      with the conditions 
      \begin{eqnarray}
      D_{1}+D_{2}=1 \label{cond01}
      \end{eqnarray}
      and
      \begin{eqnarray}
      D_{1}M_{1}^2+D_{2}M_{2}^2=m^2\label{cond02}.
      \end{eqnarray}
      The additional propagators are the regulators with "heavy masses" $M_{i}$. After the integration these conditions are directly needed.
      \begin{eqnarray}
      \frac{-i}{16\pi^2}\left(\Lambda^2-m^2\ln\left[\frac{\Lambda^2}{m^2}+1\right]-D_{1}\Lambda^2+D_{1}M_{1}^{2}\ln\left[\frac{\Lambda^2}{M_{1}^2}\right]-D_{2}\Lambda^2+D_{2}M_{2}^{2}\ln\left[\frac{\Lambda^2}{M_{2}^2}\right]\right)\nonumber\\
      =\frac{-i}{16\pi^2}\left(\Lambda^2-D_{1}\Lambda^2-D_{2}\Lambda^2-m^2\ln\left[\frac{\Lambda^2}{m^2}\right]+D_{1}M_{1}^{2}\ln\left[\frac{\Lambda^2}{M_{1}^2}\right]+D_{2}M_{2}^{2}\ln\left[\frac{\Lambda^2}{M_{2}^2}\right]\right)
      \end{eqnarray}
      Together with (\ref{cond01}) $\Lambda^2-D_{1}\Lambda^2-D_{2}\Lambda^2=0$ and
      \begin{eqnarray} D_{1}M_{1}^{2}\ln\left[\frac{\Lambda^2}{M_{1}^2}\right]+D_{2}M_{2}^{2}\ln\left[\frac{\Lambda^2}{M_{2}^2}\right]\approx(D_{1}M_{1}^{2}+D_{2}M_{2}^{2})\ln\left[\frac{\Lambda^2}{M^2}\right]=m^{2}\ln\left[\frac{\Lambda^2}{M^2}\right]
      \end{eqnarray} 
      the result for the regularized loop integral is
      \begin{eqnarray}
      \frac{-i}{16\pi^2}\left(-m^2\ln\left[\frac{M^2}{m^2}\right]\right).
      \end{eqnarray}
      In comparison with (\ref{power}) the power term $\Lambda^2$ does not appear. Therefore this term is regularization dependend.\\ 
      However, the renormalization procedure shown in the paragraph about the cut-off method holds here too. The physical quantity, the mass does not change, because the shift due to the loop is unobservable.\\
      The phenomenon just observed is of general nature and can be comprised as follows. Different regularization procedures can result in different expressions for the same diagram, but the renormalized observable quantities are not allowed to change if the regularization procedure is changed. Unless this condition is satisfied the theory used would make no sense at all. 
\item The method of dimensional regularization:\\ \\
      Another method to regularize the divergence in (\ref{tadpolefull}) is to extend the integral in 4 dimensions to an integral in $n$ dimensions. The extended integral is divergent for even $n$  with $n\geq4$. The loop integral in $n$ dimensions is given by
      \begin{eqnarray}
       \int\frac{dk^n}{(2\pi)^n}\mu^{4-n}\frac{i}{k^2-m^2}=\frac{1}{(4\pi)^{\frac{n}{2}}}\mu^{4-n}\frac{\Gamma(1-\frac{n}{2})}{\left( m^2\right)^{1-\frac{n}{2}}}=\nonumber\\\frac{m^2}{16\pi^2}\left[-\frac{2}{4-n}-\gamma-1-\ln(4\pi)+\ln\left( \frac{m^2}{\mu^2}\right) +\mathcal{O}(4-n)\right] \label{dimensional}
       \end{eqnarray}
       where $\gamma$ is the Euler-Mascheroni constant and $\mu^{4-n}$ is introduced to keep the dimension of the expression constant with respect to $n$. The integration can again be performed with the Wick rotation. Thus, the third method to regularize the integral gave another expression that differs from the results of method one and method two. All methods up to now gave expressions that where different from each other. Hence, the expressions are regularization dependend.\\ 
       The dimensional regularized loop diagram has to be renormalized. Therefore it has to be subtracted. Two schemes to do that are very common the $MS$ and the $\overline{MS}$ scheme.
       In the $MS$ scheme just the pole term $-\frac{2}{4-n}$ is subtracted while in the $\overline{MS}$ scheme $-\frac{2}{4-n}-\gamma$ is subtracted. Therefore the renormalized integrals are scheme dependent. This example shows that not only regularization but also renormalization can be scheme dependent.\\ 
       Again the physical quantities are not affected from such scheme dependences. The tadpole diagram in dimensional regularization (\ref{dimensional}) has to be summed up as shown in the cut-off method before it is subtracted. The scheme dependence does not show up in the renormalized mass. It is just shifted from the bare to the physical value.  
\end{enumerate}
The methods shown here are convenient for the calculation of one loop diagrams. A criteria which constrains the choice of the renormalization technique is if the symmetries of the theory in which the calculations are performed are still valid after renormalization.

\section{The operator product expansion (OPE) \label{OPE}}
This section provides a basic introduction of the operator product expansion. The expansion is derived in two ways where the first is the physical introduction and the second one the mathematical introduction. After that issues which are important in the application of the OPE are considered. Finally questions how the OPE should be truncated and how it can be simplified are addressed.

\subsection{Features and their Derivation}

The purpose of this section is to establish a first overview of the OPE and topics connected to it. The most important features of this expansion are explained and derived in the following sequence.
\begin{enumerate}
\item {The Operator Identity}\\ \\
      In 1969 Wilson proposed a Non-Lagrangian model of current algebra \cite{Wilson:1969zs}. The basis of his model is the 
      assumption that the singular part as $x\rightarrow y$ of the product A(x)B(y) of two operators is given by a sum  
      over local operators
      \begin{eqnarray}
      A(x)B(y)\rightarrow\sum_{n}C^{AB}_{n}(x-y)O_{n}(y) \label{ope}
      \end{eqnarray}
      where $C^{AB}_{n}(x-y)$ are singular c-number functions and $O_{n}$ are operators. This relation is derived by using the formalism of the 
      path-integral. This access to the OPE gives valuable physical insight about the OPE, in particular on the separation of 
      long and short distance fluctuations. The remarkable thing about the OPE is that it is an operator relation; this means, in 
      applying it to any matrix element 
      $\bra{\beta}A(x)B(x)\ket{\alpha}$, the same functions $C^{AB}_{n}(x-y)$ are received for all states $\ket{\alpha}$ and 
      $\ket{\beta}$.
\item{Derivation in the Operator Formalism}\\ \\
      The OPE can also be derived in the operator formalism of quantum field theory. This derivation deals with the 
      singularities of the operator product $ A(x)B(y)$ in the limit $x\rightarrow y$. These singularities have to be separated 
      from each other. This operation produces a series in which the summands are sorted by the strength of their 
      singularities which coincides with their importance. The series derived in this scheme is identical with the OPE derived 
      by using the path-integral formalism.   
\item{Important and Unimportant Summands}\\ \\
      An expansion with infinitely many summands that can not be resummed is useless if it can not be truncated. Using scale arguments it can be shown 
      that the OPE can be truncated.\\ 
      Dimensional analysis suggests that $C^{AB}_{n}(x-y)$ behaves for $x\rightarrow y$ like the power $d_{O}-d_{A}-d_{B}$ of 
      $x-y$, where $d_{i}$ is the dimension  of the operator i in powers of mass or momentum. Since $d_{i}$ increases as we 
      add more fields or derivatives to an operator $O$, the strength of the singularity of $C^{AB}_{n}(x-y)$ decreases for 
      operators C of increasing complexity.\\
      The decrease of the singularity in equation (\ref{ope}) with operators $O(y)$ of increasing complexity is the reason that 
      justifies the truncation of the OPE. Generally speaking the OPE is useful in drawing conclusions about the behavior of 
      the product A(x)B(y) for $x\rightarrow y$.\\ 
\item{Translation to Perturbation Theory}\\ \\
      For application purposes the OPE is translated into the language of Feynman diagrams. The Wilson coefficients are 
      displayed as a series of 
      graphs that contain common symbols and symbols representing the connection to the matrix elements. While the matrix elements 
      are simply multiplied 
      with the expression for the corresponding Wilson coefficient without any instruction how they should be calculated.  Special attention is payed to the fact that high and low frequencies have to be separated. This leads to the introduction of a renormalization point $\mu$.
\end{enumerate}

\subsection{The Operator Identity \label{pathderivation}}
\begin{figure}[htbp]
 \begin{center}
  \begin{picture}(0,0)%
\includegraphics{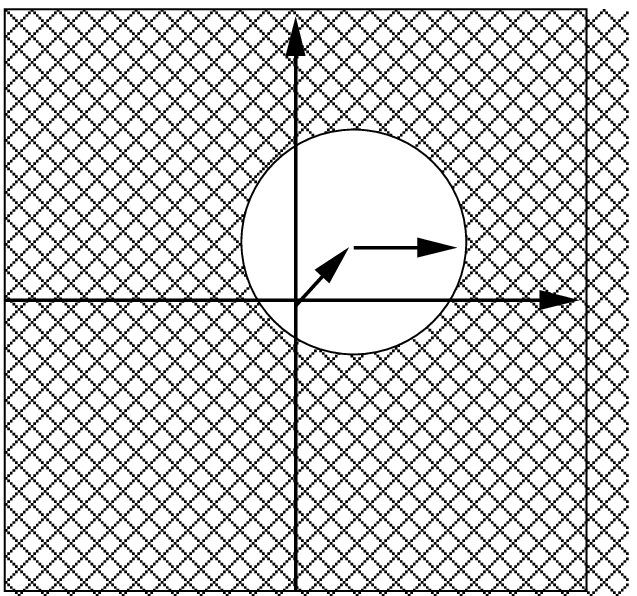}%
\end{picture}%
\setlength{\unitlength}{4144sp}%
\begingroup\makeatletter\ifx\SetFigFont\undefined%
\gdef\SetFigFont#1#2#3#4#5{%
  \reset@font\fontsize{#1}{#2pt}%
  \fontfamily{#3}\fontseries{#4}\fontshape{#5}%
  \selectfont}%
\fi\endgroup%
\begin{picture}(2777,2870)(2571,-4058)
\put(3923,-1296){$\vec{e}_{2}$}%
\put(4348,-2413){$R$}%
\put(4109,-2440){$x$}%
\put(5333,-2786){$\vec{e}_{1}$}%
\end{picture}%

  \caption{Partitioning of $\mathds{R}^2$ into a Ball of radius $R$ around the point $x$ and the remainder of the space. This picture has to be treated with care because in the 2 dimensional Minkowskispace a sphere is not a sphere. This point is going to be elaborated in section (\ref{wilsoncoefficient}).\label{partitioningR2}}
  \end{center}
 \end{figure}

In this section a generalized version of the Wilson expansion is derived. The derivation is close to \cite{Weinberg:1996kr}. For this purpose, let us consider a Greens function for local operators $A_{1}(x_{1}),A_{2}(x_{2})$, etc. whose arguments approach a point x, as well as other local operators $B_{1}(y_{1}),B_{2}(y_{2})$, etc. with fixed arguments:
\begin{eqnarray}
\bra{0}T\{A_{1}(x_{1}),A_{2}(x_{2}),...B_{1}(y_{1}),B_{2}(y_{2}),...\}\ket{0}\nonumber\\=\int\left[\Pi_{l,z} d\phi_{l}(z)\right]a_{1}(x_{1})a_{2}(x_{2})...b_{1}(y_{1})b_{2}(y_{2})...exp(iI[\phi]) \label{opepath}
\end{eqnarray}
where the lower-case letters a and b indicate replacement of the field operators, the A's and B's, with c-number fields $\phi$. The operators $A_{1}(x_{1}),A_{2}(x_{2})$, etc. form the operator product which should be expanded, while the operators $B_{1}(y_{1}),B_{2}(y_{2})$, etc. represent external states. The c-number fields $\phi$ are the field operators of the considered theory.\\  
Now surround the point x with a sphere B(R) of radius R which is much larger than the separations among the $x,x_{1},x_{2}$, etc. but much smaller than the separations between $x,y_{1},y_{2}$, etc.  (see figure \ref{partitioningR2}). Since the action is local, it may be written as 
\begin{eqnarray}
I=\int_{z\in B(R)} d^4z~\mathscr{L}(z)+\int_{z\not\in B(R)} d^4z~\mathscr{L}(z).
\end{eqnarray}
Equation (\ref{opepath}) can be put in the form
\begin{eqnarray}
\bra{0}T\{A_{1}(x_{1}),A_{2}(x_{2}),...B_{1}(y_{1}),B_{2}(y_{2}),...\}\ket{0}=\nonumber\\\int\left[\Pi_{l,z\not\in B(R)} d\phi_{l}(z)\right]b_{1}(y_{1})b_{2}(y_{2})...exp\left(i\int_{z\not\in B(R)}d^4 z\mathscr{L}(z)\right)\nonumber\\ \times\int\left[\Pi_{l,z\in B(R)} d\phi_{l}(z)\right]a_{1}(x_{1})a_{2}(x_{2})...exp\left(i\int_{z\in B(R)}d^4 z\mathscr{L}(z)\right) \label{opesep}
\end{eqnarray}
in which the only influence that the fields inside and outside the sphere have to each other is given by the boundary condition at the surface of the sphere. 
The two fields have to be connected continuously to each other. Thus, the boundary conditions are that the field or the n-th derivative of the field is continuous on the surface of the sphere. The fields inside and outside the sphere have the same value for a certain point $u$ on the sphere's surface.  Except of this, the dynamics of the fields are completely independent from each other. They could even be described by two different theories. Hence the amplitude has been split into two domains.\\
This fact is going to be exploited to simplify the expansion that was just performed. The integral over the fields inside the sphere are fully determined by the values and derivatives of the fields on the surface of the sphere, which in turn may be expressed in terms of the fields and their derivatives extrapolated from outside the sphere to the interior point x. Hence, after this conversion the integral over the sphere will not depend on specific coordinates of the spheres surface but just on its radius $R$.\\
 \begin{figure}[htbp]
 \begin{center}
  \begin{picture}(0,0)%
\includegraphics{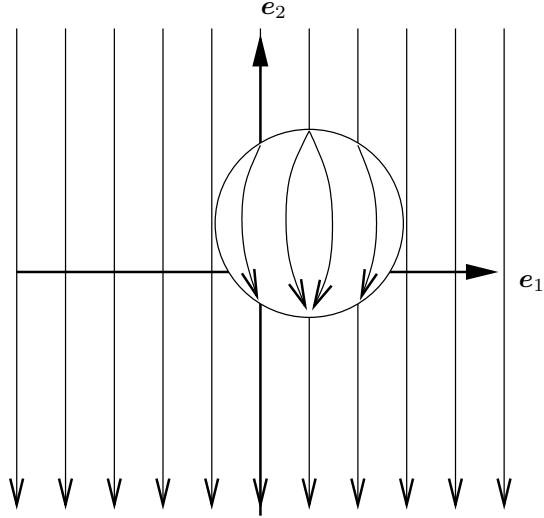}%
\end{picture}%
\setlength{\unitlength}{4144sp}%
\begingroup\makeatletter\ifx\SetFigFont\undefined%
\gdef\SetFigFont#1#2#3#4#5{%
  \reset@font\fontsize{#1}{#2pt}%
  \fontfamily{#3}\fontseries{#4}\fontshape{#5}%
  \selectfont}%
\fi\endgroup%
\begin{picture}(3070,3142)(1268,-3339)
\put(2777,-313){$\vec{e}_{2}$}%
\put(4323,-1946){$\vec{e}_{1}$}%
\end{picture}%

  \caption{A simple example of a field that can be split up as shown in figure (\ref{partitioningR2}). The field is constant outside the sphere but it has a gradient inside the sphere.\label{boundarycondition01}}
  \end{center}
  \end{figure}
In the simple example that is shown in figure (\ref{boundarycondition01}) the extrapolation is quite easy. The field is a constant outside the sphere so the extrapolation to the point x will be the same constant. This constant has already been the boundary condition on the spheres surface. In conclusion the boundary conditions stay the same.\\
The integral over the fields inside the sphere contains the part of the amplitude where short distances are involved while the integral over the fields outside the sphere contains the part of the amplitude where long distances are involved. In addition to this separation the two domains do not share any variables.\\
If we express this integral as a series in products of o(x) and the c-number fields $\phi$ and their derivatives extrapolated to x, the coefficients can only be functions $C^{A_{1},A_{2},...}_{n}(x_{1}-x,x_{2}-x,...)$ of the coordinate differences and $R$. Since the points $y_{1}$, $y_{2}$, etc. are all far outside the sphere, the sphere has no effect in the limit $R\ll \abs{x-y_{i}}$, so in this limit equation (\ref{opesep}) becomes
\begin{eqnarray}
\bra{0}T\{A_{1}(x_{1}),A_{2}(x_{2}),...B_{1}(y_{1}),B_{2}(y_{2}),...\}\ket{0}\rightarrow\nonumber\\
\int\left[\Pi_{l,z} d\phi_{l}(z)\right]b_{1}(y_{1})b_{2}(y_{2})...exp\left(i\int d^4 z\mathscr{L}(z)\right)\nonumber\\ \times
\sum_{n}C^{A_{1},A_{2},...}_{n}(x_{1}-x,x_{2}-x,...)o_{n}(x)\nonumber\\
=\sum_{n}C^{A_{1},A_{2},...}_{n}(x_{1}-x,x_{2}-x,...)\mele{T\{O_{n}(x),B_{1}(y_{1}),B_{2}(y_{2}),...\}}
\label{opelim}
\end{eqnarray}
for $x_{1}$, $x_{2}$, etc. all approaching x, where $O(x)$ is the quantum-mechanical Heisenberg-picture operator corresponding to $o_{n}(x)$. The operators that occur in this expression depend solely on x. Therefore, they are called local operators, a non-local operator would depend on x and further coordinates.\\ 
In comparison with (\ref{opesep}) the integral over the fields inside the sphere has changed drastically. It has separated into summands which are in turn separated in two factors. The first one is a coefficient function while the second one contains c-number fields and their derivatives. While the integral over the fields outside the sphere stayed nearly the same. The operator $O_{n}$ that stems from the integral over the fields inside the sphere connects the long and short distance part of the amplitude.\\ 
This yields by Fourier transforming (\ref{opelim}) with respect to the y variables and multiplying with appropriate coefficient  functions
\begin{eqnarray}
\bra{\beta}T\{A_{1}(x_{1}),A_{2}(x_{2}),...\}\ket{\alpha}\rightarrow\sum_{n}C^{A_{1},A_{2},...}_{n}(x_{1}-x,x_{2}-x,...)\bra{\beta}O_{n}(x)\ket{\alpha}
\end{eqnarray}
for arbitrary states $\ket{\alpha}$ and $\bra{\beta}$, of course the states correspond to the $B$ operators. Because this applies for arbitrary states, it is the operator product expansion in a generalized version:
\begin{eqnarray}
T\{A_{1}(x_{1}),A_{2}(x_{2}),...\}\rightarrow\sum_{n}C^{A_{1},A_{2},...}_{n}(x_{1}-x,x_{2}-x,...)O_{n}(x).
\end{eqnarray}
It is very important to understand that the OPE arises only in the limit $R\ll \abs{x-y_{i}}$. This is a special and not the general case. That means the applications of the OPE are limited to matrix elements where you have the distinct separation between the distances $\abs{x-x_{i}}$ and $\abs{x-y_{i}}$ of the form
\begin{eqnarray}
\abs{x-x_{i}}\ll\abs{x-y_{i}}.
\end{eqnarray}

\subsection{Derivation in the operator formalism \label{operatorderivation}}

The derivation in section \ref{pathderivation} provides a very intuitive introduction of the OPE. Many features of the expansion can simply be read off the proof. Some of them where already claimed. Still missing is the scheme that orders the sum appearing in the OPE. Also the analytic properties of the coefficient functions have not been analyzed. This is done in the present section via an alternative derivation of the OPE. This section is close to \cite{brandeis}.\\
A product of two operators is considered 
 \begin{eqnarray}
 P(x_{\mu},\xi_{\mu})=T\{A_{1}(x_{\mu}+\xi_{\mu}),A_{2}(x_{\mu}-\xi_{\mu})\} \label{productdef}.
 \end{eqnarray}
The vector $\xi_{\mu}$ is split into spherical coordinates:
\begin{eqnarray}
\rho=\sqrt{\xi_{\mu}\xi^{\mu}},~~~\eta_{\mu}=\frac{\xi_{\mu}}{\rho}.
\end{eqnarray}
Henceforth $P$ is considered as a function of $x_{\mu},\eta_{\mu}$ and $\rho$ :
\begin{eqnarray}
 P(x_{\mu},\xi_{\mu})=P(x_{\mu},\eta_{\mu},\rho).
 \end{eqnarray}
 If $P$ diverges for $\rho\rightarrow 0$ we define an operator 
 \begin{eqnarray}
 O_{1}(x_{\mu},\eta_{\mu})=\lim_{\rho\rightarrow 0}\frac{P(x_{\mu},\eta_{\mu},\rho)}{C_{1}(\rho)} \label{operatordef}
 \end{eqnarray}
in the weak limit, the limit performed for the matrix elements and not for the norm of the operator. This limit should be finite. Hence, $C_{1}(\rho)$ has to compensate the divergence of
 $P(x_{\mu},\eta_{\mu},\rho)$ at $\rho=0$. Therefore, $C_{1}(\rho)$ has to have a singularity at $\rho=0$ by
 itself. $C_{1}(\rho)$ is a suitable function. The singularity of $C_{1}$ is restricted by the
 condition that the result be finite and different from zero. Under the assumption that there are some matrix elements 
 of $P$ 
 \begin{eqnarray}
 \bra{\Phi'}P(x'_{\mu},\eta'_{\mu},\rho)\ket{\Psi'} \label{planewavederivative}
 \end{eqnarray}
 which near $\rho=0$ are as singular as or more singular than all other matrix elements of $P$ a 
 suitable $C_{1}$ can be found. The mathematical term for this requirement is
 \begin{eqnarray}
 \lim_{\rho\longrightarrow 
 0}\frac{\bra{\Phi}P(x_{\mu},\eta_{\mu},\rho)\ket{\Psi}}{\bra{\Phi'}P(x'_{\mu},\eta'_{\mu},\rho)\ket{\Psi'}}
 \end{eqnarray}
 exists for any $\phi,\psi,x_{\mu}$ and $\eta_{\mu}$. Choosing any of these most singular matrix elements, 
 equation (\ref{planewavederivative}), as the function $C_{1}$ the operator (\ref{operatordef}) should have 
 finite and non-vanishing matrix elements.\\
 By (\ref{productdef}) and (\ref{operatordef}) the operator $O_{1}$ is local in the sense that
 \begin{eqnarray}
 \left[ A_{j}(x),O_{1}(y,\eta)\right]_{\pm}=0 \nonumber\\
 \left[ O_{1}(x,\eta),O_{1}(y,\eta')\right]_{\pm}=0
 \label{localcond}
 \end{eqnarray}
 for $(x-y)^2<0$ with commutators or anticommutators taken appropriately.\\
 In many cases $O_{1}$ turns out to be a multiple of the identity. In perturbation theory it seems to 
 be a general rule that the most singular part of a matrix element is given by\footnote{There may be several equally singular matrix elements. This possibility is not considered here.}
 \begin{eqnarray}
 \bra{\Phi}P(x,\eta,\rho)\ket{\Psi}\approx\bra{0}P(x,\eta,\rho)\ket{0}\mele{\Phi|\Psi}
 \end{eqnarray}
 provided the vacuum expectation value does not vanish
 \begin{eqnarray}
\bra{0}P(x,\eta,\rho)\ket{0}\not = 0.
 \end{eqnarray}
 This leads to the determination of $O_{1}$
 \begin{eqnarray}
 \lim_{\rho\longrightarrow 
 0}\frac{\bra{\Phi}P(x,\eta,\rho)\ket{\Psi}}{\bra{0}P(x',\eta',\rho)\ket{0}}=\mele{\phi\psi}
 \end{eqnarray}
 or
 \begin{eqnarray}
 O_{1}=\mathds{1}.
 \end{eqnarray}
 If the vacuum expectation value vanishes
 \begin{eqnarray}
 \mele{P(x,\eta,\rho)}=0
 \end{eqnarray}
 (\ref{operatordef}) may lead to a suitable definition, but we will see that in general there are more 
 local operators which should be associated with $P$.\\
 We now turn to the second problem of analyzing the singularities of the operator product. To this end 
 the remainder is introduced
 \begin{eqnarray}
 r_{1}(x,\eta,\rho)=\frac{P(x,\eta,\rho)}{C_{1}(\rho)}-O_{1}(x,\eta) \label{remainder01}
 \end{eqnarray}
 which vanishes in the limit 
 \begin{eqnarray}
 \lim_{\rho\rightarrow 0}r_{1}(x,\eta,\rho)=0
 \end{eqnarray}
 because of (\ref{operatordef}). In order to get some information about the singularities of $P$ near 
 $\xi=0$ we multiply (\ref{remainder01}) by $C_{1}(\rho)$ and solve for $P$
 \begin{eqnarray}
  P(x,\eta,\rho)=C_{1}(\rho)O_{1}(x,\eta)+P_{2}(x,\eta,\rho)\label{a}
 \end{eqnarray}
 with
 \begin{eqnarray}
 P_{2}(x,\eta,\rho)=C_{1}(\rho)r_{1}(x,\eta,\rho)\label{P2}.
 \end{eqnarray}
 Here $C_{1}\rightarrow\infty$ and $r_{1}\rightarrow 0$ for $\rho\rightarrow 0$. Hence nothing can be 
 said in general about the limit of $P_{2}$. If 
 \begin{eqnarray}
 \lim_{\rho\rightarrow 0}P_{2}=0
 \end{eqnarray}
 (\ref{P2}) already gives complete information on the singularities of $P$ near $\xi=0$
 \begin{eqnarray}
 P(x,\xi)=C_{1}\left(\rho\right)O_{1}\left(x,\eta\right) +P_{2}(x,\eta,\rho)
 \end{eqnarray}
 with $\lim_{\rho\rightarrow 0}P_{2}\left(x,\eta,\rho\right)=0$ for $\eta$ constant. Hence the term 
 $C_{1}O_{1}$ carries the total singularity of $P$.\\ If $P_{2}$ diverges the product $P$ has 
 additional singularities near $\xi=0$. In that case its possible to repeat the procedure for $P_{2}$ that was just applied to $P$ with
 \begin{eqnarray}
 P_{2}(x,\eta,\rho)= P(x,\eta,\rho)-C_{1}(\rho)O_{1}(x,\eta).
 \end{eqnarray}
 This time the most singular matrix element $C_{2}$ of $P_{2}$ is chosen to form
 \begin{eqnarray}
 O_{2}(x,\eta)=\lim_{\rho\rightarrow 0}\frac{P_{2}(x,\eta,\rho)}{C_{2}(\rho)}.
 \end{eqnarray}
 Hence, more information on the singularities of $P$ is acquired. Introducing the remainder 
\begin{eqnarray}
r_{2}(x,\eta,\rho)=\frac{P_{2}(x,\eta,\rho)}{C_{2}(\rho)}-O_{2}(x,\eta)
\end{eqnarray}
with
\begin{eqnarray}
\lim_{\rho\rightarrow 0}=r_{2}(x,\eta,\rho)=0.
\end{eqnarray}
An Operator analogous to $P_{2}$ can be introduced 
\begin{eqnarray}
P_{3}=r_{2}C_{2}.  
\end{eqnarray}
In conclusion it is obtained
\begin{eqnarray}
P_{2}=C_{2}O_{2}+P_{3} \label{P2new}.
\end{eqnarray}
Inserting (\ref{P2new}) into (\ref{a}) it is obtained
\begin{eqnarray}
P(x,\eta,\rho)=C_{1}(\rho)O_{1}(x,\eta)+C_{2}(\rho)O_{2}(x,\eta)+P_{3}(x,\eta,\rho).
\end{eqnarray}
$C_{2}$ is less singular than $C_{1}$
\begin{eqnarray}
\lim_{\rho\rightarrow 0}\frac{C_{2}(\rho)}{C_{1}(\rho)} =0 \label{singord}.
\end{eqnarray}
This procedure has refined our analysis of the singularities of $P$, (\ref{singord}) follows from
\begin{eqnarray}
\lim_{\rho\rightarrow 0}\frac{C_{2}(\rho)}{C_{1}(\rho)}=\lim_{\rho\rightarrow 0}\frac{\bra{\phi}P_{2}(x,\eta,\rho)\ket{\psi}/C_{1}(\rho)}{\bra{\phi}P_{2}(x,\eta,\rho)\ket{\psi}/C_{2}(\rho)}=\lim_{\rho\rightarrow 0}\frac{\bra{\phi}r_{2}(x,\eta,\rho)\ket{\psi}}{\bra{\phi}P_{2}(x,\eta,\rho)\ket{\psi}/C_{2}(\rho)}=0
\end{eqnarray}
because $\lim_{\rho\rightarrow 0}r_{2}(x,\eta,\rho)=0$ and $\lim_{\rho\rightarrow 0}\frac{P_{2}(x,\eta,\rho)}{C_{2}(\rho)}=O_{2}(x,\eta)\not =0$ for at least one matrix element.\\
This process can be iterated until a $P_{n}$ is derived which vanishes at $\xi=0$. It is an assumption that the process terminates after a finite number of steps. The result of the iteration is the expansion
\begin{eqnarray}
P(x,\eta,\rho)=C_{1}(\rho)O_{1}(x,\eta)+...+C_{n}(\rho)O_{n}(x,\eta)+R(x,\eta,\rho)
\end{eqnarray}
with the following properties
\begin{eqnarray}
\lim_{\rho\rightarrow 0}\frac{C_{i+1}(\rho)}{C_{i}(\rho)}=0 \label{feature01}
\end{eqnarray}
\begin{eqnarray}
\lim_{\rho\rightarrow 0}C_{n}(\rho)=\infty~or~\lim_{\rho\rightarrow 0}C_{n}(\rho)=c\not =0
\end{eqnarray}
\begin{eqnarray}
\lim_{\rho\rightarrow 0}R(x,\eta,\rho)=0\label{feature03}.
\end{eqnarray}
The operators $O_{j}$ are given by 
\begin{eqnarray}
O_{j}=\lim_{\rho\rightarrow 0}\frac{P-\sum_{\alpha=1}^{j-1}C_{\alpha}O_{\alpha}}{C_{j}}
\end{eqnarray}
and satisfy the locality conditions (\ref{localcond}).\\ 
The choice of the matrix elements that are used as the $C_{i}$ is not unique, in turn the operators $O_{i}$ are not unique. This arbitrariness can be reduced by requiring that the Operators $O_{j}$ are linearly independent. The transformation that has to be performed to achieve this can always be arranged without changing the general features of the functions $C_{j}$. An expansion with linearly independent Operators $O_{i}$ and coefficient functions that have the features listed in (\ref{feature01})-(\ref{feature03}) is called a standard expansion.\\
This alternative derivation of the OPE gives insight in the analytic properties of the coefficient functions. Obviously the analysis of singularities in the operator product $P(x,\eta,\rho)$ at $\rho=0$ provide an alternative approach to the OPE, in addition to the separation of the amplitude in short and long distance fluctuations.

\subsection{Important and unimportant summands \label{importantunimportant}}

The operator products $P$ that are investigated have a certain dimension d. Wilson's assumption about the OPE leads to the statement that only operators $O_{n}$ with dimension $\leq d$ are relevant in the OPE. This statement is derived by scale arguments. The chain of arguments is shown in the following \cite{brandeis}. The starting point is the assumption that a scale invariant theory is investigated. This means the theory is invariant under the transformation:
\begin{eqnarray}
x\longrightarrow sx \label{invariance}.
\end{eqnarray} 
In the quantum mechanical sense this means that there is a family of unitary transformations $U(s)$ with the property:
\begin{eqnarray}
U^{-1}(s)\phi(x)U(s)=s^{d(\phi)}\phi(sx) \label{invariance2}
\end{eqnarray}
for any local field operator of the theory. The real number $d(\phi)$ is called the dimension of the operator $\phi$. Using this transformation law Wilson determined the singularities of the coefficients in the following way. Applying the transformation $U(s)$ to the expansion 
\begin{eqnarray}
A_{1}(x+\xi)A_{2}(x-\xi)=\sum_{j=1}^{\infty}C_{j}(\xi)O_{j}(x)
\end{eqnarray}
we get
\begin{eqnarray}
U^{-1}(s)A_{1}(x+\xi)A_{2}(x-\xi)U(s)=\sum_{j=1}^{\infty}U^{-1}(s)C_{j}(\xi)O_{j}(x)U(s) \nonumber\\
U^{-1}(s)A_{1}(x+\xi)A_{2}(x-\xi)U(s)=\sum_{j=1}^{\infty}C_{j}(\xi)U^{-1}(s)O_{j}(x)U(s) \nonumber\\
s^{d(A_{1})+d(A_{2})}A_{1}(sx+s\xi)A_{2}(sx-s\xi)=\sum_{j=1}^{\infty}C_{j}(\xi)s^{d(O_{j})}O_{j}(sx).
\end{eqnarray}
Expanding $A_{1}A_{2}$ on the left hand side yields
\begin{eqnarray}
s^{d(A_{1})+d(A_{2})}\sum_{j}C_{j}(s\xi)O_{j}(sx)=\sum_{j=1}^{\infty}C_{j}(\xi)s^{d(O_{j})}O_{j}(sx).
\end{eqnarray}
Since the $O_{j}$ are linearly independent the following equations are valid
\begin{eqnarray}
C_{j}(\xi)=s^{d(A_{1})+d(A_{2})-d(O_{j})}C_{j}(s\xi)
\end{eqnarray}
as the scaling law of the coefficients. The exponent of $s$ is called the dimension of $C_{j}$ and we have the result that in each term of the expansion the dimension of $C_{j}$ and the dimension of $O_{j}$ must add up to the total dimension of the left hand side
\begin{eqnarray}
d(C_{j})+d(O_{j})=d(A_{1})+d(A_{2})=d(A_{1}A_{2}).
\end{eqnarray}
The dimension of $C_{j}$ indicates the behavior for $\xi\rightarrow 0$:
\begin{eqnarray}
C_{j}(s\xi)=s^{-d(C_{j})}C_{j}(\xi)
\end{eqnarray}
or
\begin{eqnarray}
C_{j}(x)=\abs{x}^{-d(C_{j})}C_{j}(\frac{x}{\abs{x}}).
\end{eqnarray}
As a consequence the $C_{j}(x)$ are singular or $\not =0$ in the limit $x\rightarrow 0$ only if $d(O_{j})\leq d(A_{1}A_{2})$. Hence the singular and non-vanishing terms are provided by operators $O_{j}$ of dimension less or equal to the dimension of the product $A_{1}A_{2}$.\\
All the terms belonging to operators with $d(O_{j})\geq d(A_{1}A_{2})$ vanish in the limit  $x\rightarrow 0$. The conclusion is that all Operators with $d(O_{j})\geq d(A_{1}A_{2})$ are not relevant for the OPE.\\
The simple power-counting argument above is modified by renormalization effects; the expansion (\ref{ope}) must be 
formulated in terms of operators renormalized at some scale $\mu$, and then $\mu$ appears along with $x-y$ in the 
coefficient function $C^{AB}_{n}(x-y)$. An important example are asymptotically free field theories, where $C^{AB}_{n}(x-y)$ does behave like the power $d_{O}-d_{A}-d_{B}$ of $x-y$ suggested by dimensional analysis only up to a power of $ln(x-y)^2$. \\
The corresponding statement in momentum space is that for $k\rightarrow\infty$,
\begin{eqnarray}
\int d^4 xe^{-ik^{\mu}x_{\mu}}A(x)B(0)\rightarrow\sum_{n}V^{AB}_{n}(k)O_{n}(0)
\end{eqnarray}
and correspondingly 
\begin{eqnarray}
\int d^4 xe^{-ik^{\mu}x_{\mu}}T\{A(x)B(0)\}\rightarrow\sum_{n}C^{AB}_{n}(k)O_{n}(0)
 \end{eqnarray}
where $V^{AB}_{n}(k)$ and $C^{AB}_{n}(k)$ are functions of $k^{\mu}$ that for large $k^2$ decrease more and more rapidly for more and more complicated terms in the series.\\
A short remark concerning the validity of the arguments given above has to be made. The basis of all arguments have been the equations (\ref{invariance}) and (\ref{invariance2}). Unfortunately dimensional arguments do not work in every theory, in some theories they are not valid. In all realistic theories scale invariance is broken as an example any mass term breaks the scale invariance. Furthermore it can happen that the dimensionality $d$ can not be defined in a usefull way. Hence in such theories scale arguments can not be used in order to determine if the OPE can be truncated.\\
However, dimensional arguments work very well in perturbation theory. For the exact solution the situation may be quite different. This can be demonstrated in the Thirring model. There Wilson's hypothesis holds in perturbation theory with the conventional dimensions $d(A_{1}A_{2})=d(A_{1})+d(A_{2})$, but this is no longer true for the exact solution of the Thirring model \cite{Wilson:1970pq}. Hence, the conclusion is that the arguments above work in perturbation theory and perturbation theory will be used in the remainder of this thesis. Therefore, the arguments can be used.\\

\subsection{Translation to perturbation theory \label{transpert}}

The preceding sections about the OPE have been very general. Calculations in quantum field theory are mainly done via perturbative techniques. Therefore this section introduces the perturbative technique to calculate the OPE of a Greens function.\\
The first step is to define the operators in the OPE. The operators are given by combinations of the field operators appearing in the Lagrangian of the theory. In standard expansions (see section \ref{operatorderivation}) the operators are linearly independent. That means the operators in standard expansions are given by all possible combinations of field operators of the theory under consideration, that are linearly independent. In this and the following sections only standard expansions are used. These arguments fix the operators.\\
The second step is the calculation of the coefficients in Wilson's expansion, the so called Wilson coefficients. The recipe was already given in section \ref{operatorderivation}. There the Wilson coefficients have been identified with the coefficients of matrix elements in the operator product expansion. The singularities of these coefficient functions determined the operator they belonged to. This method has two weaknesses. The Wilson coefficients have not been determined uniquely. There can be several different Wilson coefficients which have singularities of the same strength. In perturbation theory the strength of the singularities of these Wilson coefficients is often not determinable. An alternative method is urgently needed.\\
The clue that provides this method is that the OPE is an operator relation and the operators $O_{n}$ are already known because we use a standard expansion. If an OPE is sandwiched between two external states
\begin{eqnarray}
\bra{\alpha}P\ket{\beta}=\bra{\alpha}\sum_{n}C_{n}O_{n}\ket{\beta}=\sum_{n}\bra{\alpha}C_{n}O_{n}\ket{\beta}= \sum_{n}C_{n}\bra{\alpha}O_{n}\ket{\beta}     
\end{eqnarray}
the Wilson coefficient $C_{n}$ is unaffected by the external states. This is the feature that is exploited in the calculation of the $C_{n}$. There are states that filter out certain coefficients when the OPE is sandwiched between them. The matrix elements of the operators $O_{n}$ with these states are all zero except of one or a few of them. In this section the perfect case in which just one matrix element is not zero is stressed. Suppose the matrix element with $O_{j}$ is not 0
\begin{eqnarray}
\bra{\alpha}P\ket{\beta}=C_{j}\bra{\alpha}O_{j}\ket{\beta}  \Leftrightarrow  C_{j}=\frac{\bra{\alpha}P\ket{\beta}}{\bra{\alpha}O_{j}\ket{\beta}}.
\end{eqnarray}
This is a perturbatively calculable expression but there is a restriction that requires a modification of the calculations usually made in perturbation theory. In the first derivation of the OPE, section \ref{pathderivation}, it was shown that the coefficients $C_{n}$ stem from a path-integral that is restricted to a Ball with radius $R$ in four dimensional space time. Therefore only short distance fluctuations play a role in the calculation of the coefficients. This is the point that allows the Wilson coefficients in many theories to be calculated via perturbative techniques. The dynamics which takes place at short distances happens at large momenta. In asymptotically free theories this is the domain where perturbative calculations are valid. The impact of this observation is that no small momenta  contributions occur in the Wilson coefficients. The scale $\mu$ that distinguishes high and low momenta is defined by additional conditions.\\
The concrete rules that follow are 
\begin{enumerate}
\item The external momenta $q^{\mu}_{i}$ must be bigger than $\mu$, $q_{i}^2>\mu$.
\item The virtual momenta in the loops of involved diagrams must be bigger than $\mu$.
\end{enumerate}
Otherwise the OPE is not applicable to the problem. The method just introduced is called the plane wave method. With these rules all Wilson coefficients are calculable, although the calculations are sometimes very difficult. There exist alternative methods but the plane wave method is very valuable because it provides a lot of physical insight in comparison with other methods.

\subsection{OPE in the real scalar $\phi^4$ theory \label{unbroken}}

In the last sections the groundwork concerning the OPE was done. This work should be substantiated in quantitative calculations. Additionally there appear complications like the question of the application in theories with broken symmetry. These points are treated in this section. The calculations should be as easy as possible. For this reason the most simple theory is used, the real scalar $\phi^4$ theory. The Lagrangian density is given by
\begin{eqnarray}
\mathscr{L}=\frac{1}{2}\left(\partial_{\mu}\phi\right)^2-\frac{1}{2}m_{0}^2\phi^2-\frac{\lambda_{0}}{4!}\phi^4
\label{phi4lagrange}
\end{eqnarray}
where the dimension of $\phi$ is energy. This theory has an intrinsic symmetry, it is a simple reflection symmetry
\begin{eqnarray}
\mathscr{L}\left(\phi\right) =\mathscr{L}\left(-\phi\right).
\end{eqnarray}
The Feynman rules are given in table \ref{erstetabelle}.
\begin{table}[htbp]
\begin{center}
\begin{tabular}{||c||}
\hline
vertex : $-i\lambda_{0}$\\
propagator : $\frac{i}{p^2-m_{0}^2+i\epsilon}$\\
\hline
\end{tabular}
\end{center}
\caption{Feynmanrules in the $\phi^4$-theory in the phase where the symmetry is not broken. 
The vertex is a four vertex.\label{erstetabelle}}
\end{table}
As an example the operator product $\phi\phi$ is expanded in a standard expansion. The set of linear independent Operators is given by
\begin{eqnarray}
\left\{\phi\phi,\phi\phi\phi\phi,\phi\phi\phi\phi\phi\phi,...\right\}.
\end{eqnarray}
Operators with an odd number of $\phi$'s can be excluded because they do not respect the reflection symmetry. The important terms in the OPE are the terms with $d\leq 2$ (see section \ref{importantunimportant}). With these informations the OPE is determined to be
\begin{eqnarray}
T\left\{\phi(q),\phi(q)\right\}=C_{1}\mathds{1}+C_{2}\phi(0)\phi(0) \label{phi2ope}.
\end{eqnarray}
This is the OPE for the propagator in the interacting $\phi^4$ theory. The OPE is going to be derived in first order in the coupling constant $\lambda_{0}$. It is convenient to start with the left hand side of the OPE (\ref{phi2ope}). The vacuum expectation value is given by
\begin{eqnarray}
D(q)=\bra{0}T\left\{\phi(q),\phi(q)\right\}\ket{0}=-i\int d^4xe^{iq^{\mu}x_{\mu}}\bra{0}T\{\phi(x)\phi(0)\}\ket{0}\nonumber\\=
\parbox[c]{2cm}{
\begin{fmffile}{phifreeprop}
\begin{fmfgraph*}(20,20)
 \fmfleft{i} 
 \fmfright{o}
 \fmf{plain}{i,o}
\end{fmfgraph*}
\end{fmffile}}+
\parbox[c]{2cm}{
\begin{fmffile}{phi4tadpole02}
\begin{fmfgraph*}(20,20)
 \fmfleft{i} 
 \fmfright{o}
 \fmf{plain}{i,v,v,o}
 \fmfdot{v}
\end{fmfgraph*}
\end{fmffile}}
=\frac{1}{q^2-m^2_{0}}\left[1+\frac{\lambda_{0}}{32\pi^2}\frac{-m^2_{0}ln\left(\frac{M^2}{m^2_{0}}\right)}{q^2-m^2_{0}}\right] \label{zz08}
\end{eqnarray}
(see section \ref{reguandreno}). This Greens function is the sum of the two terms in the OPE (\ref{phi2ope}). The OPE  can be obtained with the rules given in section \ref{transpert}. First, the requirement that q is large is exploited. The propagators are expanded in $\frac{m_{0}}{q}$  
\begin{eqnarray}
D(q^2)=\frac{i}{q^2-m_{0}^2}=\frac{i}{q^2}\cdot\frac{1}{1-\left(\frac{m_{0}}{q}\right)^2}=\underbrace{i\left(\frac{1}{q^2}+\frac{m_{0}^2}{q^4}+\frac{m_{0}^4}{q^6}+...\right.}_{geometric~series}.
\end{eqnarray}
In the propagator we keep terms, up to order $\frac{1}{q^4}$ in order to be consistent with the second term in (\ref{zz08})
 \begin{eqnarray}
D(q^2)=\frac{1}{q^2}+\frac{m^2}{q^4}+\frac{\lambda_{0}}{32\pi^2}\frac{-m^2_{0}ln\left(\frac{M^2}{m^2_{0}}\right)}{q^4}=\frac{1}{q^2}+\frac{m_{0}^2}{q^4}\left[ 1-\frac{\lambda_{0}}{32\pi^2}ln\left(\frac{M^2}{m^2_{0}}\right)\right] \label{propagator}.
\end{eqnarray}
Secondly, the loop integration has to be separated in a high and a low frequency part. This corresponds to a separation in short and long distances. Hence a scale $\mu$ is introduced which separates high and low frequencies, $high~frequencies>\mu>low~frequencies$
\begin{eqnarray}
D(q^2)=\frac{1}{q^2}+\frac{m_{0}^2}{q^4}\left[ 1-\frac{\lambda_{0}}{32\pi^2}ln\left(\frac{M^2}{\mu^2}\right)-\frac{\lambda_{0}}{32\pi^2}ln\left(\frac{\mu^2}{m^2_{0}}\right)\right] \nonumber\\=\frac{1}{q^2}+\frac{m_{0}^2}{q^4}\left[ 1-\frac{\lambda_{0}}{32\pi^2}ln\left(\frac{M^2}{\mu^2}\right)\right]-\frac{m_{0}^2}{q^4}\frac{\lambda_{0}}{32\pi^2}ln\left(\frac{\mu^2}{m^2_{0}}\right) \label{opeuntidy}.
\end{eqnarray}
This is the OPE of the propagator, although the components, Wilson coefficients and operators remain to be identified. The expansion could have been done in an alternative way by using the plane wave method, discussed in section \ref{transpert}, to calculate the Wilson coefficients $C_{1}$ and $C_{2}$. The plane wave method exploits the operator nature of the OPE. If the OPE is sandwiched between appropriate states a certain summand in the expansion is singled out. Based on that the Wilson coefficient of the summand can be calculated, but the calculations can also be comprised into diagrammatic techniques. If this is done the Wilson coefficients are expressed as diagrams. The analytic result corresponding to the diagrams are the Wilson coefficients. The states for $C_{1}$ would be $\bra{0}$ and $\ket{0}$ and for $C_{2}$ $\bra{p}$ and $\ket{p}$. The calculation of $C_{2}$ is simple, the corresponding diagram is given by
\begin{eqnarray}
\begin{fmffile}{operatorphiphi01}
\begin{fmfgraph*}(40,40)
 \fmfpen{thick} 
 \fmfleft{i} 
 \fmfright{o}
 \fmf{plain}{i,v,o}
 \fmf{phantom,tag=1}{v,v}
 \fmfposition
 \fmfipath{p[]}
 \fmfiset{p1}{vpath1(__v,__v)}
 \fmfi{plain}{subpath (0,length(p1)/4) of p1}
 \fmfiv{d.sh=cross,d.ang=0,d.siz=5thick}{point length(p1)/4 of p1}
 \fmfi{plain}{subpath (3length(p1)/4,length(p1)) of p1}
 \fmfiv{d.sh=cross,d.ang=0,d.siz=5thick}{point 3length(p1)/4 of p1}
 \fmfdot{v}
\end{fmfgraph*}
\end{fmffile}
\end{eqnarray}
where the cross denotes the contact with the operator. The calculation is simple, its just the product of two propagators and the four vertex. This product is taken in the limit where the external momentum is large. Hence, a Taylor expansion in $\frac{1}{q^2}$ is performed and the lowest order terms are kept. After all a symmetry factor of 2 has to be taken into account. The result turns out to be $C_{2}=\frac{\lambda_{0}}{2q^4}$.\\
The matrix element $\bra{0}\phi(0)\phi(0)\ket{0}$ can be calculated via known techniques. It is just the Greens function $\bra{q}\phi(0)\phi(0)\ket{q}$ in leading order $\lambda_{0}$ where the external propagators and the four vertex are amputated. The amputated part is just the Wilson coefficient $C_{2}$. An example for the contractions is given by
\begin{eqnarray}
 \frac{-i\lambda_{0}}{4!}\int d^4 p\wick{2-5,7-8}{1-6,3-4}{\bra{q},\phi(0),\phi(0),\phi(p),\phi(p),\phi(p),\phi(p),\ket{q}}.
\end{eqnarray}
The corresponding diagram is given by
\begin{eqnarray}
\begin{fmffile}{operatorphiphi02}
\begin{fmfgraph*}(40,20)
 \fmfpen{thick} 
 \fmfleft{i} 
 \fmfright{o}
 \fmf{plain}{i,v1}
 \fmf{plain}{v1,o}
 \fmf{plain,left,tag=1}{v1,v2}
 \fmf{plain,left}{v2,v1}
 \fmfv{decor.shape=circle,decor.fill=empty,decor.size=.1w,l=$\bigotimes$,label.dist=0}{v2}
 \fmfdot{v1}
 \fmffixedx{0}{v1,v2}
 \fmffixedy{1cm}{v1,v2}
\end{fmfgraph*}
\end{fmffile}
\end{eqnarray}
where the Wilson coefficient $C_{2}$ has to be amputated. The cross symbol denotes a special vertex, it is the operator $\phi(0)\phi(0)$. The result is just $\frac{-m^2_{0}\ln\frac{\mu^2}{m^2_{0}}}{16\pi^2}$, the expression for the operator in lowest order. An alternative way to derive this result would be to calculate the tadpole diagram in $\phi^4$ theory with the extension that the frequencies large than $\mu$ have to be cut off. Hence the third summand in (\ref{opeuntidy}) is $C_{2}\bra{0}\phi\phi\ket{0}$, with respect to (\ref{phi2ope}) the remainder has to be $C_{1}\mathds{1}$. Thus the following identifications can now be placed:
\begin{eqnarray}
m^2(\mu)=m^2_{0}\left[1-\frac{\lambda_{0}}{16\pi^2}\ln\frac{M}{\mu}\right] \label{comp1}
\end{eqnarray}
\begin{eqnarray}
C_{1}=\frac{1}{q^2}+\frac{m^2(\mu)}{q^4}
\end{eqnarray}
\begin{eqnarray}
C_{2}=\frac{\lambda_{0}}{2q^4}
\end{eqnarray}
\begin{eqnarray}
\mele{\phi^2}=\frac{-m^2_{0}\ln\frac{\mu^2}{m^2_{0}}}{16\pi^2} \label{phicut}.
\end{eqnarray}
A remarkable point is that a running mass arises in the OPE in first order in $\lambda_{0}$. The mass is running with respect to the scale $\mu$ which separates high and low frequencies. In the literature $\mu$ is referred to as the renormalization point (see \cite{Novikov:1984rf}). Without this separation a running mass does not appear to first order in $\lambda_{0}$, it would be constant without the separation in high and low frequencies. The ultraviolet cutoff $M$ and the naked mass $m_{0}$ are unobservable quantities and should not appear in the definition of the running mass. Therefore, the running mass is renormalized in order to eliminate the unobservable quantities.\\
The Wilson coefficient $C_{1}$ was calculated to first order in $\lambda_{0}$. The first order result can be used to calculate a part of the higher order corrections to $C_{1}$. In higher orders, diagramms appear which consist of $n$ first order results which are ordered one after another
\begin{eqnarray}
\parbox[c]{2cm}{
\begin{fmffile}{phifreeprop}
\begin{fmfgraph*}(20,20)
 \fmfleft{i} 
 \fmfright{o}
 \fmf{plain}{i,o}
\end{fmfgraph*}
\end{fmffile}}
+
\parbox[c]{2cm}{
\begin{fmffile}{phi4tadpole02}
\begin{fmfgraph*}(20,20)
 \fmfleft{i}
 \fmfright{o}
 \fmf{plain}{i,v,v,o}
 \fmfdot{v}
\end{fmfgraph*}
\end{fmffile}}
+
\parbox[c]{2cm}{
\begin{fmffile}{phi4tadpole03}
\begin{fmfgraph*}(20,20)
 \fmfleft{i}
 \fmfright{o}
 \fmf{plain}{i,v1,v1,v2,v2,o}
 \fmffixedx{0.8cm}{v1,v2}
 \fmfdot{v1,v2}
\end{fmfgraph*}
\end{fmffile}}+~~~...~
.
\end{eqnarray}
Hence, a series analogous to the geometrical series is derived
\begin{eqnarray}
\frac{1}{q^2}+\frac{1}{q^2}m_{0}^2\left[1-\frac{\lambda_{0}}{32\pi^2}ln\left(\frac{M^2}{\mu^2}\right)\right]\frac{1}{q^2}\nonumber\\+\frac{1}{q^2}m_{0}^2\left[1-\frac{\lambda_{0}}{32\pi^2}ln\left(\frac{M^2}{\mu^2}\right)\right]\frac{1}{q^2}m_{0}^2\left[1-\frac{\lambda_{0}}{32\pi^2}ln\left(\frac{M^2}{\mu^2}\right)\right]\frac{1}{q^2}+...~.
\end{eqnarray} 
This series can be summed up
\begin{eqnarray}
\frac{1}{q^2}\frac{1}{1-\frac{1}{q^2}m_{0}^2\left[1-\frac{\lambda_{0}}{32\pi^2}ln\left(\frac{M^2}{\mu^2}\right)\right]}=\frac{1}{q^2-m_{0}^2\left[1-\frac{\lambda_{0}}{32\pi^2}ln\left(\frac{M^2}{\mu^2}\right)\right]}.
\end{eqnarray}
The term
\begin{eqnarray}
m^2(\mu)=m_{0}^2\left[1-\frac{\lambda_{0}}{32\pi^2}ln\left(\frac{M^2}{\mu^2}\right)\right]
\end{eqnarray}
is identified as the mass of the particle. Obviously, the divergent parameters $m_{0}$ and $M$ still appear in the expression for the mass. These parameters have to be eliminated in order to make the theory well defined. A new scale $\mu_{0}$ is introduced
\begin{eqnarray}
m^2(\mu)=m_{0}^2\left[1-\frac{\lambda_{0}}{32\pi^2}ln\left(\frac{M^2}{\mu^2_{0}}\right)-\frac{\lambda_{0}}{32\pi^2}ln\left(\frac{\mu_{0}^2}{\mu^2}\right)\right].
\end{eqnarray}
The finite and the divergent have to be separated
\begin{eqnarray}
m^2(\mu)=m_{0}^2\left[1-\frac{\frac{\lambda_{0}}{32\pi^2}ln\left(\frac{M^2}{\mu^2_{0}}\right)}{1-\frac{\lambda_{0}}{32\pi^2}ln\left(\frac{\mu_{0}^2}{\mu^2}\right)}\right]\left[1-\frac{\lambda_{0}}{32\pi^2}ln\left(\frac{\mu_{0}^2}{\mu^2}\right)\right] .
\end{eqnarray}
If $M^2\gg\mu^2_{0}$ and $\mu^2_{0}\approx\mu^2$
\begin{eqnarray}
m^2(\mu)\approx m_{0}^2\left[1-\frac{\lambda_{0}}{32\pi^2}ln\left(\frac{M^2}{\mu^2_{0}}\right)\right]\left[1-\frac{\lambda_{0}}{32\pi^2}ln\left(\frac{\mu_{0}^2}{\mu^2}\right)\right] .
\end{eqnarray}
If $\mu=\mu_{0}$
\begin{eqnarray}
m^2(\mu_{0})= m_{0}^2\left[1-\frac{\lambda_{0}}{32\pi^2}ln\left(\frac{M^2}{\mu^2_{0}}\right)\right].
\end{eqnarray}
Hence
\begin{eqnarray}
m^2(\mu)= m^2(\mu_{0})\left[1-\frac{\lambda_{0}}{32\pi^2}ln\left(\frac{\mu_{0}^2}{\mu^2}\right)\right] .
\end{eqnarray}
Thus all divergent parameters have been eliminated. The mass is renormalized and everything is well defined. Hence, (\ref{comp1}-\ref{phicut}) is the OPE of the propagator in the $\phi^4$ theory to leading order in the coupling constant $\lambda$. \\
This simple example tells a lot about the OPE. The matrix elements are not purely non-perturbative. They receive contributions from perturbation theory. The OPE reproduces the perturbative result to a certain order in the coupling constant $\lambda$ and in $\frac{1}{q^2}$. 

\subsubsection{OPE in the phase with broken symmetry \label{broken}}

\begin{figure}[htbp]
 \begin{center}
\begin{picture}(0,0)%
\includegraphics{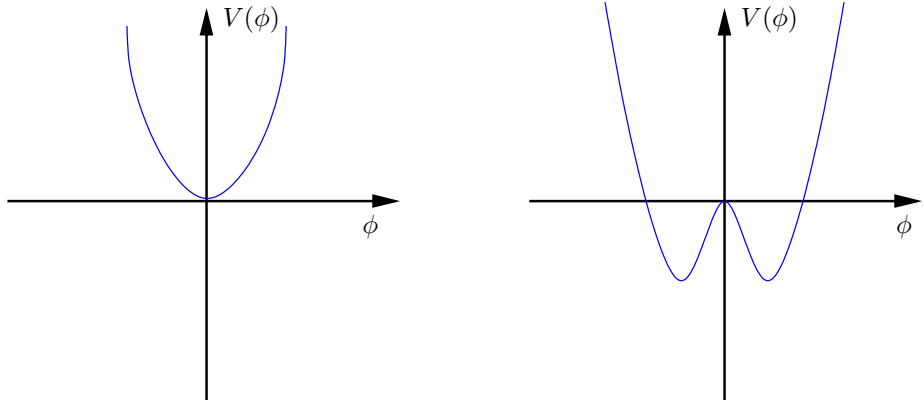}%
\end{picture}%
\setlength{\unitlength}{4144sp}%
\begingroup\makeatletter\ifx\SetFigFont\undefined%
\gdef\SetFigFont#1#2#3#4#5{%
  \reset@font\fontsize{#1}{#2pt}%
  \fontfamily{#3}\fontseries{#4}\fontshape{#5}%
  \selectfont}%
\fi\endgroup%
\begin{picture}(5550,2426)(6303,-2894)
\put(11640,-1872){$\phi$}%
\put(10711,-633){$V(\phi)$}%
\put(8447,-1872){$\phi$}%
\put(7612,-633){$V(\phi)$}%
\end{picture}%
  \caption{The potential of the Hamilton operator in $\phi^4$ theory, depending on the sign 
  of $\eta^2$. On the left hand side the case of a positive sign is shown while on the right 
  hand side the case of a negative sign is shown.\label{phipotential}}
  \end{center}
  \end{figure}

A second look at the Lagrangian (\ref{phi4lagrange}) of the $\phi^4$ theory exhibits an interesting feature of the theory if the mass is exchanged by a mass parameter with arbitrary sign
\begin{eqnarray}
\mathscr{L}=\frac{1}{2}\left(\partial_{\mu}\phi\right)^2\pm\frac{1}{2}\eta^2\phi^2-\frac{\lambda_{0}}{4!}\phi^4 \label{phi4eta}.
\end{eqnarray}
The Hamilton operator is given by
\begin{eqnarray}
H=\int d^3x\left[\frac{1}{2}\pi^2+\frac{1}{2}(\nabla\phi)^2\pm\frac{1}{2}\eta^2\phi^2+\frac{\lambda_{0}}{4!}\phi^4\right].
\end{eqnarray}
The potential is given by
\begin{eqnarray}
V(\phi)=\pm\frac{1}{2}\eta^2\phi^2+\frac{\lambda_{0}}{4!}\phi^4.
\end{eqnarray}
Depending on the sign of the $\eta^2$-term in (\ref{phi4eta}), the potential has the shape of figure \ref{phipotential}.
In the case of a negative sign in front of $\eta$ the symmetry of the system is spontaneously broken. The order parameter which determines this is $\bra{0}\phi\ket{0}$. If $\bra{0}\phi\ket{0}\not = 0$ the symmetry is broken. In the case of broken symmetry it is reasonable to expand the theory around the minima of the potential 
\begin{eqnarray}
\phi=\phi_{0}+\sigma(x),~~~\phi_{0}=\pm v=\pm\sqrt{\frac{6}{\lambda_{0}}}i\eta .
\end{eqnarray}
Where $\phi_{0}$ is given by the position of the minima in the $\phi$ space. The Lagrangian $\mathscr{L}$ after this transformation is given by
\begin{eqnarray}
\mathscr{L}=\frac{1}{2}\left(\partial_{\mu}\sigma\right)^2-\frac{1}{2}(2\eta^2)\sigma^2-\sqrt{\frac{\lambda_{0}}{6}}\eta\sigma^3-\frac{\lambda_{0}}{4!}\sigma^4 \label{shiftlagrange}.
\end{eqnarray}
The Feynman rules in this version of the theory differ from the ones in the theory where the symmetry is not broken. A comparison is given in table \ref{zweitetabelle}.
\begin{table}
\begin{center}
\begin{tabular}{||c|l||}
\hline
unshifted theory&  vertex : $-i\lambda_{0}$\\
                &  propagator : $\frac{i}{p^2-m^2}$\\
\hline
shifted theory&  vertices : $-i\lambda_{0}$,$-i\sqrt{\frac{\lambda_{0}}{3!}}\eta$ \\
                &  propagator: $\frac{i}{p^2-2\eta^2}$\\
\hline
\end{tabular}
\caption{Comparison of the Feynmanrules of $\phi^4$-theory in the phase where the symmetry is not broken with the rules in the phase where the symmetry is broken. An additional vertex arises in the phase with broken symmetry.\label{zweitetabelle}}
\end{center}
\end{table}
Where $-i\lambda_{0}$ is a four vertex and $-i\sqrt{\frac{\lambda_{0}}{3!}}\eta$ is a three vertex and is of order $\sqrt{\lambda_{0}}$. If the OPE instead of the usual techniques are used to calculate the amplitudes in the phase with broken symmetry the use of the new Feynmanrules can be avoided. There are two possibilities to incorporate the nature of the states in the calculation of the amplitudes. The first one is to shift the theory to the minima in the $\phi$ space which results in new Feynmanrules. The second is to use the OPE and incorporate the nature of the state not in the Feynmanrules but in the matrix elements. This is especially helpful if for some reason the shift which leads to new Feynmanrules is not possible. In the following it is proven, that the OPE reproduces the amplitudes in the phase of the theory with broken symmetry.\\
The OPE can directly be taken from section \ref{unbroken}. Obviously, only $m_{0}^2$ has to be replaced by $-\eta^2$ and the matrix elements change. Again, it is emphasized that the theory is not shifted. Hence, the Wilson coefficients do not change in comparison with section \ref{unbroken} except for the mass parameter. Thus the Wilson coefficients stay the same, only the matrix elements change.\\
In the following a first naive test is performed if the OPE reproduces the amplitudes in the phase where the symmetry is broken. This test will fail, but in the next section the reason for this failure will be clarified.\\
The method to calculate the matrix elements in the OPE in $\phi^4$ theory is to use the classical value for $\phi$ which is given by $\phi_{0}$. In this case 
\begin{eqnarray}
\bra{0}\phi(0)\phi(0)\ket{0}=\left( \pm\sqrt{\frac{6}{\lambda_{0}}}\eta\right)^2=\frac{6}{\lambda_{0}}\eta^2 \label{firsttry}.
\end{eqnarray}
This is the value for $\bra{0}\phi(0)\phi(0)\ket{0}$ at the scale $\mu$ at which $m^2_{physical}(\mu)=-2\eta^2$, as it is already stated in equation (\ref{shiftlagrange}) in summand 2.\\
The treatise just performed leads to the following OPE in the phase with broken symmetry
\begin{eqnarray}
i\int dx e^{-ix_{\mu}q^{\mu}}\bra{0}\phi(x)\phi(0)\ket{0}=\frac{1}{q^2}+\frac{2\eta^2}{q^4}\left(1-\frac{\lambda_{0}}{32\pi^2}\ln\left(\frac{\mu }{\mu_{0}}\right)\right)\nonumber\\=\frac{1}{q^2}-\frac{m^2_{physical}}{q^4}\left(1-\frac{\lambda_{0}}{32\pi^2}\ln\left(\frac{\mu }{\mu_{0}}\right)\right).
\end{eqnarray}
This expression should reproduce perturbation theory but how is the propagator in the phase with broken symmetry calculated ? In order to keep the calculations as simple as possible the shifted theory should be used. The diagrams that contribute to first order in $\lambda_{0}$ are
\begin{eqnarray}
\parbox{20mm}{
\begin{fmffile}{phi4shifted00}
\begin{fmfgraph*}(20,40)
 \fmfpen{thick} 
 \fmfleft{i} 
 \fmfright{o}
 \fmf{plain}{i,o}
\end{fmfgraph*}
\end{fmffile}}
~~~~
\parbox{20mm}{
\begin{fmffile}{phi4shifted01}
\begin{fmfgraph*}(20,40)
 \fmfpen{thick} 
 \fmfleft{i} 
 \fmfright{o}
 \fmf{plain}{i,v,v,o}
 \fmfdot{v}
\end{fmfgraph*}
\end{fmffile}}
~~~~
\parbox{20mm}{
\begin{fmffile}{phi4shifted02}
\begin{fmfgraph*}(20,40)
 \fmfpen{thick} 
 \fmfleft{i} 
 \fmfright{o}
 \fmf{plain}{i,v1,v2,v2,v1,o}
 \fmfdot{v1,v2}
 \fmffixedx{0}{v1,v2}
 \fmffixedy{25}{v1,v2}
\end{fmfgraph*}
\end{fmffile}}
~~~~
\parbox{20mm}{
\begin{fmffile}{phi4shifted03}
\begin{fmfgraph*}(20,40)
 \fmfpen{thick} 
 \fmfleft{i} 
 \fmfright{o}
 \fmf{plain}{i,v1}
 \fmf{plain,left,tensio=0.4}{v1,v2}
 \fmf{plain,right,tensio=0.4}{v1,v2}
 \fmf{plain}{v2,o}
 \fmfdot{v1,v2}
\end{fmfgraph*}
\end{fmffile}}~.
\end{eqnarray}
Only the last diagram contributes to the propagator the other ones vanish after renormalization (see section \ref{reguandreno}). In the limit of big momenta $q$ the expression for the propagator is
\begin{eqnarray}
D(q)=\frac{1}{q^2}+\frac{m^2_{phys}}{q^4}\left(1+\frac{3\lambda_{0}}{16\pi^2}\ln\left(\frac{\sqrt{q^2}}{m_{phys}}\right) \right) \label{propbroken}
\end{eqnarray}
and it does not coincide with (\ref{firsttry}) even after setting $\mu_{0}=q$ and $\mu=m_{phys}$. This result is of course shocking. The question appears if everything that has been stated about the OPE in the phase with broken symmetry is wrong. Fortunately it is not wrong but there is something missing in the derivation of the OPE. In the next section the riddle about the missing component will be solved.

\subsubsection{Comparison of OPE with perturbation theory in the phase with broken symmetry \label{compara}}

The question that has to be answered is if the OPE reproduces the amplitudes in the phase with broken symmetry to a given order in $\lambda_{0}$. For this purpose the propagator is expanded via perturbation theory in the phase with broken symmetry and via the OPE. The results should coincide to a given order in $\lambda_{0}$ but as shown in the preceding section they seem not to do it. The missing parts of the OPE are derived in the following.\\
As shown in section \ref{unbroken} the normalization point is crucial in the definition of the OPE. The missing agreement between the OPE and perturbation theory in the phase with broken symmetry could also stem from an incomplete implementation of the renormalization group flux. In the version of the theory with reflection symmetry the only scale that is important is the mass in the Lagrangian (\ref{phi4lagrange}) which naturally defines the scale $\mu$ which is inherent to the theory but it is not necessary to use the physical mass of the particle as the renormalization point. An other point is that the physical mass in the theory with broken symmetry is not given by the mass parameter $\eta$ in the Lagrangian (\ref{phi4eta}) but by $-2\eta$ which is the natural scale in the phase with broken symmetry.\\
These complications did not show up in the theory with unbroken symmetry where there exists no difference in the mass scale that the Wilson coefficients are calculated with and the mass scale that is set by the physical mass. Thus the flux should be less important in the theory with reflection symmetry than in the theory where this symmetry is broken.\\ 
In simple words the Wilson coefficients are calculated in a version of the theory that is defined on an other scale than the version of the theory in which the propagator 
(\ref{propbroken}) is calculated. Thus the point at which the Wilson coefficients are defined 
have to be changed to the point where the propagator (\ref{propbroken}) is defined. Unless 
this is done the effective theory (\ref{shiftlagrange}) and the OPE will not coincide.\\
In order to make the OPE independent from the normalization point $\mu$ the flux of the Wilson coefficients and all parameters has to be constructed. The method used here to achieve this is the method of renormalization group improvement. The OPE is improved by summing up the leading logarithms using the recipe that renormalization group theory proposes.\\ 
The first step is the renormalization of the coupling constant $\lambda_{0}$. Those computations are standard and can be taken from books like \cite{Peskin:1995ev}. The renormalization group equations are given by
\begin{eqnarray}
\mu\frac{d}{d\mu}C_{2}=\frac{\lambda(\mu)}{16\pi^2}C_{2} \label{problemfall}
\end{eqnarray}
\begin{eqnarray}
\mu\frac{d}{d\mu}C_{1}=\frac{m^2(\mu)}{8\pi^2}C_{2} \label{opmix}
\end{eqnarray}
\begin{eqnarray}
\mu\frac{d}{d\mu}m^2(\mu)=\frac{\lambda(\mu)}{16\pi^2}m^2(\mu)
\end{eqnarray}
\begin{eqnarray}
\mu\frac{d}{d\mu}\lambda(\mu)=3\frac{\lambda^2(\mu)}{16\pi^2} \label{letztegleich}.
\end{eqnarray}
All of these equations except of (\ref{problemfall}) are derivatives of the first order expressions. The derivation of (\ref{problemfall}) leads to new insight concerning the renormalization group flux of the OPE summands. The derivation can be found in \cite{Collins:1984xc} and is not shown here. The solutions of the equations (\ref{problemfall}-\ref{letztegleich}) are
\begin{eqnarray}
\lambda(\mu)=\frac{\lambda(\mu_{0})}{1-3\frac{\lambda(\mu_{0})}{16\pi^2}\ln\left( \frac{\mu}{\mu_{0}}\right) }
\end{eqnarray}
\begin{eqnarray}
 m^2(\mu)=\left[\frac{\lambda(\mu)}{\lambda(\mu_{0})}\right]^{\frac{1}{3}}m^2(\mu_{0})
\end{eqnarray}
\begin{eqnarray}
C_{2}\left(\mu\right)=\left[\frac{\lambda(\mu)}{\lambda(\mu_{0})}\right]^{\frac{1}{3}}C_{2}\left(\mu_{0}\right) 
\end{eqnarray}
\begin{eqnarray}
C_{1}(\mu)=C_{1}(\mu_{0})-2\frac{m^2(\mu_{0})C_{2}(\mu_{0})}{\lambda(\mu_{0})}\left(\left[\frac{\lambda(\mu)}{\lambda(\mu_{0})}\right]^{-\frac{1}{3}}-1\right).
\end{eqnarray}
This is the improved OPE in the theory with reflection symmetry, corresponding to the theory with a positive value of the squared mass parameter in the Lagrangian (\ref{phi4eta}). It has the remarkable ability that it reproduces both versions of the propagator in the $\phi^4$ theory. If $\mu=m_{0}^2$ it reproduces the propagator in the theory with reflection symmetry and if $\mu=-2m_{0}^2$ the propagator in the theory where this symmetry is broken. This assertion is proved in the following.\\
The starting conditions are chosen to be 
\begin{eqnarray}
C_{1}(\mu_{0}=q)=\frac{1}{q^2}+\frac{m^2(q)}{q^4}
\end{eqnarray}
\begin{eqnarray}
C_{2}(\mu_{0}=q)=\frac{\lambda(q)}{2q^4}
\end{eqnarray}
then the coefficients $C_{1},C_{2}$ contain no logarithms. The explicit expression for the OPE with those conditions is 
\begin{eqnarray}
D(q)=\bra{0}C_{1}(\mu)+C_{2}(\mu)\phi^2\ket{0}=\frac{1}{q^2}\nonumber\\+\frac{1}{q^4}\left(\left[\frac{\lambda(q)}{\lambda(\mu)}\right]^{\frac{1}{3}}2m^2(\mu)+\left[\frac{\lambda(q)}{\lambda(\mu)}\right]^{\frac{2}{3}}\left( \frac{1}{2}\lambda(\mu)\mele{\phi^2(\mu)}-m^2(\mu)\right) \right) .
\end{eqnarray}
With these results the validity of the OPE in $\phi^4$ theory can be tested.
\begin{enumerate}
\item{Coincidence of the improved OPE with the propagator in the theory with reflection   
      symmetry: $\mu^2=m^2$ }\\
      At this scale the matrix element given in (\ref{phicut}) vanishes which leaves the OPE as
      \begin{eqnarray}
      D(q)=\frac{1}{q^2}+\frac{1}{q^4}m^2(m)\left(2 
      \left[\frac{\lambda(q)}{\lambda(m)}\right]^{\frac{1}{3}}-\left[\frac{\lambda(q)}{
      \lambda(m)}\right]^{\frac{2}{3}}\right) \nonumber\\
      =\frac{1}{q^2}+\frac{1}{q^4}m^2(m)+\mathcal{O}(\lambda^2),
      \end{eqnarray}
      which coincides with the propagator in the theory with higher symmetry to first order 
      in $\lambda$.
\item{Coincidence of the improved OPE with the propagator in the theory with broken   
      symmetry: $\mu^2=-2m^2=m^2_{phys}$ }\\
      At this scale the matrix element is given by (\ref{firsttry}) which leaves the OPE as
      \begin{eqnarray}
      D(q)=\frac{1}{q^2}+\frac{1}{q^4}\left(-2m^2(m_{phys})\right) \left(2 
      \left[\frac{\lambda(q)}{\lambda(m_{phys})}\right]^{\frac{2}{3}}-\left[\frac{\lambda(q)}
      {\lambda(m_{phys})}\right]^{\frac{1}{3}}\right) \nonumber\\
      =\frac{1}{q^2}+\frac{1}{q^4}m^2_{phys}\left(2 
      \left[\frac{\lambda(q)}{\lambda(m_{phys})}\right]^{\frac{2}{3}}-\left[\frac{\lambda(q)}
      {\lambda(m_{phys})}\right]^{\frac{1}{3}}\right) \nonumber\\
      =\frac{1}{q^2}+\frac{m^2_{phys}}{q^4}\left(1+\frac{3\lambda}{16\pi^2}\ln\left(\frac{q}{m_{phys}}\right) \right)+\mathcal{O}(\lambda^2),
      \end{eqnarray}
      which reveals the coincidence with the theory in the phase with broken symmetry. 
\end{enumerate}
 \begin{figure}[htbp]
 \begin{center}
\begin{picture}(0,0)%
\includegraphics{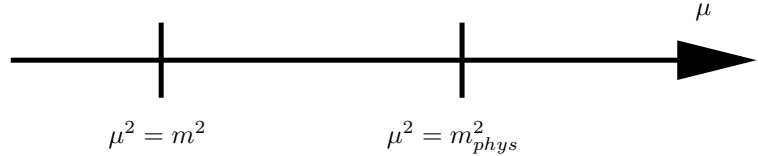}%
\end{picture}%
\setlength{\unitlength}{4144sp}%
\begingroup\makeatletter\ifx\SetFigFont\undefined%
\gdef\SetFigFont#1#2#3#4#5{%
  \reset@font\fontsize{#1}{#2pt}%
  \fontfamily{#3}\fontseries{#4}\fontshape{#5}%
  \selectfont}%
\fi\endgroup%
\begin{picture}(4566,997)(2218,-1979)
\put(6346,-1141){$\mu$}%
\put(4501,-1906){$\mu^2=m^2_{phys}$}%
\put(2836,-1906){$\mu^2=m^2$}%
\end{picture}%
  \caption{An illustration concerning the distribution of normalization points.\label{boundarycondition02}}
  \end{center}
  \end{figure}
The conclusion of this section is that the change of matrix elements, claimed in section \ref{broken}, is not sufficient to change between the phases of the theory. In addition the normalization point has also to be changed.\\
The ability of the OPE to describe processes in the phase with broken symmetry simply by changing the matrix elements and the normalization point in comparison with the OPE in the phase where the symmetry is not broken is the reason why it is employed in modern applications.\\ 
In the early days of the OPE the question arose whether the OPE is an approximation to the theory in the phase with broken symmetry or if it is exact in reproducing the amplitudes \cite{Novikov:1984rf}. The source of this problem was an approximation made by Shifman, Vainshtein and Zakharov in their original publications to QCD Sum Rules \cite{Shifman:1978bx}. This approximation is discussed in the next section.

\subsection{OPE in QCD - Wilsonian OPE versus Practical OPE \label{opeinqcd}}

The OPE has been explained in general and in terms of an example in the real scalar $\phi^4$ theory but every theory has its own peculiarities concerning the OPE. In the remainder of this thesis QCD is the theory in which calculations are performed.\\ 
Concerning the OPE in QCD the OPE exhibits big advantages. QCD also has a broken symmetry, the chiral symmetry. Unfortunately, the theory is much more difficult than the $\phi^4$ theory. In the phase with broken symmetry an additional phenomenon arises, the confinement of quarks. This means, that free quarks do not exist in that phase. Only bound states of quarks and gluons have been measured, the so called hadrons. No one knows how the transition between the phase with unbroken and broken symmetry has to be described. Thus a shift as it has been done in section \ref{broken} does not seem to be possible. Therefore, no Feynmanrules for amplitudes on the basis of quarks and gluons in the phase with broken symmetry are known. There exist effective field theories, which describe the physics in that phase on the basis of hadrons as degrees of freedom. Moreover, there exist other methods like lattice QCD which seeks for a numerical solution of the QCD Lagrangian. An additionally method to describe the physics of QCD in the broken phase is given by the OPE. \\
The OPE is for example an alternative to chiral perturbation theory. As it was stated before the Wilson coefficients can be calculated in the phase with and without broken symmetry. Hence, one can use the QCD Lagrangian, where the quark masses are current quark masses, to compute the Wilson coefficients. A repetition of the calculations performed in the $\phi^4$ theory would lead to amplitudes valid in the phase with broken symmetry. In this expansion quarks and gluons are the basic degrees of freedom. This is the way the OPE is used in actual applications.\\
There are two approximations that simplify the calculation of the OPE in QCD considerably
\begin{enumerate}
\item The perturbative parts of the matrix elements can be absorbed in the Wilson coefficients of the unit operator $\mathds{1}$. Thus, in this approximation the matrix elements are purely non-perturbative (see \cite{Shifman:1999mk}). 
\item The renormalization group flow of the Wilson coefficients can be neglected.
\end{enumerate}
In the propagator example of $\phi^4$ theory the perturbative part of the $\phi^2(0)$ matrix element has been  $\frac{-m^2_{0}}{16\pi^2}\ln\frac{\mu^2}{m^2_{0}}$. Thus in the first approximation, this term is included in $C_{1}$ and must be subtracted from the matrix element.\\
Those two approximations are applied to all OPE computations in QCD. They date back to the founders of the QCD Sum Rule method \cite{Shifman:1978bx} and have lead to missunderstandings \cite{Novikov:1984rf}.\\
The exact OPE is denoted Wilsonian OPE and the approximative one is called practical OPE or SVZ expansion/OPE. In the remainder of this thesis the practical OPE is used. \\
One generally can describe the OPE as a method to approximate the physics in the nonperturbative regime, the phase with broken symmetry, with the field theory that is valid in the perturbative regime. The big advantages in this case are that the OPE deals with current quarks and not with constituent quarks.\\
Every advantage has its price. In the OPE case it is the introduction of the matrix elements which in QCD are called condensates. Up to today there exist no method which is based on first principles to calculate these matrix elements. Chiral perturbation theory and QCD Sum Rules can be used to get an approximation for the condensates, but they are based on data from expirements and not on the QCD Lagrangian. Hence, the condensates are determined through measurements.
Nevertheless the OPE is heavily used in nonperturbative physics. In the next section the calculations of the Wilson coefficients are shown for several examples.

\section{Calculation of Wilson coefficients in QCD \label{wilsoncoefficient}}
In the last section the OPE was introduced and it was showed that the Wilson coefficients are the part of the OPE that has to be calculated if QCD amplitudes are concerned. The plane-wave method was introduced as the technique for the computations, but the details of the calculations have not been explained. Therefore, this section is devoted to the calculations of Wilson coefficients via the plane-wave method. This method is the most simple one can think of, it directly employs the operator nature of the OPE. Moreover, the first calculations of Wilson coefficients where done with this method \cite{Shifman:1978bx}. As it was quoted the practical OPE is used, thus the renormalization group flow of the coefficients is neglected.

\subsection{Equal mass case : the flavor conserving scalar current correlator \label{Wilson_Berechnung}}

Here and in further investigations products of currents are concerned. The currents are used as the interpolating field for a hadron. Unfortunately, this choice is not unique. As the approximative current for a hadron all currents have to be considered which have the same number as the hadron. Such ambiguities can have different implications. For example the ambiguity between a four quark and a two quark current has a physical implication. The quarks in the current are the valence quark. Hence, such an ambiguity is directly linked to questions concerning the quark model. Non-physical ambiguities are for example given by the number of derivatives which occur in the current. It is possible to write down n quark currents with and without derivatives which have the same quantum numbers. In this thesis mesons are analyzed. Thus, the currents have to have a even number of quark fields, but here only two quark currents are addressed. Moreover, currents with the simplest structure are used. This means for example currents with the smallest number of derivatives. The most simple current of this type is the scalar current:
\begin{eqnarray}
j(x)=\bar{q}(x)q(x)~~j^{\dagger}(x)=\bar{q}(x)q(x).
\end{eqnarray}
Hence the OPE is:
\begin{eqnarray}
j(x)j^{\dagger}(0)=\sum_{n=1}^{\infty}C_{n}O_{n}.
\end{eqnarray}
The current product has energy dimension 6, hence section (\ref{importantunimportant}) states that only operators with $dimension\leq 6$ appear in the OPE. Those operators are given by \cite{Shifman:1978bx}
\begin{eqnarray}
\mathds{1}~~dimension=0\nonumber\\
m\overline{q}q~~dimension=4\nonumber\\
G^{c}_{\mu\nu}G^{c}_{\mu\nu}~~dimension=4\nonumber\\
m\overline{q}\sigma_{\mu\nu}\lambda^{c}qG^{c}_{\mu\nu}~~dimension=6\nonumber\\
\overline{q}\Gamma_{1}q\overline{q}\Gamma_{2}q~~dimension=6\nonumber\\
f^{abc}G^{a}_{\mu\nu}G^{b}_{\nu\gamma}G^{c}_{\gamma\mu}~~dimension=6
\label{operators}.
\end{eqnarray}
To be precise there are more operators with $dimension\leq 6 $ than those just quoted, but they can be reduced to those quoted above. An example of such operators are operators with a derivative. Derivatives raise the dimension of an operator. Hence, a derivative of the $m\overline{q}q$ still appears in the OPE but it can be reduced to the $m\overline{q}q$ Operator. The calculations of Wilson coefficients gets more complex if more complex operator products are concerned, for a given operator product it gets more complex the more complex the operator is to which the Wilson coefficient belongs. Therefore, a simple example for the calculation of Wilson coefficients is the Wilson coefficient that belongs to the operator $m\overline{q}q$, the quark condensate. According to the plane-wave method this coefficient can be filtered out by sandwiching the OPE between quark states:
\begin{eqnarray}
\bra{p}j(x)j^{\dagger}(0)\ket{p}=\bra{p}\sum_{n=1}^{\infty}C_{n}O_{n}\ket{p}=\sum_{n=1}^{\infty}\bra{p}C_{n}O_{n}\ket{p}=\sum_{n=1}^{\infty}C_{n}\bra{p}O_{n}\ket{p}=\nonumber\\C_{\bar{q}q}\bra{p}O_{\bar{q}q}\ket{p}=C_{\bar{q}q}\bra{p}\bar{q}q\ket{p}=C_{\bar{q}q}\bar{u}(p)u(p)
\label{opequarkcond}.
\end{eqnarray}
The current correlator is transformed to momentum space
\begin{eqnarray}
i\int d^4xe^{iqx}\bra{0}T(j(x)j^{\dagger}(0))\ket{0}=i\int d^4xe^{iqx}\bra{0}T(\bar{q}(x)q(x)\bar{q}(0)q(0))\ket{0}.
\end{eqnarray}
After that the plane-wave method is applied for the calculation of the $\bar{q}q$ Wilsoncoefficient
\begin{eqnarray}
i\int d^4xe^{iqx}\bra{p}T(j(x)j^{\dagger}(0))\ket{p}=i\int d^4xe^{iqx}\bra{p}T(\bar{q}(x)q(x)\bar{q}(0)q(0))\ket{p}.
\end{eqnarray}
The coefficient is derived to lowest order in $\alpha_{S}$. Therefore Wick`s theorem is applied:
\begin{eqnarray}
i\int d^4xe^{iqx}\wick{1-2,5-6}{3-4}{\bra{p},\bar{q}(x),q(x),\bar{q}(0),q(0),\ket{p}}+i\int d^4xe^{iqx}\wick{1-4,3-6}{2-5}{\bra{p},\bar{q}(x),q(x),\bar{q}(0),q(0),\ket{p}}\nonumber\\=
i\parbox[c]{3cm}{
\begin{fmffile}{diagram01}
\begin{fmfgraph*}(40,35)
 \fmfleft{i1,o1} \fmfright{i2,o2}
 \fmffixedx{2cm}{v1,v2}
 \fmf{dots_arrow,label.side=left,label=$q$}{i1,v1}
 \fmf{fermion,label.side=down,label=$q+p$}{v1,v2}
 \fmf{dots_arrow,label.side=left,label=$q$}{v2,i2}
 \fmf{fermion,label.side=right,label=$p$}{o1,v1}
 \fmf{fermion,label.side=right,label=$p$}{v2,o2}
\end{fmfgraph*}
\end{fmffile}}
~~~~~~~~~+
i\parbox[c]{3cm}{
\begin{fmffile}{diagram02}
\begin{fmfgraph*}(40,35)
 \fmfleft{i1,o1} \fmfright{i2,o2}
 \fmffixedx{2cm}{v1,v2}
 \fmf{dots_arrow,label.side=left,label=$q$}{v2,i2}
 \fmf{dots_arrow,label.side=left,label=$q$}{i1,v1}
 \fmf{fermion,label.side=right,label=$p$}{o1,v2}
 \fmf{fermion,label.side=down,label=$q-p$}{v2,v1}
 \fmf{fermion,label.side=right,label=$p$}{v1,o2}
\end{fmfgraph*}
\end{fmffile}}
\nonumber\\=i\bar{u}(p,s)\frac{i(\slashed{p}+\slashed{q}+m)}{(p+q)^2-m^2}u(p,s)+i\bar{u}(p,s)\frac{i(\slashed{p}-\slashed{q}+m)}{(p-q)^2-m^2}u(p,s)\label{scalarlorentzscalar}.
\end{eqnarray}
In the diagrams dashed lines represent currents and full lines represent fermions. Hence, an expression for the left hand side of (\ref{opequarkcond}) is
\begin{eqnarray}
-\bar{u}(p)\frac{\slashed{p}+\slashed{q}+m}{(p+q)^2-m^2}u(p)-\bar{u}(p)\frac{\slashed{p}-\slashed{q}+m}{(p-q)^2-m^2}u(p).
\end{eqnarray}
The goal is to reduce those expressions to a form where one can identify the quark term $\bar{u}u$. The external particles are free particles $p^2=m^2$. Hence, the Dirac equation $\bar{u}(p)(\slashed{p}-m)=0$ or  $(\slashed{p}-m)u(p)=0$ can be applied
\begin{eqnarray}
-\bar{u}(p)\frac{m+\slashed{q}+m}{(p+q)^2-m^2}u(p)-\bar{u}(p)\frac{m-\slashed{q}+m}{(p-q)^2-m^2}u(p)=\nonumber\\
-\bar{u}(p)\frac{2m+\slashed{q}}{(p+q)^2-m^2}u(p)-\bar{u}(p)\frac{2m-\slashed{q}}{(p-q)^2-m^2}u(p)\label{scalar01}.
\end{eqnarray}
What about the $\slashed{q}$-term ? The treatment of this term is a bit more difficult ! 
\begin{eqnarray}
\bar{u}(p)\slashed{q}u(p)=\bar{u}(p)\frac{m}{m}\slashed{q}u(p)=\frac{1}{2m}\bar{u}(p)\left\{m\slashed{q}+\slashed{q}m\right\} u(p)=\frac{1}{2m}\bar{u}(p)\left\{\slashed{p}\slashed{q}+\slashed{q}\slashed{p}\right\} u(p)\nonumber\\
\frac{1}{2m}p_{\mu}q_{\nu}\bar{u}(p)\left\{\gamma^{\mu}\gamma^{\nu}+\gamma^{\nu}\gamma^{\mu}\right\} u(p)=\frac{1}{2m}p_{\mu}q_{\nu}\bar{u}(p)2g^{\mu\nu}u(p)=\frac{p_{\mu}q^{\mu}}{m}\bar{u}(p)u(p)
\end{eqnarray}
Inserting the last expression in (\ref{scalar01}) results after a trivial conversion in
\begin{eqnarray}
-\frac{1}{m^2}\frac{2m^2+p^{\mu}q_{\mu}}{(p+q)^2-m^2}\left[m\bar{u}(p,s)u(p,s)\right]-\frac{1}{m^2}\frac{2m^2-p^{\mu}q_{\mu}}{(p-q)^2-m^2}\left[m\bar{u}(p,s)u(p,s)\right]\label{scalardetermined}.
\end{eqnarray}
The Wilsoncoefficient is determined to be:
\begin{eqnarray}
-\frac{1}{m^2}\left(\frac{2m^2+p^{\mu}q_{\mu}}{(p+q)^2-m^2}+\frac{2m^2-p^{\mu}q_{\mu}}{(p-q)^2-m^2}\right)=\nonumber\\-
\frac{1}{m^2}\left(\frac{2m^2+p^{\mu}q_{\mu}}{p^2+q^2+2p_{\mu}q^{\mu}-m^2}+\frac{2m^2-p^{\mu}q_{\mu}}{p^2+q^2-2p_{\mu}q^{\mu}-m^2}\right).
\end{eqnarray}
From the derivation of the OPE we know that the Wilsoncoefficients can only depend on $q^2$. The dependence of this Wilsoncoefficient is now analyzed. The coefficient takes the form:
\begin{eqnarray}
-\frac{1}{m^2}\left(\frac{2m^2+p^{\mu}q_{\mu}}{(p+q)^2-m^2}+\frac{2m^2-p^{\mu}q_{\mu}}{(p-q)^2-m^2}\right)=\nonumber\\
-\frac{1}{m^2}\left(\frac{2m^2+p^{\mu}q_{\mu}}{q^2+2p_{\mu}q^{\mu}}+\frac{2m^2-p^{\mu}q_{\mu}}{q^2-2p_{\mu}q^{\mu}}\right) \label{unaveraged}.
\end{eqnarray}
The only p dependence enters the coefficient through the scalar products. This dependence should not occur in the coefficient after all that we know. The dependence stems from the fact that we only take the effect of a single plane-wave into account. This wave has a fixed four momentum but there is no physical circumstance that marks out a special momentum vector. This means we have to take all possible vectors into account. This is achieved by transforming the scalar product to a space with a Euclidean metric. In this space the scalarproduct is averaged over the Euclidean angle leaving the coefficient without any p dependence.

\subsubsection{Transformation to 4-dimensional space with Euclidean metric}

\begin{figure}[htbp]
 \begin{center}
\begin{picture}(0,0)%
\includegraphics{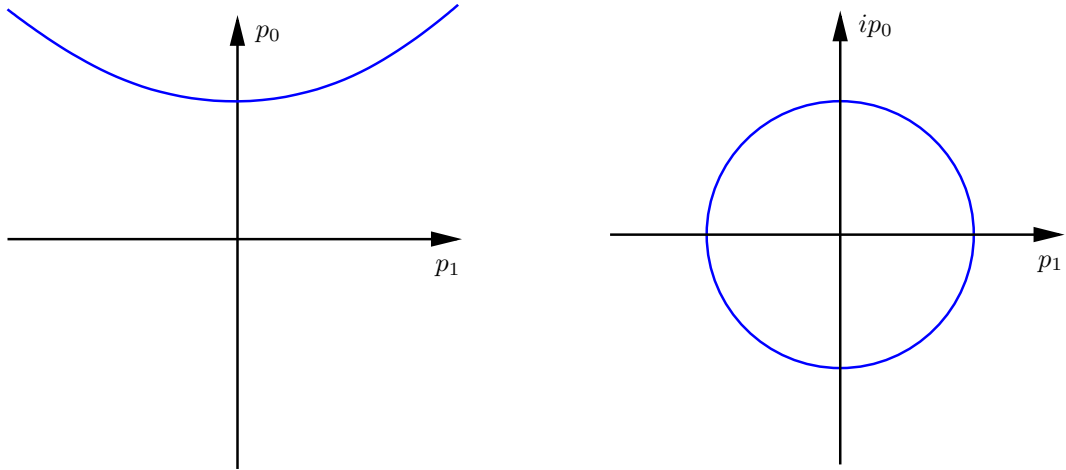}%
\end{picture}%
\setlength{\unitlength}{4144sp}%
\begingroup\makeatletter\ifx\SetFigFont\undefined%
\gdef\SetFigFont#1#2#3#4#5{%
  \reset@font\fontsize{#1}{#2pt}%
  \fontfamily{#3}\fontseries{#4}\fontshape{#5}%
  \selectfont}%
\fi\endgroup%
\begin{picture}(6402,2823)(1312,-3250)
\put(3894,-2044){$p_{1}$}%
\put(2820,-641){$p_{0}$}%
\put(7499,-2017){$p_{1}$}%
\put(6426,-614){$ip_{0}$}%
\end{picture}%
  \caption{The sphere in the 2 dimensional Minkowskispace and the sphere after applying the transformation \ref{trans}.\label{hyperbolasphere}}
  \end{center}
  \end{figure}

The most convenient way to perform the upcoming integration is to transform the problem from the Minkowski-Space to a space with a Euclidean metric. This is done by decomposing the -1 in the metric tensor of the Minkowski-Space
\begin{eqnarray}
\\
g_{\mu\nu}=
\left[ 
\begin{array}{cccc}
1&0&0&0\\
0&-1&0&0\\
0&0&-1&0\\
0&0&0&-1\\
\end{array}
\right] 
=
\left[ 
\begin{array}{cccc}
-i^2&0&0&0\\
0&-1&0&0\\
0&0&-1&0\\
0&0&0&-1\\
\end{array}
\right].
\end{eqnarray}
The $-i^2$ is subscribed to the 0 component of the 4 vectors in which the tensor is sandwiched. The effect of this procedure is that the component which was the 0 component becomes imaginary and will from now on be called the 4 component. The advantage of the effort is that in the space in which we have transformed the problem the metric is Euclidean. This enables is to perform the integration mentioned above.\\
The transformation laws are:
\begin{eqnarray}
g_{\mu\nu}\longrightarrow -\mathds{1}_{4}~~~~~~\nonumber\\
p^2\longrightarrow p_{E}^2=-p^2 \nonumber\\
p_{0}\longrightarrow p_{E,4}=ip_{0} \label{trans}.
\end{eqnarray}
The action of the transformation can be illustrated in two dimensions. The metric transforms as:
\begin{eqnarray}
\\
g_{\mu\nu}=
\left[ 
\begin{array}{cc}
1&0\\
0&-1\\
\end{array}
\right] 
\longrightarrow
-\left[ 
\begin{array}{cc}
1&0\\
0&1\\
\end{array}
\right].
\end{eqnarray}
while the vectors transform as:
\begin{eqnarray}
p_{\mu}
\left[ 
\begin{array}{c}
p_{0}\\
p_{1}\\
\end{array}
\right] 
\longrightarrow
p_{E,\mu}=
\left[ 
\begin{array}{c}
p_{1}\\
ip_{0}\\
\end{array}
\right].
\end{eqnarray}
Following the common conventions there is a zero component in the Minkowski, but not in the Euclidean space. The zero component in the Minkowski space becomes the last component in the Euclidean space.\\ 
An interesting effect concerning this transformation is the transformation of the sphere, $p^2=const.$. The sphere in the Minkowski space is a hyperbola while the sphere in a Euclidean space is a ball. The transformation (\ref{trans}) transforms these two objects into each other (see figure \ref{hyperbolasphere}).
In four dimensions the transformation causes exactly the same. Thus, from this example its plausible that the transformation enables the use of common techniques for the averaging process. 

\subsubsection{Averaging over the 4-dimensional Euclidean angle}

In the preceding section, the method used to average over all possible orientations of $p_{\mu}$ and $q_{\mu}$ was explained. In this section an example and the Wilson coefficient are computed.\\
The four dimensional Euclidean measure is:
\begin{eqnarray}
dp_{E}^4=\abs{p_{E}}^3sin^2(\theta_{2})sin(\theta_{1})d\theta_{2}d\theta_{1}d\phi dp_{E}.
\end{eqnarray}
The solid angle in 4 dimensions is:
\begin{eqnarray}
d\Omega_{4}=sin^2(\theta_{2})sin(\theta_{1})d\theta_{2}d\theta_{1}d\phi.
\end{eqnarray}
While the surface of the unit sphere is given by
\begin{eqnarray}
\int_{\theta_{2}=0}^{\pi}\int_{\theta_{1}=0}^{\pi}\int_{\phi=0}^{2\pi}d\Omega_{4}=2\pi^2.
\end{eqnarray}
If the average over the scalar product of two vectors $p_{E,\mu}$,$q_{E,\mu}$ in the 4 dimensional Euclidean space should be calculated it is very convenient to define the 4 axis in the direction of the vector $q_{E,\mu}$. Then the scalar product reads:
\begin{eqnarray}
p^{E,\mu}q_{E,\mu}=cos(\theta_{2})\abs{p_{E}}\abs{q_{E}}.
\end{eqnarray}
Therefore averaging over arbitrary orientations of these two vectors means to perform the following integration:
\begin{eqnarray}
\frac{\int_{\theta_{2}=0}^{\pi}\int_{\theta_{1}=0}^{\pi}\int_{\phi=0}^{2\pi}p^{E,\mu}q_{E,\mu}d\Omega_{4}}{\int_{\theta_{2}=0}^{\pi}\int_{\theta_{1}=0}^{\pi}\int_{\phi=0}^{2\pi}d\Omega_{4}}\nonumber\\=\frac{\abs{p_{E}}\abs{q_{E}}}{2\pi^2}\int_{\theta_{2}=0}^{\pi}\int_{\theta_{1}=0}^{\pi}\int_{\phi=0}^{2\pi}cos(\theta_{2})sin^2(\theta_{2})sin(\theta_{1})d\theta_{2}d\theta_{1}d\phi=0
\label{averaging}.
\end{eqnarray}
This trivial example leads to the formula for averaging $\left(p_{E,\mu}q^{\mu}_{E}\right)^n$:
\begin{eqnarray}
<\left(p_{\mu}q^{\mu}\right)^n >=\frac{\left(\abs{p_{E}}\abs{q_{E}}\right)^n}{2\pi^2}\int_{\theta_{2}=0}^{\pi}\int_{\theta_{1}=0}^{\pi}\int_{\phi=0}^{2\pi}cos^n(\theta_{2})sin^2(\theta_{2})sin(\theta_{1})d\theta_{2}d\theta_{1}d\phi .
\end{eqnarray}
For odd n the result vanishes. For even n some results are shown in table \ref{average}.
\begin{table}
\begin{tabular}[htbp]{||c|c||}
\hline
n& $\frac{<\left(p_{\mu}q^{\mu}\right)^n >}{\left(\abs{p}\abs{q}\right)^n}$\\
\hline 
2&$\frac{1}{4}$\\
\hline  
4&$\frac{1}{8}$\\ 
\hline 
6&$\frac{5}{64}$\\
\hline                                 
\end{tabular}
\caption{Averaging the scalar product.\label{average}}
\end{table}
This example directly shows how (\ref{unaveraged}) can be averaged over all possible orientations of $p_{\mu}$ and $q_{\mu}$. After the transformation to Euclidean space the procedure just outlined is repeated. The only difference is the integrand in (\ref{averaging}) it is replaced by (\ref{unaveraged}). This integration is not trivial but possible, algebraic computer tools can perform it easily. The result is
\begin{eqnarray}
C_{\overline{q}q}=-\frac{1}{m^2}\frac{(1-v)(1+2v)}{1+v}
\end{eqnarray}
where $v=\sqrt{1-\frac{4m^2}{q^2}}$. Some authors have an additional power of m in the coefficient \cite{Bagan:1985zp} because they take $\bar{q}q$ and not $m\bar{q}q$ as the condensate.\\
An important remark still has to be stated. The integration that has to be performed during the averaging process could here be done in closed form. In general this is not possible. Thus in such cases the integrand has to be expanded in the scalar product. The term in the expansion can all be averaged as it was shown in (\ref{averaging}). Therefore an expansion of the Wilson coefficient is calculable. At this point the problems begin. The summands are in principle given by powers of $\frac{m}{q^2}$ if $m$ is small compared to $q^2$ the expansion can be truncated. The first summand is often a satisfying approximation. In the case of big masses $m$ this can not be done, each summand is important. Hence, the series has to be resummed. The resummation is a difficult process. During the investigations made for this thesis no systematic method was found to achieve the resummation. However, the authors of \cite{Shifman:1978bx} somehow solved this problem.\\
The renormalization group flow of the coefficient is neglected as it was explained in section \ref{opeinqcd}. Thus the calculations are finished and can be comprised as the sequence: 
\begin{enumerate}
\item Apply the plane-wave method with appropriate external states.
\item Bring all free spinors to the right of the expression and identify them as placeholders for the 
      condensates.
\item Eliminate the external momenta, either by setting their square equal to the corresponding masses       or by using an appropriate average method. 
\end{enumerate}

\subsection{Equal mass case : the flavor conserving pseudoscalar current correlator \label{pseudoscalar-quark}}

In this case everything is analogous to the scalar current correlator, except of a trivial point concerning $\gamma_{5}$ matrices. The current is given by
\begin{eqnarray}
j(x)=\bar{q}(x)\left(i\gamma_{5}\right)q(x)~~j^{\dagger}(x)=\bar{q}(x)\left(i\gamma_{5}\right)q(x)
\end{eqnarray}
where the $i$ is necessary in order to make the current self adjoint. Again the coefficient is calculated to lowest order $\alpha_{S}^0$. The application of Wick's theorem results in
\begin{eqnarray}
i\int d^4xe^{iqx}\wick{1-2,7-8}{4-5}{\bra{p},\bar{q}(x),\left(i\gamma_{5}\right),q(x),\bar{q}(0),\left(i\gamma_{5}\right),q(0),\ket{p}}+i\int d^4xe^{iqx}\wick{1-5,4-8}{2-7}{\bra{p},\bar{q}(x),\left(i\gamma_{5}\right),q(x),\bar{q}(0),\left(i\gamma_{5}\right),q(0),\ket{p}}\nonumber\\=
\parbox[c]{3cm}{
\begin{fmffile}{diagram01}
\begin{fmfgraph*}(40,35)
 \fmfleft{i1,o1} \fmfright{i2,o2}
 \fmffixedx{2cm}{v1,v2}
 \fmf{dots_arrow,label.side=left,label=$q$}{i1,v1}
 \fmf{fermion,label.side=down,label=$q+p$}{v1,v2}
 \fmf{dots_arrow,label.side=left,label=$q$}{v2,i2}
 \fmf{fermion,label.side=right,label=$p$}{o1,v1}
 \fmf{fermion,label.side=right,label=$p$}{v2,o2}
\end{fmfgraph*}
\end{fmffile}}
~~~~~~~~~+
\parbox[c]{3cm}{
\begin{fmffile}{diagram02}
\begin{fmfgraph*}(40,35)
 \fmfleft{i1,o1} \fmfright{i2,o2}
 \fmffixedx{2cm}{v1,v2}
 \fmf{dots_arrow,label.side=left,label=$q$}{v2,i2}
 \fmf{dots_arrow,label.side=left,label=$q$}{i1,v1}
 \fmf{fermion,label.side=right,label=$p$}{o1,v2}
 \fmf{fermion,label.side=down,label=$q-p$}{v2,v1}
 \fmf{fermion,label.side=right,label=$p$}{v1,o2}
\end{fmfgraph*}
\end{fmffile}}
\nonumber\\=-\bar{u}(p)\left( i\gamma_{5}\right) \frac{\slashed{p}+\slashed{q}+m}{(p+q)^2-m^2}\left( i\gamma_{5}\right) u(p)-\bar{u}(p)\left( i\gamma_{5}\right) \frac{\slashed{p}-\slashed{q}+m}{(p-q)^2-m^2}\left( i\gamma_{5}\right) u(p)\label{pseudoscalarquark}.
\end{eqnarray}
Step 1 is this time a bit longer because the $\gamma_{5}$ matrices have to be accounted for. Their action is the change of the sign in front the slashed quantities. The expression which determines the Wilson coefficient is altered into
\begin{eqnarray}
\bar{u}(p)\frac{-\slashed{p}-\slashed{q}+m}{(p+q)^2-m^2}u(p)+\bar{u}(p)\frac{-\slashed{p}+
\slashed{q}+m}{(p-q)^2-m^2}u(p).
\end{eqnarray}
Now, using the same manipulations as for the scalar current the spinors can be transferred to the right
\begin{eqnarray}
\frac{-m-\frac{p_{\mu}q^{\mu}}{m}+m}{(p+q)^2-m^2}\bar{u}(p)u(p)+\frac{-m+
\frac{p_{\mu}q^{\mu}}{m}+m}{(p-q)^2-m^2}\bar{u}(p)u(p)=\nonumber\\
\frac{1}{m^2}\frac{-p_{\mu}q^{\mu}}{q^2+2p_{\mu}q^{\mu}}m\bar{u}(p)u(p)+\frac{1}{m^2}\frac{
p_{\mu}q^{\mu}}{q^2-2p_{\mu}q^{\mu}}m\bar{u}(p)u(p)\label{pseudodetermined}.
\end{eqnarray}
In comparison with (\ref{scalardetermined}) the numerator of (\ref{pseudodetermined}) has obviously changed. Hence, the expression to be averaged is given by 
\begin{eqnarray}
\frac{1}{m^2}\frac{-p_{\mu}q^{\mu}}{q^2+2p_{\mu}q^{\mu}}+\frac{1}{m^2}\frac{
p_{\mu}q^{\mu}}{q^2-2p_{\mu}q^{\mu}}.
\end{eqnarray}
The relevant integral is of the same type as discussed in section \ref{Wilson_Berechnung} and can be obtained in closed form
\begin{eqnarray}
C_{\overline{q}q}=\frac{1}{m^2}\frac{1-v}{1+v}.
\end{eqnarray}

\subsection{Unequal mass case : the flavor changing scalar current correlator \label{hl-scalar-quark}}

From the OPE point of view mesonic systems consisting of two quarks with unequal masses are more complicated. In the case of scalar mesons such systems are interpolated by the current
\begin{eqnarray}
j(x)=\bar{q}_{1}(x)q_{2}(x)~~j^{\dagger}(x)=\bar{q}_{2}(x)q_{1}(x).
\end{eqnarray}
In order to calculate a Wilson coefficient of such systems the sequence just quoted has to be passed trough as it was done in the case of equal mass systems. The principle of the calculation remains unchanged, but the steps are more complicated. The vacuum polarization transformed to momentum space is:
\begin{eqnarray}
i\int d^4xe^{iqx}\bra{0}T(j(x)j^{\dagger}(0))\ket{0}=i\int d^4xe^{iqx}\bra{0}T(\bar{q}_{1}(x)q_{2}(x)\bar{q}_{2}(0)q_{1}(0))\ket{0}.
\end{eqnarray}
For the calculation of the $m\bar{q}q$ Wilson coefficient the plane-wave method is applied
\begin{eqnarray}
i\int d^4xe^{iqx}\bra{p_{1}}T(j(x)j^{\dagger}(0))\ket{p_{1}}=i\int d^4xe^{iqx}\bra{p_{1}}T(\bar{q}_{1}(x)q_{2}(x)\bar{q}_{2}(0)q_{1}(0))\ket{p_{1}}.
\end{eqnarray}
The plane-wave method discussed in section \ref{Wilson_Berechnung} must be altered a bit to account for the two different types of quark fields. Thus, two quark condensates enter the calculation, one for each field. Correspondingly, the external states of the amplitude are either of type 1 or type 2. This is symbolized by the 1 at the external states. Again the Wilson coefficient is derived to lowest order
\begin{eqnarray}
i\int d^4xe^{iqx}\wick{1-2,5-6}{3-4}{\bra{p,1},\bar{q}_{1}(x),q_{2}(x),\bar{q}_{2}(0),q_{1}(0),\ket{p,1}}=
\parbox[c]{3cm}{
\begin{fmffile}{unequal01}
\begin{fmfgraph*}(40,35)
 \fmfleft{i1,o1} \fmfright{i2,o2}
 \fmffixedx{2cm}{v1,v2}
 \fmf{dots_arrow,label.side=left,label=$q$}{i1,v1}
 \fmf{fermion,label.side=down,label=$q+p$,width=2}{v1,v2}
 \fmf{dots_arrow,label.side=left,label=$q$}{v2,i2}
 \fmf{fermion,label.side=right,label=$p$}{o1,v1}
 \fmf{fermion,label.side=right,label=$p$}{v2,o2}
\end{fmfgraph*}
\end{fmffile}}\nonumber\\
=-\bar{u}_{1}(p)\frac{\slashed{p}+\slashed{q}+m_{2}}{(p+q)^2-m_{2}^2}u_{1}(p).
\end{eqnarray}
Note that the exchange term does not contribute to this amplitude. Using the Dirac Equation  $\bar{u}(p)(\slashed{p}-m_{1})=0$ or $\bar{u}_{1}(p,s)(\slashed{p}-m_{1})=0$ leads to
\begin{eqnarray}
-\bar{u}_{1}(p)\frac{m_{1}+\slashed{q}+m_{2}}{(p+q)^2-m_{2}^2}u_{1}(p).
\end{eqnarray}
The $\slashed{q}$-term is treated as in section \ref{Wilson_Berechnung}. The amplitude is given by
\begin{eqnarray}
-\frac{1}{m_{1}}\frac{m^2_{1}+p_{\mu}q^{\mu}+m_{1}m_{2}}{(p+q)^2-m_{2}^2}\bar{u}_{1}(p)u_{1}(p)\label{unequalscalar}.
\end{eqnarray}
In subsequent applications the limit $m_{1}\rightarrow 0$ is of interest. Therefore it will be calculated here 
\begin{eqnarray}
-\frac{1}{m_{1}}\frac{m^2_{1}+m_{1}m_{2}+p_{\mu}q^{\mu}}{q^2-m_{2}^2+m_{1}^2+2q_{\mu}p^{\mu}}\bar{u}_{1}(p)u_{1}(p).
\end{eqnarray}
The expression for the Wilson coefficient is expanded in a series 
\begin{eqnarray}
-\frac{1}{m_{1}(q^2-m_{2}^2)}\frac{m^2_{1}+m_{1}m_{2}+p_{\mu}q^{\mu}}{1+\frac{m_{1}^2+2qp}{q^2-m_{2}^2}}=-\frac{m^2_{1}+m_{1}m_{2}+p_{\mu}q^{\mu}}{m_{1}(q^2-m_{2}^2)}\left[1- \frac{m_{1}^2+2qp}{q^2-m_{2}^2}+...\right. .
\end{eqnarray}
In the limit $m_{1}\rightarrow 0$ only the leading term survives, remember $p_{\mu}q^{\mu}=m_{1}\sqrt{q_{\mu}q^{\mu}}cos(\theta)$ because $p_{E,i}p_{E}^{i}=p_{\mu}p^{\mu}$ per definition
\begin{eqnarray}
-\frac{m_{2}+\abs{q}cos(\theta)}{q^2-m_{2}^2}.
\end{eqnarray}
After averaging over the four dimensional Euclidean angle $\theta$ the term is given by
\begin{eqnarray}
-\frac{m_{2}}{q^2-m_{2}^2}.
\end{eqnarray}
Hence the Wilson coefficient in this approximation is given by
\begin{eqnarray}
C_{\bar{q}q}=\frac{-m_{2}}{q^2-m_{2}^2}=\frac{m_{2}}{m_{2}^2-q^2}\label{unequalscalarresult}.
\end{eqnarray}

\subsection{Unequal mass case : the flavor changing pseudoscalar current correlator}

The calculation goes along the same lines as in sections \ref{pseudoscalar-quark} and \ref{hl-scalar-quark}. The computations are illustrated here:
\begin{eqnarray}
j(x)=\bar{q}_{1}(x)\left(i\gamma_{5}\right)q_{2}(x)~~j^{\dagger}(x)=\bar{q}_{2}(x)\left(i\gamma_{5}\right)q_{1}(x).
\end{eqnarray}
Hence, the Wilson coefficient is determined by
\begin{eqnarray}
i\int d^4xe^{iqx}\wick{1-2,7-8}{4-5}{\bra{p}_{1},\bar{q}_{1}(x),\left(i\gamma_{5}\right),q_{2}(x),\bar{q}_{2}(0),\left(i\gamma_{5}\right),q_{1}(0),\ket{p}_{1}}=
\parbox[c]{3cm}{
\begin{fmffile}{unequal01}
\begin{fmfgraph*}(40,35)
 \fmfleft{i1,o1} \fmfright{i2,o2}
 \fmffixedx{2cm}{v1,v2}
 \fmf{dots_arrow,label.side=left,label=$q$}{i1,v1}
 \fmf{fermion,label.side=down,label=$q+p$,width=2}{v1,v2}
 \fmf{dots_arrow,label.side=left,label=$q$}{v2,i2}
 \fmf{fermion,label.side=right,label=$p$}{o1,v1}
 \fmf{fermion,label.side=right,label=$p$}{v2,o2}
\end{fmfgraph*}
\end{fmffile}}\nonumber\\
=-\bar{u}_{1}(p)\left(i\gamma_{5}\right)\frac{\slashed{p}+\slashed{q}+m_{2}}{(p+q)^2-m_{2}^2}\left(i\gamma_{5}\right)u_{1}(p)=-\frac{1}{m_{1}}\frac{m^2_{1}-m_{1}m_{2}+p_{\mu}q^{\mu}}{q^2-m_{2}^2+m_{1}^2+2qp}\bar{u}_{1}(p)u_{1}(p).
\end{eqnarray}
In comparison with (\ref{unequalscalar}) only the sign of the $m_{1}m_{2}$-term in the numerator is changed. Thus, the result in the approximation where $m_{1}=0$ is given by (\ref{unequalscalarresult}) with $m_{2}$ replaced my $-m_{2}$
\begin{eqnarray}
C_{\bar{q}q}=\frac{m_{2}}{q^2-m_{2}^2}=\frac{-m_{2}}{m_{2}^2-q^2}\label{unequalpseudoscalarresult}.
\end{eqnarray}

\subsection{Equal mass case : the flavor conserving vector current correlator}

Vector mesons where the valence quark and antiquark have the same flavor are extrapolated by the current:
\begin{eqnarray}
j_{\mu}(x)=\bar{q}(x)\gamma_{\mu}q(x)~~j^{\dagger}_{\mu}(x)=\bar{q}(x)\gamma_{\mu}q(x).
\end{eqnarray}
The standard plane-wave analysis discussed above leads to the expression:
\begin{eqnarray}
\bra{p}j_{\mu}(x)j^{\dagger}_{\nu}(0)\ket{p}=\nonumber\\
i\int d^4xe^{iqx}\wick{1-2,7-8}{4-5}{\bra{p},\bar{q}(x),\gamma_{\mu},q(x),\bar{q}(0),\gamma_{\nu},q(0),\ket{p}}+i\int d^4xe^{iqx}\wick{1-7}{4-5,2-8}{\bra{p},\bar{q}(x),\gamma_{\mu},q(x),\bar{q}(0),\gamma_{\nu},q(0),\ket{p}}\nonumber\\
=\bar{u}(p,s)\gamma_{\mu}\frac{\slashed{p}+\slashed{q}+m}{(p+q)^2-m^2}\gamma_{\nu}u(p)+\bar{u}(p)\gamma_{\nu}\frac{\slashed{p}-\slashed{q}+m}{(p-q)^2-m^2}\gamma_{\mu}u(p) \label{vectorcoefficient}.
\end{eqnarray}
The occurrence of open Dirac indices is new. Such indices did not occur in the calculations to scalar particles. Fortunately the expressions for vector particels can be factorized in Lorentz tensors and Lorentz scalar parts as:
\begin{eqnarray}
\Pi_{\mu\nu}\left(q^2\right) =\left(g_{\mu\nu}-\frac{q_{\mu}q_{\nu}}{q^2}\right)\Pi_{T}\left(q^2\right)-\frac{q_{\mu}q_{\nu}}{q^2}\Pi_{L}\left( q^2\right). 
\end{eqnarray}
Due to current conservation only the transversal part contributes to the amplitudes, the 
longitudinal part can be neglected. Therefore it is removed from the amplitudes. After this subtraction only the transversal part times a Lorentz tensor is left. The transversal part is a Lorentz scalar. In order to keep the calculations simple only the transversal part is calculated. In comparison with the scalar and pseudoscalar coefficients the additional work is given solely by the procedure which extracts the transversal part of the coefficient from the full expression of the coefficient. Three steps are necessary in order to accomplish the extraction.
\begin{enumerate}
\item Contract the amplitude with $g_{\mu\nu}$, and average this expression over the four dimensional Euclidean angle. The result is $3C_{\bar{q}q,T}\left(q^2\right)-C_{\bar{q}q,L}\left( q^2\right)$.
\item Contract the amplitude with $\frac{q_{\mu}q_{\nu}}{q^2}$, and average this expression over the four dimensional Euclidean angle. The result is $-C_{\bar{q}q,L}\left( q^2\right)$.
\item Subtract those two expressions and divide by 3. The result is  $C_{\bar{q}q,T}\left(q^2\right)=C_{\bar{q}q}\left(q^2\right)$.
\end{enumerate}
Some comments on the calculation of Wilson coefficients can also be found in  \cite{Bagan:1985zp}. As an example the Wilson coefficient for the quark condensate in the vector current correlator case is calculated to lowest order in the quark mass.
\begin{enumerate}
\item Contraction of (\ref{vectorcoefficient}) with $g_{\mu\nu}$
\begin{eqnarray}
\bar{u}(p)\frac{\left( \slashed{p}+\slashed{q}\right) +m}{(p+q)^2-m^2}u(p)+\bar{u}(p)\frac{-2\left( \slashed{p}-\slashed{q}\right) +4m}{(p-q)^2-m^2}u(p)=\nonumber\\\bar{u}(p)\frac{2}{m}\left[\frac{m^2-pq}{(p+q)^2-m^2}+\frac{m^2+pq}{(p-q)^2-m^2}\right]u(p)\label{contractcoef01}.
\end{eqnarray}
The expressions have to be averaged over the four dimensional Euclidean angle. This can be done directly with the expression in (\ref{contractcoef01}) or with each term of its Taylor expansion. The expansion is done in the scalar product $pq$, the result to lowest order in the mass is $\frac{4m}{q^2}$.
\item Contraction of (\ref{vectorcoefficient}) with $\frac{q_{\mu}q_{\nu}}{q^2}$
\begin{eqnarray}
\frac{\bar{u}(p)}{q^2}\frac{\slashed{q}\left(\slashed{p}+\slashed{q} +m\right)\slashed{q}}{(p+q)^2-m^2}u(p)+\frac{\bar{u}(p)}{q^2}\frac{\slashed{q}\left(\slashed{p}-\slashed{q} +m\right)\slashed{q}}{(p-q)^2-m^2}u(p)
=\nonumber\\
\frac{\bar{u}(p)}{q^2}\frac{\left(2qp\slashed{q}+q^2(-\slashed{p}+\slashed{q}+m\right)}{(p+q)^2-m^2}u(p)+\frac{\bar{u}(p)}{q^2}\frac{\left(2qp\slashed{q}+q^2(-\slashed{p}-\slashed{q}+m) \right)}{(p-q)^2-m^2}u(p)=\nonumber\\\bar{u}(p)
\frac{1}{m}\left[\frac{2(pq)^2+q^2pq}{(p+q)^2-m^2}+\frac{2(pq)^2-q^2pq}{(p-q)^2-m^2}\right]u(p)\label{contractcoef02}.
\end{eqnarray}
Again the Taylor expansion is considered. In lowest order of the mass the longitudinal part of the Wilson coefficient is $\frac{-2m}{q^2}$.
\item The last step is simple and the result is 
\begin{eqnarray}
C_{\bar{q}q}=\frac{2m}{q^2}+\mathcal{O}\left(\frac{1}{q^4}\right) \label{lightvector}.
\end{eqnarray}
\end{enumerate}
The solution quoted in (\ref{lightvector}) can be used for light quark systems. In the case of heavy quark systems all orders in he Taylor expansion have to be considered. Hence, the expansion has to be summed up or the average has to be taken direct without doing a Taylor expansion. The result where all orders in mass are included is:
\begin{eqnarray}
C_{\bar{q}q}(q^2)=-\frac{2}{3m}\frac{(1-v)(2+v)}{(1+v)}.
\end{eqnarray}
The examples just shown illustrates the plane-wave method for the computation of Wilson coefficients. They have been arranged from simple to difficult but it does not end at the vectorparticle level. Tensor particels of arbitrary rank, can also be considered. With increasing rank the Dirac structure and the corresponding projector are more and more complex. Additional complications arise in loop diagrams and in diagrams involving gluons. Furthermore, there are ambiguities in the interpolating currents (see section \ref{gluonmix}).\\
The plane-wave method is the most rudimentary method for the computations of Wilson coefficients. Every method posses advantages and disadvantages. The advantages of the plane-wave method are that it can be used for the calculation of every Wilson coefficient without having to learn a great amount of new formalisms. The disadvantage is the complexity in the calculations which have to be performed. There exist methods where the calculations are significantly less complex. The most important one besides the plane-wave method is the fixed point gauge technique, also called background field method. There are many examples concerning the application of this method in \cite{Narison:1989aq}.

\subsection{Diagrammatic representation of an OPE}
  
An OPE of the two point correlator in a meson channel can be graphically represented. This is illustrated below for the first two terms, in the OPE
\begin{eqnarray}
\bra{p}j_{\mu}(x)j^{\dagger}_{\nu}(0)\ket{p}=
\parbox[c]{3cm}{
\begin{fmffile}{loop}
\begin{fmfgraph*}(30,30)

 \fmfleft{i} 
    \fmfright{o}
    \fmf{dots,label.side=down}{i,v1} 
    \fmf{dots,label.side=down}{v2,o}
    \fmf{plain,label.side=left,left,tension=.3}{v1,v2}
    \fmf{plain,label.side=left,left,tension=.3}{v2,v1}
    \fmfdotn{v}{2}

\end{fmfgraph*}
\end{fmffile}}
\mathds{1}+
\parbox[c]{3cm}{
\begin{fmffile}{quark}
\begin{fmfgraph*}(30,30)

 \fmfleft{i} 
    \fmfright{o}
    \fmf{dots,label.side=down}{i,v1} 
    \fmf{dots,label.side=down}{v2,o}
    \fmf{phantom,label.side=left,left,tension=.3,tag=1}{v1,v2}
    \fmf{plain,label.side=left,left,tension=.3}{v2,v1}
    \fmfdotn{v}{2}
    \fmfposition
    \fmfipath{p[]}
    \fmfiset{p1}{vpath1(__v1,__v2)}
    \fmfi{plain}{subpath (0,length(p1)/4) of p1}
    \fmfiv{d.sh=cross,d.ang=0,d.siz=5thick}{point length(p1)/4 of p1}
    \fmfi{plain}{subpath (3length(p1)/4,length(p1)) of p1}
    \fmfiv{d.sh=cross,d.ang=0,d.siz=5thick}{point 3length(p1)/4 of p1}
\end{fmfgraph*}
\end{fmffile}}
\mele{m\bar{q}q}.
\end{eqnarray}
This includes all terms in the OPE to lowest order in $\alpha_{S}$. The term in front of the unit operator is referred to as the  perturbative term. This terminology stems from the fact that the practical OPE (see section \ref{opeinqcd}) is used where the Wilson coefficient of the unit operator is given by perturbative QCD.\\
The term in front of the quark condensate is similar to the scattering diagram used in the calculation of the corresponding Wilson coefficient above. In fact the diagrammatic representation can be understood as stemming from the calculations done in the plane-wave method. The crosses which do not appear in the scattering diagram symbolize the contact of the Wilsoncoefficient with the quark condensate.\\
Note that the connection between the scattering diagrams and those representing the Wilson coefficients is non-trivial due to the angle average in Euclidean space discussed above.

\subsection{Quark mass effects \label{gluonmix}}

The last sections have been very general and mathematical but for applications of the OPE also physical boundary conditions have to be taken into account. A very important boundary condition is given by the mass of the quarks that are chosen to build up the interpolating currents. They are important for Wilson coefficients involving the gluon condensate.\\
Suppose the Wilson coefficient for the gluon condensate is calculated for the case of a current which is build up of two quarks with different flavor. Then two masses $m_{1}$ and $m_{2}$ enter the Wilson coefficient. Depending on the meson which should be approximated by the current  various limits for the masses are interesting. In the case of a heavy-light system the limit $m_{2}\rightarrow 0$ is interesting. Where $m_{1}$ should be the mass of the heavy and $m_{2}$ the mass of the light quark. Surprisingly this limit exhibits a serious problem. The Wilson coefficient in this limit diverges! According to physical and mathematical arguments this divergence should not occur. Hence, further considerations are necessary. Methods to handle them where already starting to develop in the original publications to the OPE \cite{Shifman:1978bx}, became fully developed in \cite{Reinders:1984sr} and where completed by \cite{Bagan:1985zp}.\\ 
Here a short introduction is given following \cite{Reinders:1984sr}. Sandwiching the OPE between one gluon states gives a suprising result: 
\begin{eqnarray}
\sum_{n}C_{n}\bra{k}O_{n}\ket{k}=C_{m_{1}}\bra{k}m_{1}\bar{q}_{1}q_{1}\ket{k}+C_{m_{2}}\bra{k}m_{2}\bar{q}_{2}q_{2}\ket{k}+C_{G}\bra{k}G^2\ket{k}+... \label{ope1gluon}.
\end{eqnarray}
Calculating the matrix element $\bra{k}m\bar{q}q\ket{k}$, that means writing down its OPE leads to
\begin{eqnarray}
\bra{k}m\bar{q}q\ket{k}=\frac{1}{12}\frac{\alpha_{s}}{\pi}\bra{k}G^2\ket{k}+...~\label{qqope}.
\end{eqnarray}
We see from this expression that the quark matrix element is of the same order in $\alpha_{s}$ as $C_{G}$, since in lowest order the coefficient $C_{m}$ is of zeroth order in $\alpha_{s}$. Inserting (\ref{qqope}) into (\ref{ope1gluon}) the following expression is obtained
\begin{eqnarray}
\sum_{n}C_{n}\bra{k}O_{n}\ket{k}=\left[C_{G}+\frac{1}{12}\frac{\alpha_{s}}{\pi}C_{m_{1}}+\frac{1}{12}\frac{\alpha_{s}}{\pi}C_{m_{2}}\right]\bra{k}G^2\ket{k}+...~.
\end{eqnarray}
Therefore the coefficient of the gluon condensate differs from what it was expected to be. The physical version is given by:
\begin{eqnarray}
C_{G,physical}=C_{G}+\frac{1}{12}\frac{\alpha_{s}}{\pi}\left(C_{m_{1}}+C_{m_{2}}\right).
\end{eqnarray}
The physical interpretation of this mixing is that the quark condensation has also to be taken into account in the Wilsoncoefficient of the gluon condensate. The mass of the quarks determines if their condensation has to be taken into account in the Wilson coefficient of the gluon condensate. A simple rule can be formulated. In the case of heavy quarks the quark condensation has not to be taken into account in the Wilson coefficient of the gluon condensate.  Moreover, the quark condensation can totally be neglected in the corresponding OPE. In the case of light quarks the situation is reversed. The quark condensation has to be taken into account in the gluon condensate coefficient and in the corresponding OPE.\\
These rules can be comprised in the statements that heavy quarks condense mainly trough gluons while light quarks condense mainly through quarks. The most drastic example is the $G^3$ condensate. Light quarks totally decouple from this condensate (see \cite{Bagan:1985zp} and references herein). Decoupling means that the Wilson coefficient is zero.\\
The effects just described depend on the relative size of the quark mass compared to the renormalization point $\mu$, that was introduced to separate long and short distance fluctuations. If the quark mass is smaller than $\mu$ quark condensation is favored, if it is bigger than $\mu$ gluon condensation of quarks is favored.\\
One way to understand the effect mathematically is to analyze the gluon condensate coefficient. The loop integral in the coefficient at large momentum $q^2$ receives its main contribution from two regions of virtual momenta. The first is $p^2\approx q^2$ and the second is $p^2\approx m^2$. For small quark masses the latter region can`t be included in the coefficient  $C_{G}$ since some of the quark propagators soft. This piece must be subtracted and absorbed into the matrix element $\bra{0}m\bar{q}q\ket{0}$. While for the heavy quarks the propagators are not soft in that region.\\

In summary this observation leads to the following expressions.
\begin{enumerate}
\item{For pure light quark systems the nonperturbative corrections are}
 \begin{eqnarray}
 C_{m_{1}}\bra{0}m_{1}\bar{q}_{1}q_{1}\ket{0}+C_{m_{2}}\bra{0}m_{2}\bar{q}_{2}q_{2}\ket{0}+C_{G,physical}\bra{0}G^2\ket{0}.     
 \end{eqnarray}
\item{For heavy-light systems $(m_{2}\gg m_{1})$ the corrections are}
 \begin{eqnarray}
 C_{m_{1}}\bra{0}m_{1}\bar{q}_{1}q_{1}\ket{0}+C_{m_{2}}\bra{0}m_{2}\bar{q}_{2}q_{2}\ket{0}+C_{G,physical}\bra{0}G^2\ket{0}
 =C_{m_{1}}\bra{0}m_{1}\bar{q}_{1}q_{1}\ket{0}\nonumber\\+\left(C_{G}+\frac{1}{12}\frac{\alpha_{s}}{\pi}C_{m_{1}}\right)\bra{0}G^2\ket{0}+ C_{m_{2}}\left( \bra{0}m_{2}\bar{q}_{2}q_{2}\ket{0}+\frac{1}{12}\frac{\alpha_{s}}{\pi}\bra{0}G^2\ket{0}\right).
 \end{eqnarray}
  The heavy quark mass expansion for the heavy quark condensate $\bra{0}\bar{h}h\ket{0}=-\frac{1}{m_{h}}\frac{\alpha_{s}}{12\pi}\bra{0}G^2\ket{0}+...$ is used to eliminate the $C_{m_{2}}$ term. The result for the non perturbative corrections is finally
  \begin{eqnarray}
  C_{m_{1}}\bra{0}m_{1}\bar{q}_{1}q_{1}\ket{0}+\left(C_{G}+\frac{1}{12}\frac{\alpha_{s}}{\pi}C_{m_{1}}\right)\bra{0}G^2\ket{0}.
  \end{eqnarray}
 \item{For pure heavy quark systems the relevant expression is}
  \begin{eqnarray}
  C_{m_{1}}\bra{0}m_{1}\bar{q}_{1}q_{1}\ket{0}+C_{m_{2}}\bra{0}m_{2}\bar{q}_{2}q_{2}\ket{0}+C_{G,physical}\bra{0}G^2\ket{0}=
   C_{G}\bra{0}G^2\ket{0}\nonumber\\+C_{m_{1}}\left(m_{1}\bra{0}\bar{q}_{1}q_{1}\ket{0}+\frac{1}{12}\frac{\alpha_{s}}{\pi}\bra{0}G^2\ket{0}\right)+C_{m_{2}}\left(\bra{0}m_{2}\bar{q}_{2}q_{2}\ket{0}+\frac{1}{12}\frac{\alpha_{s}}{\pi}\bra{0}G^2\ket{0}\right) .
  \end{eqnarray}
  Again the heavy quark mass expansion for the heavy quark condensate is used. The result is
   \begin{eqnarray}
    C_{G}\bra{0}G^2\ket{0}.
   \end{eqnarray}
   It is identical with the one calculated without taking into account the coefficients of the $m_{i}\bar{q}_{i}q_{i}$ terms. The conclusion is that the heavy quark condensate has practically no effect on the polarization functions. An example is the charmonium system, which in the OPE is determined exclusively by gluonic effects. 
\end{enumerate}
In retrospective this section improves the naive picture of the gluon Wilson coefficient. The condensation via quarks or gluons is entangled and can not be treated separately. 

\section{The condensates \label{condensates}}
In the last sections the Wilson coefficients were in principle determined, but what about the condensates? There exists no method to calculate them from first principles. All values found in the literature have been extracted from experiments. Some condensates have been directly extracted from the data, others involve additional estimations. Hence it is logical to start with the ones that have directly been extracted from the data. These are the matrix elements of the light quark condensates, some of  the four quark condensates, the gluon condensate and the mixed gluon condensate. The triple gluon condensate has also been directly extracted from data, but is not treated here.
\begin{itemize}
\item $\bra{0}\overline{u}u\ket{0}=-(0.250GeV)^3=-1.5625\cdot10^{-2}GeV^3$
\item $\bra{0}\overline{d}d\ket{0}=-(0.250GeV)^3=-1.5625\cdot10^{-2}GeV^3$
\item $\bra{0}\overline{s}s\ket{0}=0.8\times\bra{0}\overline{u}u\ket{0}=-1.25\cdot10^{-2}GeV^3$
\end{itemize}
Which have been computed in the formalism of Gell-Mann, Oaks,Renner \cite{Narison:1989aq}. Then there is a set of matrix elements which have computed with the help of QSRs.
\begin{itemize}
\item $\bra{0}\frac{\alpha_{s}}{\pi}G_{\mu\nu}^{c}G_{\mu\nu}^{c}\ket{0}=0.012~GeV^4=1.2\cdot10^{-2}~GeV^4$
\item $g\mele{\overline{q}\sigma^{\mu\nu}\frac{\lambda_{a}}{2}qG^{a}_{\mu\nu}}=M_{0}^{2}\mele{\overline{q}q}=0.80GeV^2\cdot(-1.5625\cdot10^{-2}GeV^3)=-0.0125GeV^5$
\end{itemize}
The four quark condensate and the triple gluon condensate have also been computed by using QSRs. All other matrix elements have been extracted from the data with less direct methods. In the case of the heavy quark condensates the expansion of the condensate reads
\begin{eqnarray}
\bra{0}\overline{h}h\ket{0}=-\frac{1}{12m_{h}}\bra{0}\frac{\alpha_{s}}{\pi}G_{\mu\nu}^{c}G_{\mu\nu}^{c}\ket{0}+...~.
\end{eqnarray}
The next to leading terms are multiplied by higher powers of $\frac{1}{m_{h}}$. Hence, they are suppressed in the heavy quark case. The estimates for the heavy quark condensates in first order are:
\begin{itemize}
\item $m_{c}=001.30GeV \rightarrow \bra{0}\overline{c}c\ket{0}=-7.693\cdot10^{-4}~GeV^3$
\item $m_{b}=004.50GeV \rightarrow \bra{0}\overline{b}b\ket{0}=-2.2\cdot10^{-4}~GeV^3$
\item $m_{t}=176.20GeV \rightarrow \bra{0}\overline{t}t\ket{0}=-5.675\cdot10^{-6}~GeV^3$.
\end{itemize}
Thus the heavy quark condensates are negligible compared to the light quark condensates. The estimates of higher dimensional quark condensates goes along different lines. The basic idea that stands behind the procedure is the assumption of vacuum state dominance. This idea is illustrated by factorization of a simplified 4 quark condensate  
\begin{eqnarray}
\bra{0}\bar{q}q\bar{q}q\ket{0}=\bra{0}\bar{q}q\mathds{1}\bar{q}q\ket{0}=\bra{0}\bar{q}q\sum_{n}\ket{n}\bra{n}\bar{q}q\ket{0}=\sum_{n}\bra{0}\bar{q}q\ket{n}\bra{n}\bar{q}q\ket{0}\approx\bra{0}\bar{q}q\ket{0}\bra{0}\bar{q}q\ket{0}.
\end{eqnarray}
After the insertion of a complete set of states it is assumed that everything except the term with the vacuum states can be neglected. Hence, the vacuum is assumed to be the numerically biggest contribution. The result is of course only an estimate, but a very handy one. The simplified four quark condensate has been factorized in the product of two two-quark condensates. These condensates have been estimated as shown before.\\ 
The four quark condensate as defined in (\ref{operators}) has an interior structure which changes the factorization procedure slightly, but the calculations give insight in an interesting fact and will therefore be shown here.\\
In many calculations $\Gamma_{1}$ and $\Gamma_{2}$ turn out to be $\lambda_{a}\gamma_{\mu}$ where $\lambda_{a}=2t_{a}$ and the $t_{a}$s are the Gell-Mann flavor matrices. The condensate for this case is factorized here. Here only quarks of one color are considered
\begin{eqnarray}
\bra{0}\bar{q}\lambda_{a}\gamma_{\mu}q\bar{q}\lambda_{a}\gamma_{\mu}q\ket{0}=
\bra{0}\bar{q}_{i,\alpha}\gamma^{\alpha\beta}_{\mu}\lambda_{a}^{ij}q_{j,\beta}\bar{q}_{k}^{\eta}\gamma^{\mu}_{\eta\nu}\lambda_{a}^{kl}q_{l}^{\nu}\ket{0}=\gamma^{\alpha\beta}_{\mu}\lambda_{a}^{ij}\gamma^{\mu}_{\eta\nu}\lambda_{a}^{kl}\bra{0}\bar{q}_{i,\alpha}q_{j,\beta}\bar{q}_{k}^{\eta}q_{l}^{\nu}\ket{0}
\label{4quarkfacto}.
\end{eqnarray}
The interior structure is already factored out, but in this example a non-trivial index structure of the quark fields is given and there remains one step before the vacuum saturation assumption can be applied. Two kinds of indices have to be considered next. They are the Dirac indices $\alpha$, $\beta$, $\eta$ and $\nu$ and the flavor indices $i$,$j$,$k$ and $l$. Vacuum expectation values of operators do not have spin or flavor and of course no color. Therefore the flavors and the anti-flavors have to cancel and also the operators must be a scalar. To achieve this the fields are grouped in two pairs in each pair is a $q$ and $\bar{q}$ field operator. The indices in these pairs have to be equal in order to cancel the properties they correspond to. Hence two combinations are possible.
\begin{itemize}
\item $\bar{q}_{i,\alpha}q_{j,\beta}\bar{q}_{k}^{\eta}q_{l}^{\nu}$
\item $\bar{q}_{i,\alpha}q_{l}^{\nu}\bar{q}_{k}^{\eta}q_{j,\beta}$
\end{itemize}
Therefore (\ref{4quarkfacto}) can be written as
\begin{eqnarray}
\gamma^{\alpha\beta}_{\mu}\lambda_{a}^{ij}\gamma^{\mu}_{\eta\nu}\lambda_{a}^{kl}\bra{0}\bar{q}_{i,\alpha}q_{j,\beta}\bar{q}_{k}^{\eta}q_{l}^{\nu}-\bar{q}_{i,\alpha}q_{l}^{\nu}\bar{q}_{k}^{\eta}q_{j,\beta}\ket{0}\label{4quarkgeordnet}.
\end{eqnarray}
The double summation in the right summand is cancelled after application of the vacuum states and the minus sign stems from the anti-commutator relation that has to be used in order to change the position of the quarkfields. As the last step the summation has to be performed. It is achieved by employing the relation
\begin{eqnarray}
\bra{0}\bar{q}_{A}q_{B}\ket{0}=\frac{\delta_{AB}}{N}\bra{0}\bar{q}q\ket{0}\label{superindex}
\end{eqnarray}
where the quark fields carry a superindex which collects all possible indices and the quarkfields without indices are the summed up fields in which the indices are always equal. The normalization factor counts the summands in this sum. In the actual example it is $4\times3=12$. Application of the relation (\ref{superindex}) to (\ref{4quarkgeordnet}) results in:
\begin{eqnarray}
\frac{\gamma^{\alpha\alpha}_{\mu}\lambda_{a}^{ii}\gamma^{\mu}_{\eta\eta}\lambda_{a}^{kk}}{N^2}\bra{0}\bar{q}_{i,\alpha}q_{i,\alpha}\bar{q}_{k}^{\eta}q_{k}^{\eta}\ket{0}-\frac{\gamma^{\alpha\eta}_{\mu}\lambda_{a}^{ik}\gamma^{\mu}_{\eta\alpha}\lambda_{a}^{ki}}{N^2}\bra{0}\bar{q}_{i,\alpha}q_{i}^{\alpha}\bar{q}_{k}^{\eta}q_{k,\eta}\ket{0}.
\end{eqnarray}
The corresponding index free notation is
\begin{eqnarray}
\frac{1}{N^2}\left[tr\left(\gamma_{\mu}\lambda_{a}\right)tr\left(\gamma^{\mu}\lambda_{a}\right)-tr\left(\gamma_{\mu}\lambda_{a}\gamma^{\mu}\lambda_{a} \right)\right] 
\bra{0}\bar{q}q\bar{q}q\ket{0}.
\end{eqnarray}
At this stage its recommendable to apply the vacuum saturation hypothesis. The result is
\begin{eqnarray}
\frac{1}{N^2}\left[tr\left(\gamma_{\mu}\lambda_{a}\right)tr\left(\gamma^{\mu}\lambda_{a}\right)-tr\left(\gamma_{\mu}\lambda_{a}\gamma^{\mu}\lambda_{a} \right)\right] 
\bra{0}\bar{q}q\ket{0}^2=-\frac{16}{9}\bra{0}\bar{q}q\ket{0}^2
\end{eqnarray}
where also the numerical result is displayed. The full result also has to incorporate the possibility of different colors, but the structure of the factorization formula stays the same 
\begin{eqnarray}
\bra{0}\bar{\psi}\Gamma_{1}\psi\bar{\psi}\Gamma_{2}\psi\ket{0}=
\frac{1}{N^2}\left[ tr\Gamma_{1}tr\Gamma_{2}-
tr\Gamma_{1}\Gamma_{2}\right] 
\bra{0}\bar{\psi}\psi\ket{0}^2
\end{eqnarray}
as shown in \cite{Shifman:1978bx}. Here the normalization factor is $4\times3\times3=36$ and $\bra{0}\bar{\psi}\psi\ket{0}=\bra{0}(\bar{u}u+\bar{d}d+\bar{s}s)\ket{0}$. The factorization can also be achieved by using Fierz transformations. An important example for a factorization of a four quark condensate is:
\begin{eqnarray}
\alpha_{s}\mele{\overline{q}\gamma_{\mu}\lambda_{a}q\sum_{n=u,d,s}(\overline{n}\gamma^{\mu}\lambda_{a}n)}=-\frac{16}{9}\alpha_{s}\mele{\overline{q}q}^2=-\frac{16}{9}1.75\cdot 10^{-4}GeV^6=-3.11\cdot 10^{-4}GeV^6
\end{eqnarray}
(see \cite{Narison:1989aq}). Along these lines all quark condensates can be factorized in the approximation where the vacuum saturates the whole condensate but if this assumption is valid or not is another question. There hadrons in QCD which are very sensitive to the four quark condensate. If the Sum Rules are fitted to the data, it turns out that the vacuum saturation hypothesis fails. In order to incorporate the violation of the vacuum hypothesis a factor $k$ is introduced. The factorized four quark condensate is multiplied by this factor
\begin{eqnarray}
\bra{0}\bar{\psi}\Gamma_{1}\psi\bar{\psi}\Gamma_{2}\psi\ket{0}=
k\cdot\frac{1}{N^2}\left[ tr\Gamma_{1}tr\Gamma_{2}-
tr\Gamma_{1}\Gamma_{2}\right] 
\bra{0}\bar{\psi}\psi\ket{0}^2.
\end{eqnarray}
This factor takes into account the violation of the vacuum saturation hypothesis. If $k=1$ the hypothesis holds, if it is not the hypothesis is violated. The four quark condensate is still subject to new estimations. The value for $k$ seems to change steadily \cite{Narison:1990cy,Leupold:1997dg}. However, in this work the validity of the vacuum saturation hypothesis is assumed.\\
Estimates of higher gluon condensates exist, but are not reliable. Therefore, they have been extracted from the data.\\
The last remaining condensate on the list is
 $g\mele{\overline{q}\sigma^{\mu\nu}\frac{\lambda_{a}}{2}qG^{a}_{\mu\nu}}$ here also no 
satisfactory method for the estimation or calculation exists and this condensate has again been extracted from data \cite{Narison:1990cy}.\\
In conclusion it can be said that in contrast to the Wilson coefficients the knowledge of the condensates is poor and relatively high errors are connected with their values. Clearly, the reason for this is that they involve non-perturbative physics, while the Wilson coefficients can be computed perturbatively.

\section{The dispersion relations\label{dispersionrelations}}
The central elements in QCD Sum Rules are the OPE and the dispersion relations. Above the OPE has been introduced and here the dispersion relations will be reviewed. First the basic formulas are introduced and different regularization methods are discussed. It turns out that everything is based on only one dispersion relation. The derivation of this relation will be explained in the subsequent section.
\subsection{The formulae used \label{formulae}}
The relation on which all other relations are based on is given by
\begin{eqnarray}
\Pi\left(q^2\right)=\frac{1}{\pi}\int_{0}^{\infty}\frac{Im\Pi(s)}{(s-q^2)} ds \label{dispersion}
\end{eqnarray}
and states that a correlator is known if its imaginary part along the positive real axis is known. In the derivation only causality has been assumed. In practical applications (\ref{dispersion}) has to be checked for convergence. $Im\Pi(s)$ can be very steep and the integral can be divergent. In such a case the integral has to be subtracted. Therefore, a Taylor expansion of the correlator is performed
\begin{eqnarray}
\Pi\left(q^2\right)=\Pi\left(0\right)+\left[\frac{d}{dq^2}\Pi\left(q^2\right)\right]_{q^2=0}q^2+...\nonumber\\+\frac{1}{(n-1)!}\left[\left( \frac{d}{dq^2}\right) ^{n-1}\Pi\left(q^2\right)\right]_{q^2=0}\left( q^2\right)^{n-1}+\frac{(q^2)^n}{\pi}\int\frac{Im\Pi(s)}{s^n(s-q^2)} ds \label{subtracted}.
\end{eqnarray}
The first n-1 terms are the expansion while the last term is the remainder term of the expansion. Thus, the integral is split in two parts a polynomial in $q^2$ and a subtracted dispersion integral. Divergences of the correlator are eliminated by renormalizing the coefficients of the polynomial. For a logarithmic divergence one substitution is sufficient
\begin{eqnarray}
\Pi\left(q^2\right)=\Pi\left(0\right)+\frac{q^2}{\pi}\int\frac{Im\Pi(s)}{s(s-q^2)} ds \label{onesubtraction}.
\end{eqnarray}
After renormalization of the amplitude ensures that $\Pi\left(0\right)$ is either finite or 0. If  $\Pi\left(0\right)=0$ 
\begin{eqnarray}
\Pi\left(q^2\right)=\frac{q^2}{\pi}\int\frac{Im\Pi(s)}{s(s-q^2)} ds.
\end{eqnarray}
The divergent integral in (\ref{dispersion}) is finite after one subtraction due to the $\frac{1}{s}$ factor. Another way of treating such integrals is the method of power moments $M_{n}\left(q^2\right)$, also called Hilbert moments. These moments are derivatives of the dispersion relation (\ref{dispersion}) with respect to $q^2$
\begin{eqnarray}
\frac{d}{dq^2}\Pi\left(q^2\right)=\frac{1}{\pi}\int\frac{Im\Pi(s)}{(s-q^2)^2} ds
\end{eqnarray}
\begin{eqnarray}
\left(\frac{d}{dq^2}\right)^2\Pi\left(q^2\right)=\frac{2}{\pi}\int\frac{Im\Pi(s)}{(s-q^2)^3} ds
\end{eqnarray}
\begin{eqnarray}
\left(\frac{d}{dq^2}\right)^3\Pi\left(q^2\right)=\frac{2\cdot 3}{\pi}\int\frac{Im\Pi(s)}{(s-q^2)^4} ds
\end{eqnarray}
\begin{eqnarray}
\left(\frac{d}{dq^2}\right)^4\Pi\left(q^2\right)=\frac{2\cdot 3\cdot 4}{\pi}\int\frac{Im\Pi(s)}{(s-q^2)^5} ds.
\end{eqnarray}
Or in compact form: 
\begin{eqnarray}
M_{n}\left(q^2\right)=\frac{1}{n!}\left(\frac{d}{dq^2}\right)^n\Pi\left(q^2\right)=\frac{1}{\pi}\int\frac{Im\Pi(s)}{(s-q^2)^{(n+1)}} ds\qquad n\geq 1 .\label{diff}
\end{eqnarray}
At $q^2=0$ $M_{n}$ coincides with the coefficients of the subtracted dispersion relation. This time not the correlator itself is computed but its derivatives, this approach will also prove useful in further considerations. Again the convergence of the integral is improved with increasing $n$ until the $n$ is reached at which it becomes finite.\\
\begin{figure}[htbp]
 \begin{center}
 \begin{picture}(0,0)%
\includegraphics{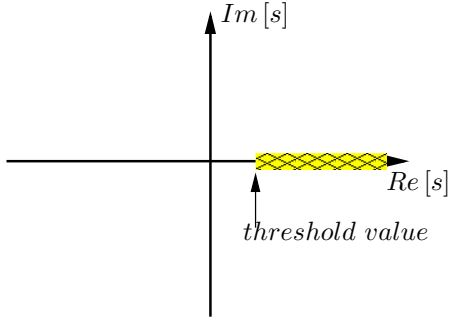}%
\end{picture}%
\setlength{\unitlength}{4144sp}%
\begingroup\makeatletter\ifx\SetFigFont\undefined%
\gdef\SetFigFont#1#2#3#4#5{%
  \reset@font\fontsize{#1}{#2pt}%
  \fontfamily{#3}\fontseries{#4}\fontshape{#5}%
  \selectfont}%
\fi\endgroup%
\begin{picture}(2488,1907)(2928,-2476)
\put(5223,-1697){$Re\left[s\right]$}%
\put(4366,-1995){$threshold~ value$}%
\put(4231,-691){$Im\left[s\right]$}%
\end{picture}%

  \caption{The analytic structure of the correlators used in this work. On the positive real axis above the threshold begins a cut on which additional poles occur. The cut corresponds to physical processes.\label{cutpoles}}
  \end{center}
  \end{figure}
A further method to improve the convergence of the dispersion relation (\ref{dispersion}) is the Borel transformation. This method is also using derivatives but not a finite number of them but an infinite number. The transformation is accomplished by applying the Borel operator 
 \begin{eqnarray}
\widehat{\mathcal{B}}=\frac{1}{(n-1)!}(-q^2)^n\left(\frac{d}{dq^2}\right)^n,-q^2\rightarrow\infty,~n\rightarrow\infty,\qquad\frac{-q^2}{n}=M^2~fixed.
\end{eqnarray}
The action of the Operator is to change the kernel of the integral
\begin{eqnarray}
\widehat{\mathcal{B}}\Pi\left(q^2\right)=\widehat{\mathcal{B}}\frac{1}{\pi}\int_{0}^{\infty}\frac{Im\Pi(s)}{(s-q^2)}ds= \frac{1}{\pi}\int_{0}^{\infty}e^{-\frac{s}{M^2}}Im\Pi(s)ds.
\end{eqnarray}
Hence, after the Borel transformation high momentum components are exponentially suppressed for $q^2<0$. For many QSR applications, the Borel transformation is the optimal choice. 

\subsection{Derivation of the dispersion relation}

\begin{figure}[htbp]
 \begin{center}
\begin{picture}(0,0)%
\includegraphics{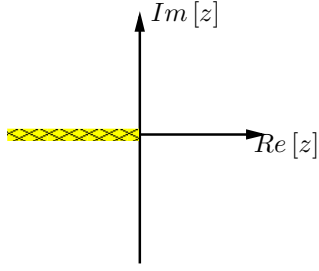}%
\end{picture}%
\setlength{\unitlength}{4144sp}%
\begingroup\makeatletter\ifx\SetFigFont\undefined%
\gdef\SetFigFont#1#2#3#4#5{%
  \reset@font\fontsize{#1}{#2pt}%
  \fontfamily{#3}\fontseries{#4}\fontshape{#5}%
  \selectfont}%
\fi\endgroup%
\begin{picture}(1631,1589)(2689,-1823)
\put(3568,-349){$Im\left[z\right]$}%
\put(4187,-1136){$Re\left[z\right]$}%
\end{picture}%

  \caption{The analytic structure of the square root. The cut of the square root is defined to be on the negative real axis. \label{rootcut}}
  \end{center}
  \end{figure}

\begin{figure}[htbp]
 \begin{center}
\begin{picture}(0,0)%
\includegraphics{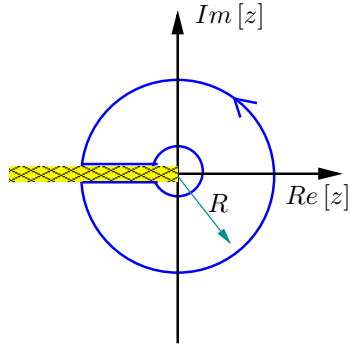}%
\end{picture}%
\setlength{\unitlength}{4144sp}%
\begingroup\makeatletter\ifx\SetFigFont\undefined%
\gdef\SetFigFont#1#2#3#4#5{%
  \reset@font\fontsize{#1}{#2pt}%
  \fontfamily{#3}\fontseries{#4}\fontshape{#5}%
  \selectfont}%
\fi\endgroup%
\begin{picture}(2073,2074)(2737,-1818)
\put(3946,-1006){$R$}%
\put(4411,-961){$Re\left[z\right]$}%
\put(3871,119){$Im\left[z\right]$}%
\end{picture}%

  \caption{The integration contour for the dispersion relation before the limit $R\rightarrow\infty$ is taken.\label{rawcontour}}
  \end{center}
  \end{figure}

The analytic properties of the correlators together with the residue theorem \cite{Bronstein} form the basis of the derivation. In a sense the analytic structure of the correlators used in this work is simple. They have a cut and poles on the positive real axis above some threshold. The remaining part of the complex plane, over which they are defined is free of singularities (see figure \ref{cutpoles}).
Hence, according to the residue theorem an integration over a contour which excludes the shaded region in figure \ref{cutpoles} vanishes. However, a look at the integral in (\ref{dispersion}) reveals that the integrand has an additional pole at $s=q^2$. The residue of this pole is the reason why the dispersion integral is $\Pi\left(q^2\right)$. Instead of a general derivation of the dispersion relation a simple illustrative example is discussed. These illustrations mimic the analytic structure of the correlators that are concerned here. The particular example chosen is the square root, which has a cut on the negative real axis. The analytic structure of the square root is given in figure \ref{rootcut}.
The only difference to the correlators is that there are no poles on the cut of the square root. Everything that will follow from here on is simply an application of the residue theorem, which is given by
\begin{eqnarray}
\sum_{k=1}^{n}Res(f,z_{k})=\frac{1}{2\pi i}\oint f(z)dz \label{residuetheorem}.
\end{eqnarray}
In this example the integral equation is given by
\begin{eqnarray}
\sum_{k=1}^{n}Res\left(\frac{\sqrt{z}}{z-z_{0}},z_{k}\right)=\frac{1}{2\pi i}\oint \frac{\sqrt{z}}{z-z_{0}}dz .
\end{eqnarray}
The choice of the integration contour is restricted by three requirements
\begin{enumerate}
\item Exclusion of the singularities along the negative real axis. 
\item Inclusion of the pole given by $\frac{1}{z-z_{0}}$.
\item Reduction of the integral to the form given in (\ref{dispersion}).
\end{enumerate}
Thus, the integration contour can be chosen as in figure \ref{rawcontour}.
The next goal is to eliminate the integrations along the circles. Therefore, the limits $R\rightarrow\infty$ and $r\rightarrow 0$ where $R$ is the radius of the big circle and $r$ is the radius of the small circle in figure \ref{rawcontour} are taken. In many cases the integrand vanishes along the circles in this limit. In this example the integrand does not vanish on the circle whose radius had been set to infinity, but a simple subtraction eliminates this problem (see equation (\ref{subtracted}))   
\begin{eqnarray}
\sqrt{z}=\sqrt{0}+\frac{z_{0}}{\pi}\int\frac{\sqrt{z}}{z(z-z_{0})} dz=\frac{z_{0}}{\pi}\int\frac{\sqrt{z}}{z(z-z_{0})} dz.
\end{eqnarray}
On the circles the integral is given by
\begin{eqnarray}
i\frac{z_{0}}{\pi}\int\frac{\sqrt{R}}{Re^{i\phi}-z_{0}}e^{i\frac{\phi}{2}} dz.
\end{eqnarray}
Hence, the integrand vanishes for $R\rightarrow\infty$ and $R\rightarrow 0$. Therefore only the contour along the negative axis survives after the limit is taken. This is a huge step forward but the integration still runs over the real and complex part of the square root whereas the one in (\ref{dispersion}) runs only over the imaginary part.\\
The solution of this puzzle is given by another feature of the square root. On the integration contour that runs above the negative real axis the square root has the value $+i\sqrt{\abs{z}}$ on the one below its value is $-i\sqrt{\abs{z}}$. Such discontinuities also occur in correlators but in general their real part is not zero. However, the square root is purely imaginary on both contours. Moreover, the the integrations along the negative real axis can be comprised to a single integration because the integration on the contour where the square root is $-i\sqrt{\abs{z}}$ runs in the opposite direction to the one where the root is $+i\sqrt{\abs{z}}$. Therefore the two integrations are equal and can be added. Thus the integration takes the form
\begin{eqnarray}
\sum_{k=1}^{n}Res(f,z_{k})=\frac{z_{0}}{\pi i}\int_{-\infty}^{0} \frac{\sqrt{z}}{z(z-z_{0})} dz =\frac{z_{0}}{\pi}\int_{-\infty}^{0} \frac{Im\left[ \sqrt{z}\right] }{z(z-z_{0})} dz\qquad\sqrt{z}=iIm\left[ \sqrt{z}\right]\label{rootdisp}.
\end{eqnarray}
Only the left hand side has to be modified. It is given by the residues of the integrand in (\ref{rootdisp}), but the integrand has only one singularity inside the integration region. Hence, there is only one residue. The calculation of this residue is simple and can be performed as follows, the square root has to be expanded around $z_{0}$, the residue is the coefficient of the $\frac{1}{z-z_{0}}$-term
\begin{eqnarray}
\sqrt{z}=\sqrt{z_{0}}+\frac{1}{2}\frac{1}{\sqrt{z_{0}}}(z-z_{0})-\frac{1}{8}\frac{1}{z_{0}^{3/2}}(z-z_{0})^2+...
\end{eqnarray}
\begin{eqnarray}
\frac{\sqrt{z}}{z-z_{0}}=\frac{\sqrt{z_{0}}}{z-z_{0}}+\frac{1}{2}\frac{1}{\sqrt{z_{0}}}-\frac{1}{8}\frac{1}{z_{0}^{3/2}}(z-z_{0})+...~.
\end{eqnarray}
The analysis of the Laurent expansion that has just been derived shows that the residue is given by $\sqrt{z_{0}}$ and the integration (\ref{rootdisp}) finally shows up to be
\begin{eqnarray}
\sqrt{z_{0}}=\frac{z_{0}}{\pi}\int_{-\infty}^{0}\frac{Im\left[\sqrt{z}\right] }{z(z-z_{0})} dz \label{finalroot}
\end{eqnarray}
and for the example of the root the dispersion relation has been valid. Though, there are a few remarks that are necessary in order to extend the derivation to correlators. The integration contour for correlators is mirrored with respect to the imaginary axes and then the zero point is shifted to the threshold value shown in figure \ref{cutpoles}. This is done because the positive real axis should be excluded from the inside of the integration contour. Hence, it has to be justified why the integration along two contours can be replaced along the integration over one contour and why the integration over real and imaginary part can be replaced with an integration over the imaginary part. \\
The solution was already given when the root was concerned but there is a property of correlators that has not been explained yet. Along the integrations observed here the following relations hold
\begin{eqnarray}
Im\left[\Pi\left(q^2+i\epsilon\right) \right]=-Im\left[\Pi\left(q^2-i\epsilon\right) \right] 
\end{eqnarray}
\begin{eqnarray}
Re\left[\Pi\left(q^2+i\epsilon\right) \right]=Re\left[\Pi\left(q^2-i\epsilon\right) \right] 
\label{realpart}
\end{eqnarray}
where $q^2$ is real and above the threshold. The case of the imaginary part is already known from the root example and the result was that the integrations just add up to one integration. From \ref{realpart} it follows that the real parts cancel. Thus only the integration over the imaginary part survives.\\
As the last step the residues of the correlators have to be calculated but this is a difficult task that is not going to be performed here. Instead of another theoretical derivation some practical applications are going to be performed. The integral in  (\ref{finalroot}) is now calculated.\\
There are two cases that have to be distinguished. 
\begin{enumerate}
\item $z_{0}$ does not lie in the region that has been excluded from the integration.
\item $z_{0}$ lies in the region that has been excluded from the integration.
\end{enumerate}
The first case is the one which is interesting in applications. Therefore it is going to be explained for the root example. Hence in the first case the pole is located inside the integration contour indicated in figure (\ref{rawcontour}). Then relation (\ref{finalroot}) can be used and only a real integration has to be performed
\begin{eqnarray}
\sqrt{z_{0}}=\frac{z_{0}}{\pi}\int_{-\infty}^{0}\frac{\sqrt{\abs{z}}}{z(z-z_{0})}dz=-i\frac{2\sqrt{z_{0}}}{\pi}\left(ArTanh\left(i\frac{\sqrt{\abs{z}}}{\sqrt{z_{0}}}\right) \right)_{-\infty}^{0}=-i\frac{2\sqrt{z_{0}}}{\pi}\left(i\frac{\pi}{2}\right)=\sqrt{z_{0}}.
\end{eqnarray}
Thus, the dispersion relation in subtracted form is verified in the case where $z_{0}$ does not lie on the cut of the square root. \\
The lesson that can be learned from this section is that a correlator is fully determined by its imaginary part above some threshold. In the following sections this part of the correlator will be identified with the non-perturbative resonance physics, which cannot be calculated by perturbative QCD. Hence, if the full imaginary part is known and the correlator is calculated at some point using a dispersion relation, the correlator at this point also includes non-perturbative information.\\
Dispersion relations are long known and have been used in electrodynamics under the name Kramers-Kroning relations. A nice review from a modern perspective is given in \cite{Weinberg:1996kr}.  

\section{The QCD Sum Rule (QSR) Method \label{sumrules}}
In the preceding sections the operator product expansion and the dispersion relations were discussed. By combining those tools one obtains the so called QCD Sum Rules. The QSR allow calculations of properties which are connected to n-point correlators, in a sense these correlators can be calculated by QSR. This thesis is restricted to correlators where the external states, are vacuum states, but the QSR can deal with arbitrary external states, addressing e.g. the in-medium properties of hadrons.\\
The objects which are investigated are mainly 2-point correlators. Such correlators contain much physical information. The first example for a 2-point correlator that occurs in QFT is the propagator, for example $\bra{0}T\left\{\bar{q}(0),q(x)\right\}\ket{0}$. The propagator has a clear physical interpretation. It is the amplitude for the propagation of a particle from the space-time point $0$ to $x$. In the literature \cite{Peskin:1995ev} two cases are analyzed. One without interaction and one with interaction. Without interaction everything gets simpler and the propagator is fully calculable and has just one singularity. In momentum space it is a pole connected with the mass of the particle. This pole occurs when the particle is on shell $q^2=m^2$. With interactions the singularity structure of the propagator is much richer. In addition to the pole corresponding to the mass of the particle, poles corresponding to bound states and/or resonances occur.\\
With QCD Sum Rules one can calculate the properties of bound states by matching the correlation functions at a scale between the perturbative and non-perturbative regimes. However, the poles can not be calculated in a perturbative approach to the problem. Hence, non-perturbative methods are needed. There is a domain where the OPE agrees with the correlation function. In this domain where the correlators have no poles. Thus, an ansatz for the imaginary part of the propagator can be made. Furthermore, the space-like region can be connected to the time-like region, where the physical singularities of the correlator are located, by means of a dispersion relation. Then the propagator on the real axis below threshold can be calculated from the ansatz using a dispersion relation. The result can be matched with the OPE in the space like region. This is how QCD Sum Rules work.\\
Such a procedure allows a comparison between phenomenology and theory because for some correlators the imaginary parts can be measured. Therefore, in such cases one side of the dispersion relation is given by the phenomenology and the other one by theory. These remarks are characteristic for the way in which QSRs are calculated and evaluated. 

\subsection{The connection between the correlator and the spectral density}

There is a direct connection between a spectral function, sometimes called spectral density, and the correlator in the corresponding channel. The vector-current correlator is of particular interest. It is factorize into a Lorentz tensor and a Lorentz scalar part
\begin{eqnarray}
\Pi_{\mu\nu}(q_{\alpha})=\left(g_{\mu\nu}-\frac{q_{\mu}q_{\nu}}{q^2}\right)\Pi(q^2)=\left(g_{\mu\nu}q^2-q_{\mu}q_{\nu}\right)\rho(q^2).
\end{eqnarray}
The spectral function is proportional to the imaginary part of $\Pi(q^2)$. The spectral function in the vector channel can be extracted from data on $e^+e^-$- or $\mu^+\mu^-$- annihilation  \cite{Peskin:1995ev}. In other channels no direct measurements exist.

\subsection{The model for the 2-point correlator}

\begin{figure}[htbp]
 \begin{center}
  \includegraphics[scale=0.9]{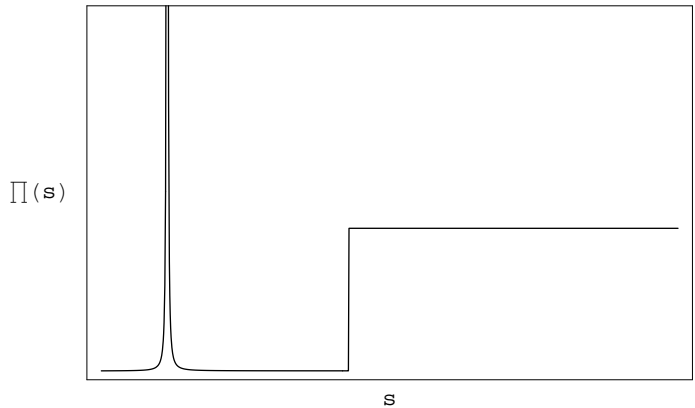}
  \caption{Model for the phenomenological part of the sum rules. \label{oneresonance}}
  \end{center}
  \end{figure}

As discussed in the previous section, a QCD Sum Rule is a dispersion relation  connecting the phenomenological spectral density with the OPE
\begin{eqnarray}
\frac{1}{\pi}\int_{0}^{\infty}\frac{Im\left[\Pi_{pheno}(s)\right] }{(s-q^2)} ds =OPE\left(q^2\right).\label{dispersiontheory}
\end{eqnarray}
The momentum domain has to be restricted to large space-like momenta. Hence, the substitution $Q^2=-q^2$ is convenient to avoid the unimportant minus sign.\\
Obviously a QSR can be used in two ways. The first one is to define the right hand side as given and to calculate the properties of the left hand side or vice versa. Both possibilities are used and have to be used, as will be discussed below. The evaluation of a QSR is a chapter for itself and is best illustrated through examples. The classical way to use QSR is to use an ansatz for the phenomenological part given by a narrow resonance and a continuum, where the resonance is given by a delta function and the continuum by a step function
\begin{eqnarray}
\Pi_{pheno}(s)=\bra{0}j\ket{\psi}^2\delta(s-M_{res})+const.\cdot\theta(s-t_{c}).
\end{eqnarray}  
If a tiny width is given to the $\delta$-functions the phenomenological part in the form sketched in figure \ref{oneresonance}.
Obviously such a model is very simplified, since the particle that is represented by the $\delta$-resonance is usually a resonance with a non-zero width. Thus, a resonance with a finite lifetime is represented by a stable particle. Moreover the radial excitations are all missing. The step function is an approximation for the continuum. The first step to make calculations handy is to use the imaginary part of the perturbative expression for the correlator in the domain where $q^2$ is positive instead of a step function. Especially because the perturbative expression is a part of the OPE which belongs to the system and is therefore known. Others improvements will be shown later.\\
The right hand side is the theoretical side which is given by the OPE. However, the OPE can not be calculated exactly, it has to be truncated. In Section \ref{importantunimportant} the criteria for convergence of the OPE were discussed. Furthermore, the terms in the truncated OPE are not known exactly, but to a certain order in $\alpha_{S}$ or to a certain accuracy of estimation, as discussed sections \ref{wilsoncoefficient} and \ref{condensates}. Finally the exact Wilsonian OPE is not employed but rather an approximation, the so called practical OPE or SVZ expansion (see section \ref{opeinqcd}). Hence, also the theoretical side of the QSR contains uncertainties and simplifications. However, QSRs have been and are heavily used in non-perturbative physics with big success and with a satisfactory accuracy. In many cases QSR are the only tools available on the market. \\
A well known applications of QSRs are the $\rho$- and the $J/\psi$-meson. The analysis of these two systems is instructive and will be presented in the following section.

\section{The $J/\psi$ QSR as a classical QSR \label{charmonium}}
In this section QSRs for the $J/\psi$-meson are employed to determine the gluon condensate $\alpha_{S}\bra G_{\mu\nu}^a G_{\mu\nu}^a\ket{0}$.

\subsection{The spectrum of $c\overline{c}$-mesons}

\begin{table}[htbp]
\begin{tabular}{||c|l|l|l|l|l|l||}
\hline
$I^G\left(J^{PC}\right)$& groundstate & 1.excitation& 2.excitation& 3.excitation& 4.excitation& 5.excitation\\
\hline 
$0^+\left(0^{-+}\right)$&$\eta_{c}$&&&&&\\
\hline  
$0^-\left(1^{--}\right)$& $J/\psi$ &$\psi(2S)$&$\psi(3770)$&$\psi(4040)$&$\psi(4160)$&$\psi(4415)$\\ 
\hline 
$0^+\left(0^{++}\right)$&$\chi_{c0}$&&&&&\\
\hline
$0^+\left(1^{++}\right)$&$\chi_{c1}$&&&&&\\
\hline
$0^+\left(2^{++}\right)$&$\chi_{c2}$&&&&&\\
\hline                                 
\end{tabular}
\caption{The known charmonium spectrum. The excited states correspond to radial excitations.}\label{ccradial}
\end{table}
In table \ref{ccradial} the known spectrum of $c\overline{c}$-meson (Charmonium) is shown. In the vector channel $(J^{PC}=1^{--})$ the ground state and several radial excitations are known while in the other channels only the ground states are known. Such a situation is characteristic for hadron spectroscopy. In many cases only the ground state corresponding to given quantum numbers is known. Therefore the calculation of the ground states properties would already be a remarkable success and for this system QSR do the job very well. SVZ realized during the work to their paper \cite{Novikov:1977dq} that $c\overline{c}$-mesons are bound exclusively by gluons. Thus, an attempt to calculate the mass of a $c\overline{c}$-meson with a QSR in which an OPE is used which contains only the perturbative and the lowest order gluon condensate term is reasonable, although the rules of section \ref{importantunimportant} say that all terms up to dimension six should be included. It turns out that the attempt is successfull, but to achieve the goal a sacrifice has to be made. Section \ref{condensates} states the value of the gluon condensate $\bra{0}\frac{\alpha_{s}}{\pi}G_{\mu\nu}^{c}G_{\mu\nu}^{c}\ket{0}$ and as the source of it the $J/\Psi$ sum rule. The gluon condensate has not been calculated up to now and therefore it has to be measured. The $c\overline{c}$-mesons are the best systems for it because of the structure of the OPE and because the spectrum of the $J/\Psi$-meson can and has been measured to a high accuracy.\\

\subsubsection{The phenomenological part of the sum rules}

\begin{figure}[htbp]
 \begin{center}
  \includegraphics[scale=0.9]{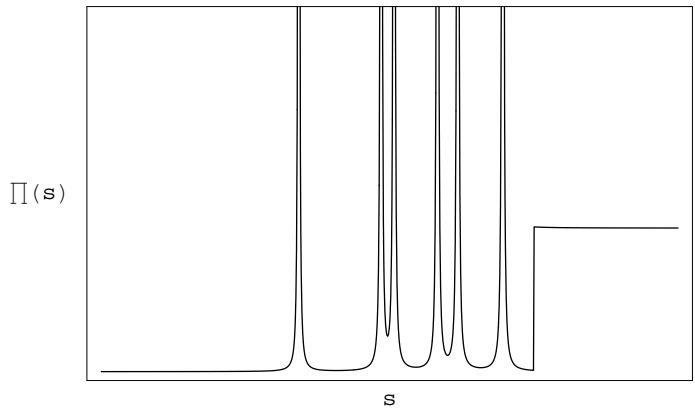}
  \caption{Sketch of the spectral function of the $J/\Psi$-meson as it is known today. \label{manyresonance}}
  \end{center}
  \end{figure}

\begin{figure}[htbp]
 \begin{center}
  \includegraphics[scale=0.9]{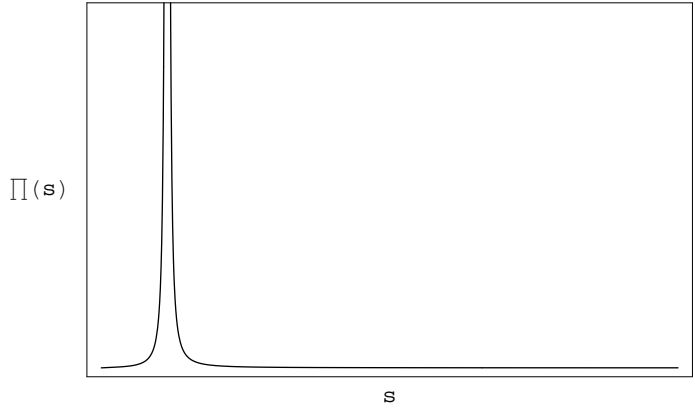}
  \caption{Approximation for the spectral function of the $J/\Psi$-meson used in the sum rules for the calculation of the gluon condensate. \label{jpsiapproxspec}}
  \end{center}
  \end{figure}

The spectral function of the $J/\Psi$-meson and its radial excitations has been measured in $\mu^+ \mu^-$-annihilation (see for example \cite{Peskin:1995ev}). The resonances can be approximated by $\delta$-functions and the spectral function shown in figure \ref{manyresonance}.
The parameterization of the spectral function is very intuitive:
\begin{eqnarray}
Im\Pi(s)=\sum_{resonances}\bra{0}j_{\mu}\ket{n_{res}}^2\delta\left(s-m^2_{R}\right)+\frac{1}{4\pi}\left(1+\frac{\alpha_{S}(s)}{\pi}\right)\Theta(s-t_{c}) \label{fullspectrum}.
\end{eqnarray}
In the sum rule for the ground state of the states with $J^P=1^-$ an approximation to the spectrum is used where all excited states are neglected and the ground state is approximated by a $\delta$-function. Thus, the effective spectral function is given by figure \ref{jpsiapproxspec}.
The evaluation of the sum rule will show that this approximation is reasonable. In the evaluation process the contribution of the neglected parts of the sum rules are suppressed.
 
\subsubsection{The theoretical (QCD) part of the sum rules}

The current which approximates the $J/\Psi$-meson is given by
\begin{eqnarray}
j_{\mu}(x)=\overline{c}(x)\gamma_{\mu}c(x)
\end{eqnarray}
and has exactly the quark structure and quantum numbers of the $J/\Psi$. In the OPE for the 2-point correlator of this current only two terms are kept 
\begin{eqnarray}
i\int d^4x e^{-iq^{\mu}x_{\mu}}\mele{j_{\mu}(x)j_{\nu}(0)}=\left( g_{\mu\nu}q^2-q_{\mu}q_{\nu}\right) \left[ C_{pert}\cdot I+C_{G^2}\cdot\bra{0}\frac{\alpha_{s}}{\pi}G_{\mu\nu}^{c}G_{\mu\nu}^{c}\ket{0}\right] \label{opejpsi},
\end{eqnarray}
where the coefficients are given by:
\begin{enumerate}
\item{perturbative contribution $C_{pert}$}\\ \\
     The approximation to the perturbative coefficient used here contains the bare loop and 
     $\alpha_{S}$ contributions
     \begin{eqnarray}
     Im\left[ 
      \parbox[c]{3cm}{
      \begin{fmffile}{QCDMesonpert01}
       \begin{fmfgraph*}(30,30)
        \fmfleft{i} 
        \fmfright{o}
        \fmf{dots}{i,v1} 
        \fmf{dots}{v2,o}
        \fmf{plain,left,tension=.2}{v1,v2}
        \fmf{plain,left,tension=.2}{v2,v1}
        \fmfdotn{v}{2}
       \end{fmfgraph*}
     \end{fmffile}}+
     \parbox[c]{3cm}{
       \begin{fmffile}{QCDMesonpert02}
      \begin{fmfgraph*}(30,30)
        \fmfleft{i} 
        \fmfright{o}
        \fmf{dots}{i,v1} 
        \fmf{dots}{v2,o}
        \fmf{plain,left,tension=0.2,tag=1}{v1,v2}
        \fmf{plain,left,tension=0.2,tag=2}{v2,v1}
        \fmfdot{v1,v2}
        \fmfposition
        \fmfipath{p[]}
        \fmfiset{p1}{vpath1(__v1,__v2)}
        \fmfiset{p2}{vpath2(__v2,__v1)}
        \fmfi{gluon,left,tension=0.2}{point length(p1)/2 of p1 -- point length(p2)/2 of p2}
      \end{fmfgraph*}
     \end{fmffile}}+
     \parbox[c]{3cm}{
      \begin{fmffile}{QCDMesonpert03}
      \begin{fmfgraph*}(30,30)
       \fmfleft{i} 
       \fmfright{o}
       \fmf{dots}{i,v1} 
       \fmf{dots}{v2,o}
       \fmf{plain,left,tension=0.2,tag=1}{v1,v2}
       \fmf{plain,left,tension=0.2,tag=2}{v2,v1}
       \fmfdot{v1,v2}
       \fmfposition
       \fmfipath{p[]}
       \fmfiset{p1}{vpath1(__v1,__v2)}
       \fmfiset{p2}{vpath2(__v2,__v1)}
       \fmfi{gluon,right,tension=1.5}{point length(p1)/5 of p1 -- point 4length(p1)/5 of p1}  
      \end{fmfgraph*}
     \end{fmffile}}
     \right] 
     \nonumber\\
      =Im\left[C_{pert}(s)\right]=\Theta(s-4m^2)\frac{1}{8\pi}v(s)(3-v^2(s))~~~~~~~~~~~~~~~~~~~~~~~~~~~~~~~
      \nonumber\\\times\left\{1+\frac{4}{3}\alpha_{s}\left[
      \frac{\pi}{2v(s)}-\frac{v(s)+3}{4}\left(
      \frac{\pi}{2}-\frac{3}{4\pi}\right)\right]\right\}~~~~~~~~~
      v(s)=\sqrt{1-\frac{4m^2}{q^2}}
      \label{vectorpert}.
     \end{eqnarray}
  Dispersion relations can be used to calculate the real part of the amplitude. It is instructive to plot the perturbative part of the OPE, the plot is given in figure \ref{vectorpertplot} and shows only the amplitude corresponding to the first diagram in (\ref{vectorpert}), the one loop diagram.    
 \begin{figure}[htbp]
 \begin{center}
  \includegraphics[scale=0.9]{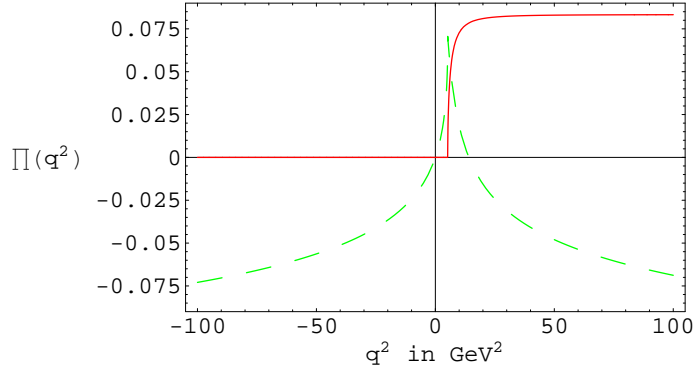}
  \caption{The one loop contribution as a function of the total momentum squared. The dashed line is the real part and the full line the imaginary part.\label{vectorpertplot}}
  \end{center}
  \end{figure}
Higher order corrections do not change the overall structure of the plot. Thus, the lowest order contribution shows the important features of the perturbative part of the 2-point correlator. The imaginary part is going to be used as an approximation for the continuum in the model for the spectral function. The amplitude is real below the threshold, above threshold the imaginary part is non-zero.

\item{nonperturbative contribution $C_{G^2}$}\\ \\
The only nonperturbative contribution stems from the gluon condensate term. The coefficient is determined by the bare loop with two external gluon legs (see section \ref{wilsoncoefficient})
\begin{eqnarray}
C_{G^2}\left(q^2\right) =\left[ 
\parbox[c]{3cm}{
 \begin{fmffile}{QCDMesongluon01}
      \begin{fmfgraph*}(30,30)
        \fmfleft{i} 
        \fmfright{o}
        \fmf{dots}{i,v1} 
        \fmf{dots}{v4,o}
        \fmf{plain,left,tension=0.0001,tag=1}{v1,v4}
        \fmf{plain,left,tension=0.0001,tag=2}{v4,v1}
	
	\fmf{phantom}{v1,v2}
	\fmf{phantom,tag=3}{v2,v3}
	\fmf{phantom}{v3,v4}
 
	\fmffixedx{0}{v2,v3}
	\fmffixedy{0.4cm}{v2,v3}
	\fmffixedy{0}{v1,v4}
	
        \fmfposition
        \fmfipath{p[]}
        \fmfiset{p1}{vpath1(__v1,__v4)}
        \fmfiset{p2}{vpath2(__v4,__v1)}
	 \fmfiset{p3}{vpath3(__v2,__v3)}
	\fmfi{gluon,right}{point 0 of p3 -- point length(p2)/2 of p2} 
       \fmfi{gluon,right}{point length(p3) of p3 -- point length(p1)/2 of p1} 
       
       \fmfiv{d.sh=cross,d.ang=0,d.siz=5thick}{point 0 of p3}
	 \fmfiv{d.sh=cross,d.ang=0,d.siz=5thick}{point length(p3) of p3} 
	 \fmfdot{v1}
	\fmfdot{v4}
      \end{fmfgraph*}
     \end{fmffile}}+
     \parbox[c]{3cm}{      
     \begin{fmffile}{QCDMesongluon02}
       \begin{fmfgraph*}(30,30)
        \fmfleft{i} 
        \fmfright{o}
        \fmf{dots}{i,v1} 
        \fmf{dots}{v4,o}
	\fmf{plain,left,tension=0.01,tag=1}{v1,v4}
       \fmf{plain,left,tension=0.01,tag=2}{v4,v1}
       \fmf{phantom}{v1,v2}
        \fmf{phantom,tag=3}{v2,v3}
	\fmf{phantom}{v3,v4}
	\fmffixedx{0.4cm}{v2,v3}
	\fmffixedy{0cm}{v2,v3}
        \fmffixedy{0}{v1,v4}
       \fmfposition
       \fmfipath{p[]}
       \fmfiset{p1}{vpath1(__v1,__v4)}
       \fmfiset{p2}{vpath2(__v4,__v1)}
       \fmfiset{p3}{vpath3(__v2,__v3)}
       \fmfi{gluon,right}{point length(p1)/5 of p1 -- point 0 of p3} 
       \fmfi{gluon,right}{point 4length(p1)/5 of p1 -- point length(p3) of p3}  
        \fmfdot{v1}
	\fmfdot{v4}
	 \fmfiv{d.sh=cross,d.ang=0,d.siz=5thick}{point 0 of p3}
	 \fmfiv{d.sh=cross,d.ang=0,d.siz=5thick}{point length(p3) of p3}
       \end{fmfgraph*}
     \end{fmffile}}
\right] 
\nonumber\\=\frac{\alpha_{S}}{48\pi q^4}\left\lbrace \frac{3(v^2(s)+1)(v^2(s)-1)^2}{v^4(s)}\frac{1}{2v(s)}\ln\left(\frac{v(s)+1}{v(s)-1}\right)-\frac{v^4(s)-2v^2(s)+3}{v^4(s)} \right\rbrace\nonumber\\
v(s)=\sqrt{1-\frac{4m^2}{q^2}}.~~~~~~~~~~~~~~~~~~~~~~~~~~~~~~~~~~~~~~~~~~~~~~~~~~~~~~~~~~~~~~~~~~~~~~~~~~~~~~
\label{jpgluon}
\end{eqnarray}
In contrast to the diagrammatic description the coefficient is not a simple diagram. Its plot is given in figure \ref{wilsongluonplot}. 
 \begin{figure}[htbp]
 \begin{center}
  \includegraphics[scale=0.9]{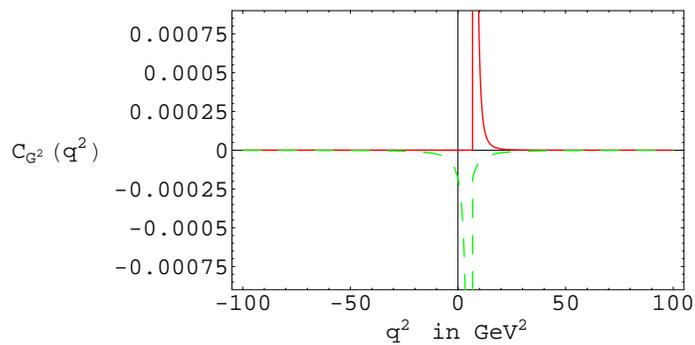}
  \caption{Plot of $C_{G}\left(q^2\right)$, the dashed line is the real part and the full line the 
  imaginary part. The singularities above $q^2=0$ occur at $4m^2$.\label{wilsongluonplot}}
  \end{center}
  \end{figure}
The coefficient $C_{G}(q)$ in (\ref{jpgluon}) is valid in the limit $\abs{q}\gg0$. Hence, it is expected to find an unreasonable behavior of the coefficient if $q$ approaches zero. In figure  \ref{wilsongluonsingularitiesplot} the coefficient is plotted in the momentum region close to zero. There a highly oscillating behavior is found, which signals the invalidity of the coefficient in that momentum domain. Unfortunately, this effect is not visible in figure \ref{wilsongluonplot}. 
 \begin{figure}[htbp]
 \begin{center}
  \includegraphics[scale=0.9]{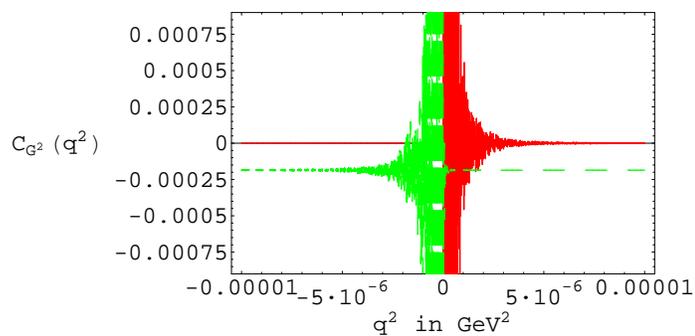}
  \caption{Plot of $C_{G}\left(q^2\right)$ showing highly oscillating singularities at $q^2=0$, the 
  dashed line is the real part and the full line the 
  imaginary part.\label{wilsongluonsingularitiesplot}}
  \end{center}
  \end{figure}
\end{enumerate} 
Thus, the sum rule has been constructed and can now be evaluated. Momentum Sum Rules are employed for this task, see section \ref{dispersionrelations} for the basis that is needed for further considerations.

\subsubsection{Evaluation of the Sum Rule}

\begin{figure}[htbp]
 \begin{center}
  \includegraphics[scale=0.9]{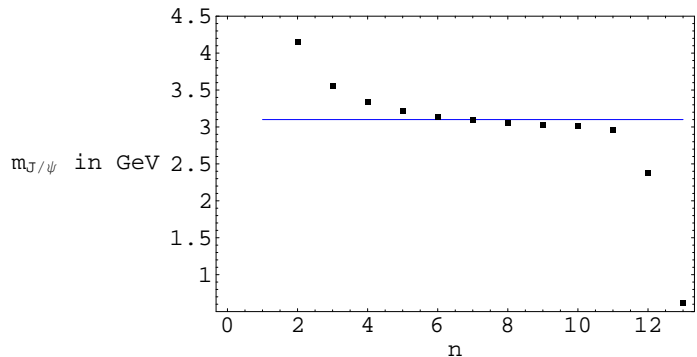}
  \caption{Plot of the moments $M_{n}(-1GeV^2)$ belonging to the QSR of the $J/\Psi$-meson. The full line is the measured mass of the $J/\Psi$, while the dots are the moments at a squared momentum of -1. The zeroth and the first momentum do not converge due to a dispersion integral used to calculate the perturbative part of the Wilson coefficient. In the domain of the plateau, the contributions of the perturbative and non-perturbative terms in the OPE are optimized.\label{jpsimoments}}
   \end{center}  
  \end{figure}

If (\ref{fullspectrum}) is used in the dispersion the following equation is obtained
\begin{eqnarray}
\frac{1}{n!}\left(-\frac{d}{dQ^2}\right)^n\Pi\left(Q^2\right)=\sum_{res}\frac{\bra{0}j_{\mu}\ket{n_{res}}^2}{(m_{R}^2+Q^2)^{(n+1)}}+\frac{1}{4\pi^2}\int\frac{1+\frac{\alpha_{S}(s)}{\pi}}{(s+Q^2)^{(n+1)}}\Theta(s-t_{c})ds~~n\geq 1 \label{disp}
\end{eqnarray}
where $n$ has to be bigger than 0 to make the integral convergent. This relation allows the calculation of the $J/\psi$ mass. The mass of the ground state with the quantum numbers $J^P=1^-$. 
A simple conversion will show that this is true 
\begin{eqnarray}
M_{n}(Q^2)=
\sum_{res}\frac{\bra{0}j_{\mu}\ket{n_{res}}^2}{(m_{R}^2+Q^2)^{(n+1)}}+\frac{1}{4\pi^2}\int\frac{1+\frac{\alpha_{S}(s)}{\pi}}{(s+Q^2)^{(n+1)}}\Theta(s-t_{c})ds.
\end{eqnarray}
The $M_{n}$'s are called the Moments. First of all the term corresponding to the $J/\psi$-resonance is factored out
\begin{eqnarray}
M_{n}(Q^2)=\frac{\bra{0}j_{\mu}\ket{n_{J/\psi}}^2}{(m_{J/\psi}^2+Q^2)^{(n+1)}}\nonumber\\
\times\left( 1+\frac{(m_{J/\psi}^2+Q^2)^{(n+1)}}{\bra{0}j_{\mu}\ket{n_{J/\psi}}^2}
 \left\{\sum_{res>J/\psi}\frac{\bra{0}j_{\mu}\ket{n_{res}}^2}{(m_{res}^2+Q^2)^{(n+1)}}+\frac{1}{4\pi^2}\int\frac{1+\frac{\alpha_{S}(s)}{\pi}}{(s+Q^2)^{(n+1)}}\Theta(s-t_{c})ds\right\}\right).
\end{eqnarray}
The quantity
\begin{eqnarray}
\delta_{n}(Q^2)=\frac{(m_{J/\psi}^2+Q^2)^{(n+1)}}{\bra{0}j_{\mu}\ket{n_{J/\psi}}^2}\nonumber\\              
\times \left\{\sum_{res>J/\psi}\frac{\bra{0}j_{\mu}\ket{n_{res}}^2}{(m_{res}^2+Q^2)^{(n+1)}}+\frac{1}{4\pi^2}\int\frac{1+\frac{\alpha_{S}(s)}{\pi}}{(s+Q^2)^{(n+1)}}\Theta(s-t_{c})ds\right\} )
\end{eqnarray}
is convenient because $\delta_{n}$ is suppressed for big $n$
\begin{eqnarray}
\left(\frac{m_{J/\psi}^2+Q^2}{m_{res}^2+Q^2}\right)^{n+1}\longrightarrow 0,~ n\longrightarrow \infty~~~~~~~~~\nonumber\\
\left(\frac{m_{J/\psi}^2+Q^2}{s+Q^2}\right)^{n+1}\longrightarrow 0,~ n\longrightarrow \infty, s>t_{c}.
\end{eqnarray}
This means that:
\begin{eqnarray}
M_{n}(Q^2)=\frac{\bra{0}j_{\mu}\ket{n_{res}}^2}{(m_{J/\psi}^2+Q^2)^{(n+1)}}\left( 1+\delta_{n}(Q^2)\right).
\end{eqnarray}
Now the ratio
\begin{eqnarray}
r_{n}(Q^2)=\frac{M_{n}}{M_{n-1}}=\frac{1}{m_{J/\psi}^2+Q^2}\cdot\frac{1+\delta_{n}}{1+\delta_{n-1}}
\end{eqnarray}
is considered. Depending on the channel, the factor $\frac{1+\delta_{n}}{1+\delta_{n-1}}$ can sometimes be replaced by 1. The spectrum of $J^P=1^-$ $c\overline{c}$-meson is such a case. This can be tested because the spectrum has been measured. This is equivalent to the replacement of the spectrum in figure \ref{manyresonance} with the one of figure \ref{jpsiapproxspec}. If the $r_{n}$'s are also computed by the OPE, the right hand side of the dispersion relation, theoretical predictions are possible. \\
A simple example would be the calculation of the $J/\psi$ mass. The c-quark mass is known sufficiently well to expose bounds on it and therefore there is only one really free parameter remaining, the gluon condensate. Hence, the mass calculation is possible, but the c-quark mass and the gluon condensate have to be fitted. Thus, the gluon condensate can be calculated by adjusting it to a value which reproduces the mass of the $J/\psi$. The result is shown in figure \ref{jpsimoments}.
At small $n$ the continuum is not suppressed strongly enough and gives the dominant contribution to the phenomenological side. On the theoretical side the perturbative part of the OPE gives the main contribution to the Moments. Fortunately, with growing $n$ the continuum is suppressed and the resonance gives the dominant contribution to the phenomenological side. Unfortunately, with further increasing $n$ the contribution of the non-perturbative terms to the theoretical side becomes the dominant contribution. The plateau lies in the region where the resonance dominates the phenomenological part of the sum rules and the perturbative part of the OPE dominates the theoretical side.\\ 
The plateau determines the mass of the $J/\psi$. In the $n$ region of the plateau, the contribution of the perturbative term to the moments dominates and the non-perturbative term of the OPE (\ref{opejpsi}) is smaller by factors of 10. The plateau starts when the non-perturbative contribution to the moments is a 100 times smaller than the perturbative and ends when the non-perturbative contribution is equal to the perturbative one. Therefore left to the plateau everything is perturbation theory and there are no bound states. Hence, there is no plateau. While on the right of the plateau the non-perturbative terms dominate, but they are not known accurately enough and therefore the sum rule does not converge any more. In the domain of the plateau the convergence of the sum rule is optimized between convergence of perturbation theory and the lack of knowledge concerning the higher terms in the OPE. The c-quark mass and the gluon condensate are fitted in order to reproduce the mass of the system. The result is $m_{c}=1.4~GeV$ and the gluon condensate has the value quoted in \ref{condensates}, in this example the moments have been evaluated at $q^2=-1~GeV^2$  other values are of course also allowed.\\
Historically this was the starting point for the calculations of QSR \cite{Shifman:1978bx}. The gluon condensate has up to now never been calculated from first principles. It is multiplied by the strong coupling constant in order to form a renormalization group invariant.\\
Now this value of the gluon condensate can be used to calculate QSR for other systems. Reinders, Rubinstein and Yazaki performed a rigorous test of the gluon condensate on the $c\overline{c}$ spectrum. They repeated the calculation just performed for every quantum number $J^P$ for which the ground state was measured. These calculations confirmed the gluon condensate given in section \ref{condensates} (see \cite{Reinders:1984sr}). Moreover many other systems have been analyzed during these times.\\
Many systems containing two quarks with nearly equal masses have been calculated often using Borel transformed QSR (Borel Sum Rules) in order to improve the convergence of the Sum Rules. The mass of the $\rho$-meson given by QSR is remarkably close to the value measured and the plateau is very long, (see \cite{Shifman:1978bx}). In the period from 1978 to 1988 hadrons seemed to be describable in the way just outlined, but with proceeding time problems occurred. They occurred in exotic systems like hybrids or glue balls and in heavy-light systems. In the case of the D-meson, the sum rules converged very badly, and a broad plateau is not given.

\section{Borel transformation \label{borel}}
Section \ref{charmonium} discussed the general pattern how QSR are used. It is allways a dispersion relation consisting of a phenomenological and a theoretical (QCD) part. These parts are the left and the right hand side of the dispersion relation (see for example (\ref{dispersiontheory})). The phenomenological part of the QSR can be determined by the intuition of the user, by input from alternative theoretical approaches or from measurements, while the theoretical part is given by the OPE. The input from both sources is afflicted with errors. The spectral functions are in most cases approximations and can not be directly measured with the exception of spectral functions of the vector-current correlators. Furthermore the OPE is also an approximation because it is a truncated expansion. This is a truncation on several levels. First, it is an expansion in $\frac{1}{Q^2}$. Second, the Wilson coefficients are approximated. Finally, the condensates are approximate. It is difficult to judge the combined effect of the approximations. Nevertheless, the QSR are in many cases very successful.\\
The circumstance that ensures the reliability of the QSR is the existence of a working window. If the OPE would be known completely for the given spectrum the dots in figure \ref{jpsimoments} would not change much for small $n$ but they would drastically change for the $n$ to the right of the plateau. They would stay close to the line which shows the value of the mass of the system which is observed and would approach it. There exist exactly solvable models which substantiate this statement, one is given in \cite{Narison:1989aq}. In the ideal case just mentioned the OPE would not be truncated in any instance and its the truncation that is responsible for the break down of the plateau. To the right of the plateau the non-perturbative terms in the OPE are more important than the perturbative ones, but they are not known well enough for reliable calculations and therefore the results become unreliable and the plateau breaks down. The working window is therefore defined by the criteria of the dominance of perturbation theory or in other words the perturbative part of the OPE. This part is known to a sufficient accuracy and the errors that stem from the non-perturbative part are negligible as long as the perturbative part dominates in the evaluation of the sum rule. A common abbreviation used for the requirement is that asymptotic freedom still holds in the domain where QSR are used, which can be very misleading. The existence of such a windows is a hypothesis and has to be checked from case to case.\\ 
Many publications just use a guess for the spectral function. Hence, the situation can arise that the OPE is under control, but the form of the spectrum is incorrect. In this scenario the sum rules may be completely arbitrary. There may be a plateau, but the properties corresponding to it are incorrect or the plateau can simply be missing and no statement is made even if the OPE is correct. In this case there would be no working window and the OPE would be discarded.\\
These two cases have always to be concerned in the evaluation of a sum rule. Hence, a possibility to reduce the sensitivity of the sum rules would be more than welcome, and sometimes mandatory. Fortunately such a possibility exists, the Borel transformation which was introduced in section \ref{dispersionrelations} in order to improve the convergence of a dispersion integral.\\
The Borel transformation improves the QSR in three ways.
\begin{enumerate}
\item The dispersion integral is regularized because the Borel transformed sum rule is exponentially suppressed at high energies.
\item The exponential suppression also minimizes the influence of excited states.
\item The unknown terms in the OPE are suppressed. 
\end{enumerate}
Hence, the Borel transformation improves the convergence of the sum rule if the ansatz for the spectral function is a good approximation for the corresponding OPE. The Borel transformation is performed by applying the operator
\begin{eqnarray}
\widehat{\mathcal{B}}=\frac{1}{(n-1)!}(Q^2)^n\left(-\frac{d}{dQ^2}\right)^n,\qquad Q^2\rightarrow\infty,\qquad n\rightarrow\infty,\qquad\frac{Q^2}{n}=M^2~fixed
\end{eqnarray}
where $M^2$ is the so called Borel parameter. 
The improving features of the OPE can be illustrated by two simple calculations. 
\begin{enumerate}
\item
Application of $\widehat{\mathcal{B}}$ to $\left(\frac{1}{s+Q^2}\right)$:
\begin{eqnarray}
\widehat{\mathcal{B}}\left(\frac{1}{s+Q^2}\right)=\frac{1}{(n-1)!}(Q^2)^n\left(-\frac{d}{dQ^2}\right)^n\frac{1}{s+Q^2}=\frac{1}{(n-1)!}(Q^2)^n\frac{n!}{(s+Q^2)^{n+1}}\nonumber\\=\frac{n}{Q^2}\frac{(Q^2)^{n+1}}{(s+Q^2)^{n+1}}
=\frac{n}{Q^2}\frac{1}{(1+\frac{s}{Q^2})^{n+1}}=\frac{n}{Q^2}\frac{1}{(1+\frac{s\frac{Q^2}{n}}{n})^{n+1}}\rightarrow\frac{1}{M^2}e^{-\frac{s}{M^2}}
\label{borel2}.
\end{eqnarray}
An exponential kernel in the dispersion integral suppresses the integrand for large s and makes the integral finite. At the same time the spectral function is suppressed for large s.
\item
Application of $\widehat{\mathcal{B}}$ to $\left(\frac{1}{Q^2}\right)^m$:
\begin{eqnarray}
\widehat{\mathcal{B}}\left(\frac{1}{Q^2}\right)^m=\frac{1}{(n-1)!}(Q^2)^n\left(-\frac{d}{dQ^2}\right)^n\left(\frac{1}{Q^2}\right)^m\nonumber\\
\frac{1}{(n-1)!}(Q^2)^n\cdot m(m+1)\cdot ...\cdot(m+n-1)\cdot\left(Q^2\right)^{-m-n}\nonumber\\=\frac{1}{(m-1)!}(Q^2)^{-m}\frac{(m+n-1)!}{(n-1)!}\nonumber\\
\underbrace{=}_{n-1\rightarrow n}\frac{1}{(m-1)!}(Q^2)^{-m}\frac{(m+n)!}{n!}\underbrace{=}_{Stirling}\frac{1}{(m-1)!}(Q^2)^{-m}\frac{\left(\frac{m+n}{e}\right)^{m+n}\sqrt{2\pi (m+n)}}{\left(\frac{n}{e}\right)^{n}\sqrt{2\pi n}}\nonumber\\
=\frac{1}{(m-1)!}(Q^2)^{-m}\frac{(m+n)^{m}}{e^{m}}\frac{(m+n)^{n}}{n^{n}}\sqrt{\frac{m+n}{n}}\nonumber\\=\frac{1}{(m-1)!}\left(\frac{m+n}{Q^2}\right)^{m}\frac{\left(1+\frac{m}{n}\right)^n}{e^{m}}\sqrt{1+\frac{m}{n}}
\rightarrow\frac{1}{(m-1)!}\left(\frac{1}{M^2}\right)^{m}.
\end{eqnarray}
Clearly the Borel transformation leads to an improved convergence of the OPE.
\end{enumerate}
For the perturbative part of the OPE the Borel transformation of the logarithm is needed.\\ 
Application of $\widehat{\mathcal{B}}$ to $ln\left( Q^2\right)$ yields
\begin{eqnarray}
\widehat{\mathcal{B}}\cdot\ln\left( Q^2\right) =\frac{1}{(n-1)!}(Q^2)^n\left(-\frac{d}{dQ^2}\right)^n \ln\left( Q^2\right) =\frac{1}{(n-1)!}(Q^2)^n\left(-\frac{(n-1)!}{(Q^2)^n}\right)=-1.
\end{eqnarray}
From the point of view of mathematical consistency it would be better if the evaluation of a QSR would work by moment sum rules where the improving features are much weaker. Generally a sum rule is more reliable if the moment sum rule converges. However, there are sum rules where the moment sum rules do not converge but the Borel sum rules do. A nice pedagogical review to this topic is given in \cite{Shifman:1999mk}, where a sum rule is constructed which does not converge without being Borel transformed.

\section{Heavy-Light Systems \label{hl_systems}}
\begin{figure}[htbp]
 \begin{center}
\begin{picture}(0,0)%
\includegraphics{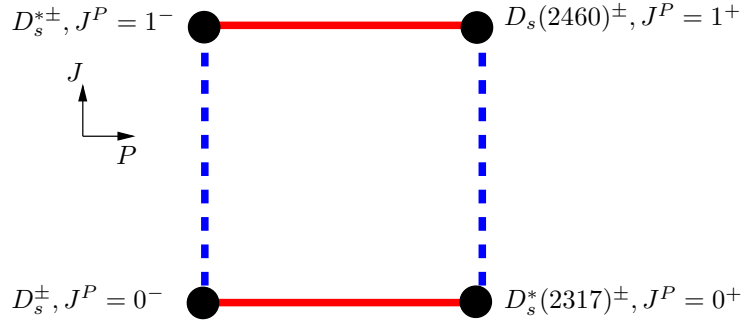}%
\end{picture}%
\setlength{\unitlength}{4144sp}%
\begingroup\makeatletter\ifx\SetFigFont\undefined%
\gdef\SetFigFont#1#2#3#4#5{%
  \reset@font\fontsize{#1}{#2pt}%
  \fontfamily{#3}\fontseries{#4}\fontshape{#5}%
  \selectfont}%
\fi\endgroup%
\begin{picture}(2979,1910)(616,-2563)
\put(958,-1132){$J$}%
\put(3580,-800){$D_{s}(2460)^{\pm},J^P=1^+$}%
\put(3580,-2493){$D^*_{s}(2317)^{\pm},J^P=0^+$}%
\put(1257,-1630){$P$}%
\put(631,-826){$D^{*\pm}_{s},J^P=1^-$}%
\put(631,-2491){$D^{\pm}_{s},J^P=0^-$}%
\end{picture}%
  \caption{Hypermultiplett build up by the most prominent $D_{s}$-states. \label{hypermulti}}
  \end{center}
  \end{figure}

In the sector of heavy-light systems great progress has been made in recent times. A major break through was a discovery made in 2004. In the sector of particels with charm and strangeness two new particels where found, the $D^*_{s}(2317)^{\pm}$ with quantum numbers $J^{P}=0^+$ and the $D_{s}(2460)^{\pm}$ with quantum numbers $J^{P}=1^+$. This discovery confirmed theoretical predictions already made in 1992 \cite{Nowak:1992um}. Together with the already known states $D^{\pm}_{s}$ with $J^P=0^-$ and $D^{*\pm}_{s}$ with $J^P=1^-$ a hypermultiplett is formed which consist of two doublets (see figure \ref{hypermulti}). 
The doublets are given by two points which are connected via a dashed line, the coordinates of the dots are their quantum numbers $J^P$. The quantum numbers are determined by the total angular momentum, the orbital angular momentum and the coupling of the orbital angular momentum and the spins. The parity of a D-meson is given by $P=(-1)^{l+1}$, where $l$ is the quantum number of orbital angular momentum of the valence quarks. Hence negative parity is given by even $l$ and positive by odd $l$. Therefore the orbital angular momentum of the left doublet is odd and of the right one even. Moreover the particels of figure \ref{hypermulti} should be as lowest lying in the spectrum as possible. Hence, the smallest $l$ possible is assumed for them. This results in $l=0$ for the left doublet and $l=1$ for the right doublet. The spin configuration for the left doublet is antiparallel for the the $0^-$ state and parallel for the $1^-$ state, while for the right doublet the spins are parallel for the $0^+$ state and anti-parallel for the $1^+$ state.\\
A descriptive reason for the existence of such particles is given by symmetry arguments. There are two symmetries that determine the structure of the hypermultiplett, the chiral symmetry and the heavy quark symmetry. In the limit where the symmetries are exact and not spontaneously broken the four particels of figure \ref{hypermulti} are degenerate in mass. Both symmetries are explicitly broken, and restored in two totally different limits. Chiral symmetry in the limit of vanishing mass and heavy quark symmetry in the limit of infinitely large mass. If chiral symmetry is assumed to be exact but spontaneously broken for the light quarks and heavy quark symmetry to be exact for the heavy quarks, the main contributions to the mass degeneracy and therefore the overall structure of the masses of the hypermultiplet can be interpreted in terms of the light quark condensates, as discussed by \cite{Bardeen:2003kt}.\\
Historically the states with negative parity have been known for a long time before the states with positive parity were discovered. A theoretical link to the existence of the positive parity states is given by chiral symmetry arguments. The Lagrangian which describes only the negative parity states of a heavy-light system does not posses chiral symmetry in the limit of vanishing light quark masses. However, a Lagrangian which describes the hypermultiplett where both, the negative and the positive parity states are included, exhibits chiral symmetry in the chiral limit. Such symmetry arguments do not prove the existence of the positive parity states, but they are strong arguments and show that theory is on the right way.\\
Heavy quark symmetry links the $J=0$ states directly to the $J=1$ states. In the heavy quark effective theory, the theory in the limit of infinitely heavy quarks, the spin decouples from the dynamics of the system. In that limit a heavy-light system with parallel spins is completely degenerate with a heavy-light system with anti-parallel spins. Hence, the states in the doublets are linked to each other.\\
These considerations have raised a lot of interest for the spectroscopy of $D_{s}$-mesons. One expects similar arguments to hold also for other heavy-light systems. However, the spectrum can not be reproduced exactly by such methods, because only the main contributions to the mass spectrum are included and the fine structure of the spectrum is missing. One method to calculate the mass spectrum more accurately is provided by the QSR. In the following, the QSR in the heavy quark effective theory (HQET) and the results for the masses of the particels in the hypermultiplett are shown. After that the full relativistic theory is going to be reviewed as it has been used and accepted for a long time. The shortcomings of such approaches concerning the calculation of hadron properties will lead to the next topic.

\subsection{Heavy flavor sum rules for the D-meson \label{heavyflavorsr}}
All of the arguments just made can be checked by QSR calculations. This section concerns the question whether the doublets are degenerate in the heavy-quark limit or not. The masses of the $J^P=0^+$ and $J^P=1^+$ and those of the $J^P=0^-$ and $J^P=1^-$ states respectively are expected to be degenerate in the heavy quark limit.\\
In the remainder of the section a mini review of the HQET is given in order to illustrate the most important computation tools. The basis of all calculations is the Lagrangian of the system which is in the actual case split into two parts. The light quark dynamics is described by the ordinary QCD Lagrangian while the heavy quark dynamics is described the heavy quark effective theory Lagrangian
\begin{eqnarray}
\mathscr{L}=\mathscr{L}_{QCD}+\mathscr{L}_{HQET}=\mathscr{L}_{light}+\mathscr{L}_{heavy}.
\end{eqnarray}
Those Lagrangians are given by 
\begin{eqnarray}
\mathscr{L}_{QCD}=\bar{q}\left(i\slashed{D}-m_{q}\right)q-\frac{1}{4}G^a_{\mu\nu}G^a_{\mu\nu}~~~~\mathscr{L}_{HQET}=\bar{h}i\slashed{D}h=\bar{h}iv_{\mu}D^{\mu}h \label{hqetLagrangian}.
\end{eqnarray}
All that is necessary for further considerations are the Feynman rules of the theory. The propagator for the heavy quark is given by 
\begin{eqnarray}
\frac{i}{v_{\mu}p^{\mu}}\frac{\slashed{v}+1}{2}=\frac{i}{\omega}\left(\slashed{v}+1\right) 
\end{eqnarray}
and the vertex for the heavy quark by
\begin{eqnarray}
ig_{s}v_{\mu}t_{a}.
\end{eqnarray}
The rules for the light quark are the ones of ordinary QCD. The heavy quark vertex does not contain any Dirac matrices. Hence the interaction is independent of the spin. The momentum in the heavy quark theory is split up in two parts
\begin{eqnarray}
q_{\mu}=m_{heavy}v_{\mu}+p_{\mu}~~~v_{\mu}v^{\mu}=1,
\end{eqnarray}
where $mv_{\mu}$ is the on-shell part of the momentum $q_{\mu}$ and  $p_{\mu}$ the off-shell part. $p_{\mu}$ is also called the residual momentum. The splitting of the heavy quark momentum is useful because the heavy quark is expected to be non-relativistic. The huge mass is much larger than the typical momentum in a hadron. Hence, higher order terms in $p^{\mu}$ can be neglected. The dependence   on $v_{\mu}$ and $p_{\mu}$ can be comprised in a dependence on $\omega=2v_{\mu}p^{\mu}$.\\
The mass of the heavy quark does not appear in the Lagrangian of the HQET, but in the finite quark mass corrections. A reflection of the "missing" mass parameter is found on the hadron side of the sum rule. The spectral function runs over $\omega$ and the resonances occur at the points $2\bar{\Lambda}=M_{hl-system}-m_{heavy~quark}$, where $2\bar{\Lambda}$ is the value where the heavy quark mass is set to infinity. In this limit the heavy light system also contains a infinitely heavy quark.\\
In a effective theory the fields get modified which is reflected by the Lagrangian (\ref{hqetLagrangian}) where the heavy quark fields $h$ occur, but also the current operators change. Unfortunately the situation here gets very difficult in comparison with the situation in QCD. The currents are not unique in heavy flavor sum rules (see for example \cite{Dai:1996yw}). However only the following currents in table \ref{hqetcurrents} are used in the sections below.
\begin{table}[htbp]
\begin{tabular}{||c|l|l|l|l|l|l||}
\hline
$J^P$& current\\
\hline 
$0^+$&$\bar{q}h$\\
\hline  
$0^-$&$\bar{q}\gamma_{5}h$ \\ 
\hline 
$1^+$&$\bar{q}\gamma_{5}\left(\gamma_{\mu}-\slashed{v}v_{\mu}\right)h $\\
\hline
$1^-$&$\bar{q}\left(\gamma_{\mu}-\slashed{v}v_{\mu}\right)h$\\
\hline                      
\end{tabular}
\caption{Table with the currents which are used for the approximation of the analyzed mesons in the HQET.}\label{hqetcurrents}
\end{table}
These currents are used to compute sum rules for the D-meson hypermultiplett which reproduce the expected mass degeneracy in the heavy quark limit. The calculations in the HQET are illustrated in the following.

\subsubsection{Calculation of the heavy-light quark loop}

The calculation for the pseudo-scalar and scalar current correlator are demonstrated. In the scalar case the current is
\begin{eqnarray}
j(x)=\bar{q}(x)h_{Q}(x),~j^{\dagger}(x)=\bar{h}_{Q}(x)q(x)\label{hqetscalar}.
\end{eqnarray}
The current correlator is given by
\begin{eqnarray}
\Pi_{pert}(q^2)=i\int d^4xe^{iqx}\bra{0}T\left(j(x),j^{\dagger}(x)\right)\ket{0}=i\int d^4xe^{iqx}\wick{2-5}{3-4}{\bra{0},\bar{q}(x),h_{Q}(x),\bar{h}_{Q}(x),q(x),\ket{0}}\nonumber\\=
\parbox[c]{5cm}{
\begin{fmffile}{heavylighthqet01}
 \begin{fmfgraph*}(45,40)
    \fmfleft{i} 
    \fmfright{o}
    \fmf{dots_arrow,label.side=down,label=$q$}{i,v1} 
    \fmf{dots_arrow,label.side=down,label=$q$}{v2,o}
    \fmf{fermion,label.side=left,label=$p$,left,tension=.3}{v1,v2}
    \fmf{dbl_plain_arrow,label.side=left,label=$p+q$,left,tension=.3}{v2,v1}
 \end{fmfgraph*}
\end{fmffile}}
=\int\frac{d^4p}{(2\pi)^4}\left( -tr\left[\frac{i(\slashed{p}+m)}{p^2-m^2}\frac{i}{v^{\mu}(q_{\mu}+p_{\mu})}\frac{\slashed{v}+1}{2}\right]\right) \nonumber\\
=\int\frac{d^4p}{(2\pi)^4}\frac{4v^{\mu}p_{\mu}+4m}{\left( p^2-m^2\right)2\left( v^{\mu}(q_{\mu}+p_{\mu})\right)}=4\int\frac{d^4p}{(2\pi)^4}\frac{v^{\mu}p_{\mu}+m}{\left( p^2-m^2\right)\left(\omega+2v^{\mu}p_{\mu}\right)}
\label{hqetpertscalar},
\end{eqnarray}
where the double line represents the propagator in the heavy quark effective theory. The expression for the correlator is renormalized by using the $\overline{MS}$-scheme. Therefore it has to be regularized by dimensional regularization. To begin a Feynman parameter is introduced and the integration is transformed to a spheric symmetric integration
\begin{eqnarray}
4\int\frac{d^4p}{(2\pi)^4}\int_{0}^{1}dx\frac{v^{\mu}p_{\mu}+m}{\left[x\left( p^2-m^2\right)+(1-x)\left(\omega+2v^{\mu}p_{\mu}\right)\right]^2}=4\int\frac{d^{4}l}{(2\pi)^4}\int_{0}^{1}dx\frac{-\frac{1-x}{x}+m}{x^2\left[l^2-\Delta\right]^2 },
\end{eqnarray}
with $l_{\mu}=p_{\mu}+v_{\mu}\frac{1-x}{x}$ and $\Delta=\left(\frac{1-x}{x}\right)^2+m^2-\frac{1-x}{x}\omega$. The $dp^4$ integral is extended to d dimensions and the integration is performed
\begin{eqnarray}
d\left(\mu^2\right)^{\left(2-\frac{d}{2}\right)} \int_{0}^{1}dx\frac{-\frac{1-x}{x}+m}{x^2}\int\frac{d^{d}l}{(2\pi)^d}\frac{1}{\left[l^2-\Delta\right]^2}\nonumber\\=id\int_{0}^{1}dx\frac{-\frac{1-x}{x}+m}{x^2}\frac{1}{\left(4\pi\right)^{d/2}}\frac{\Gamma(2-\frac{d}{2})}{\Gamma(2)}\left(\frac{\mu^2}{\Delta}\right)^{2-\frac{d}{2}}\nonumber\\\longrightarrow\frac{di}{(4\pi)^2}\int_{0}^{1}dx\frac{-\frac{1-x}{x}+m}{x^2}\left(\frac{2}{4-d}-\log\left( \frac{\Delta}{\mu^2}\right)-\gamma+log(4\pi)+\mathcal{O}(4-d)\right).
\end{eqnarray}
After the subtraction and in the limit $d\rightarrow 4$ the following expression is derived
\begin{eqnarray}
\Pi_{pert}(q^2)=-\frac{i}{4\pi^2}\int_{0}^{1}dx\frac{-\frac{1-x}{x}+m}{x^2}\log\left( \frac{\Delta(q^2)}{\mu^2}\right).
\end{eqnarray}
This form of the amplitude is very inconvenient because of the Feynman parameter integral. One method to get rid of it is to consider only the imaginary part of the amplitude. Which is sufficient for the derivation of the whole amplitude (see section \ref{dispersionrelations}). The logarithm determines the imaginary part of the correlator. It is imaginary if the argument is negative $\frac{\Delta(q^2)}{\mu^2}<0$. This domain is located in the interval between the roots of $\Delta(q^2)$. Hence, the imaginary part of the correlator is
\begin{eqnarray}
-\frac{i}{4\pi^2}\int_{x_{1}}^{x_{2}}dx\frac{-\frac{1-x}{x}+m}{x^2}\log(\frac{\Delta(q^2)}{\mu^2}),
\end{eqnarray}
where $x_{1/2}=\frac{\omega\pm\sqrt{\omega^2-4m^2}+2}{2(m^2+\omega+1)}$ are the roots of $\Delta(q^2)$. In this interval  $\log(\frac{\Delta(q^2)}{\mu^2})=i\pi$
\begin{eqnarray}
Im\Pi_{pert}(q^2)=\frac{1}{4\pi}\int_{x_{1}}^{x_{2}}dx\frac{-\frac{1-x}{x}+m}{x^2}=\frac{1}{8\pi}(\omega-2m)\sqrt{\omega^2-4m^2}.
\end{eqnarray}
Consequently the imaginary part of the correlator is determined, but the color structure of the problem has not been considered. A factor 3 has to be added
\begin{eqnarray}
Im\Pi_{pert}(q^2)=\frac{3}{4\pi}\int_{x_{1}}^{x_{2}}dx\frac{-\frac{1-x}{x}+m}{x^2}=\frac{1}{8\pi}(\omega-2m)\sqrt{\omega^2-4m^2}.
\end{eqnarray}
The next step is to calculate the two point correlator of the pseudo-scalar current correlator, where the current is
\begin{eqnarray}
j(x)=\bar{q}(x)i\gamma_{5}h_{Q}(x),~j^{\dagger}(x)=-\bar{h}_{Q}(x)i\gamma_{5}q(x)
\end{eqnarray}
\begin{eqnarray}
\Pi_{pert,5}(q^2)=\parbox[c]{5cm}{
\begin{fmffile}{heavylighthqet02}
 \begin{fmfgraph*}(45,40)
    \fmfleft{i} 
    \fmfright{o}
    \fmf{dots_arrow,label.side=down,label=$q$}{i,v1} 
    \fmf{dots_arrow,label.side=down,label=$q$}{v2,o}
    \fmf{fermion,label.side=left,label=$p$,left,tension=.3}{v1,v2}
    \fmf{dbl_plain_arrow,label.side=left,label=$p+q$,left,tension=.3}{v2,v1}
 \end{fmfgraph*}
\end{fmffile}}
=4\int\frac{d^4p}{(2\pi)^4}\frac{-v^{\mu}p_{\mu}+m}{\left( p^2-m^2\right)\left(\omega+2v^{\mu}p_{\mu}\right)}
\label{hqetpertpseudoscalar}.
\end{eqnarray}
The only difference between (\ref{hqetpertscalar}) and (\ref{hqetpertpseudoscalar}) is the minus sign in front of $v^{\mu}p_{\mu}$. The result for $\Pi_{pert,5}(q^2)$ is 
\begin{eqnarray}
Im\Pi_{pert,5}(q^2)=\frac{3}{8\pi}(\omega+2m)\sqrt{\omega^2-4m^2}.
\end{eqnarray}
This is an example for a interesting rule. The results for chiral partners are obtained from each other by exchanging $m$ with $-m$.

\subsubsection{Calculation of the quark condensate coefficient $C_{\bar{q}q}$}

Again the example of the pseudo-scalar current is considered (see table \ref{hqetcurrents})
\begin{eqnarray}
C_{\bar{q}q}\mele{\bar{q}q}=i\int d^4xe^{iqx}\wick{1-2,5-6}{3-4}{\bra{p,s},\bar{q}(x),h_{Q}(x),\bar{h}_{Q}(x),q(x),\ket{p,s}}=
\parbox[c]{3cm}{
\begin{fmffile}{heavylighthqet03}
\begin{fmfgraph*}(40,35)
 \fmfleft{i1,o1} \fmfright{i2,o2}
 \fmffixedx{2cm}{v1,v2}
 \fmf{dots_arrow,label.side=left,label=$q$}{i1,v1}
 \fmf{dbl_plain_arrow,label.side=down,label=$q+p$}{v1,v2}
 \fmf{dots_arrow,label.side=left,label=$q$}{v2,i2}
 \fmf{fermion,label.side=right,label=$p$}{o1,v1}
 \fmf{fermion,label.side=right,label=$p$}{v2,o2}
\end{fmfgraph*}
\end{fmffile}}
\nonumber\\=i\bar{u}(p,s)\frac{i}{v^{\mu}(q_{\mu}+p_{\mu})}\frac{\slashed{v}+1}{2}u(p,s).
\end{eqnarray}
The results of section \ref{Wilson_Berechnung} are used to derive
\begin{eqnarray}
C_{\bar{q}q}\mele{\bar{q}q}=-\frac{1}{m}\left( \frac{1}{v^{\mu}(q_{\mu}+p_{\mu})}\frac{v^{\mu}p_{\mu}+m}{2}\right)\bar{u}(p,s)u(p,s)
=-\frac{1}{m}\frac{v^{\mu}p_{\mu}+m}{\omega+2v^{\mu}p_{\mu}}\bar{u}(p,s)u(p,s).
\end{eqnarray}
The expression
\begin{eqnarray}
-\frac{1}{m}\left(\frac{v^{\mu}p_{\mu}+m}{\omega+2v^{\mu}p_{\mu}}\right)
\end{eqnarray}
has to be averaged over the fourdimensional Euclidean angle as it is shown in section \ref{Wilson_Berechnung}. The result is
in the limit $m\rightarrow 0$ is
\begin{eqnarray}
C_{\bar{q}q}= -\frac{1}{\omega}\label{hqetquarkpseudo}.
\end{eqnarray}
The scalar current correlator is determined by the expression
\begin{eqnarray}
C_{\bar{q}q}\mele{\bar{q}q}=\parbox[c]{3cm}{
\begin{fmffile}{heavylighthqet03}
\begin{fmfgraph*}(40,35)
 \fmfleft{i1,o1} \fmfright{i2,o2}
 \fmffixedx{2cm}{v1,v2}
 \fmf{dots_arrow,label.side=left,label=$q$}{i1,v1}
 \fmf{dbl_plain_arrow,label.side=down,label=$q+p$}{v1,v2}
 \fmf{dots_arrow,label.side=left,label=$q$}{v2,i2}
 \fmf{fermion,label.side=right,label=$p$}{o1,v1}
 \fmf{fermion,label.side=right,label=$p$}{v2,o2}
\end{fmfgraph*}
\end{fmffile}}~~~
=-\frac{1}{m}\left(\frac{v^{\mu}p_{\mu}-m}{\omega+2v^{\mu}p_{\mu}}\right)\bar{u}(p,s)u(p,s).
\end{eqnarray}
The change in the expression in comparison with the pseudo-scalar case is just the minus sign in front of the $m$-term. The result in the limit $m\rightarrow 0$ is
\begin{eqnarray}
C_{\bar{q}q}= \frac{1}{\omega}
\end{eqnarray}  
where the difference to the pseudo-scalar case (\ref{hqetquarkpseudo}) is just a flip in the sign of the coefficient.

\subsubsection{The Borel transformed heavy flavor sum rule}

\begin{figure}[htbp]
 \begin{center}
  \includegraphics{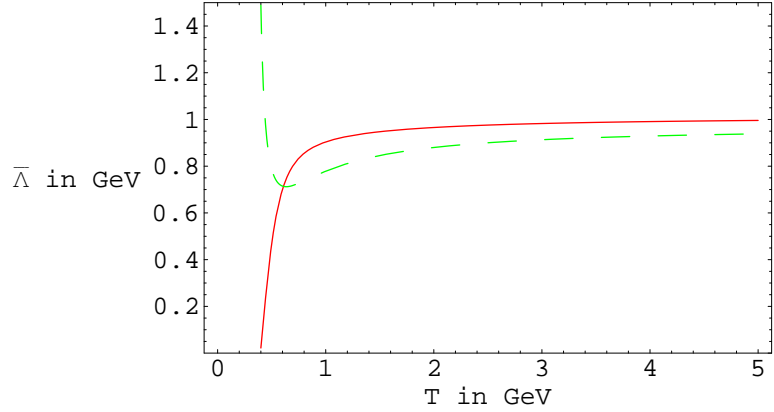}
  \caption{This figure shows the Borel curves for $\overline{\Lambda}$, the mass parameter in the HQET, for the states in the hypermultiplet of figure \ref{hypermulti}. The full line is the curve for the $P=+$ states while the dashed line is the curve for the $P=-$ states.\label{hqetborel}}
  \end{center}
  \end{figure}

The Borel transformed heavy flavor sum rule for the $D-mesons$ with isospin or strangeness, in the limit $m_{light}=0$ reads for the positive parity doublet
\begin{eqnarray}
\bra{0}j\ket{0}^2e^{-2\frac{\overline{\Lambda}}{T}}=\frac{3}{16\pi^2}\int_{0}^{\omega_{c}}\omega^2e^{-\frac{\omega}{T}}d\omega+\frac{\bra{0}\bar{q}q\ket{0}}{2}-\frac{1}{8T^2}M^2_{0}\bra{0}\bar{q}q\ket{0} \label{hqet_positive}
\end{eqnarray}
and for the negative parity doublet
\begin{eqnarray}
\bra{0}j\ket{0}^2e^{-2\frac{\overline{\Lambda}}{T}}=\frac{3}{16\pi^2}\int_{0}^{\omega_{c}}\omega^2e^{-\frac{\omega}{T}}d\omega-\frac{\bra{0}\bar{q}q\ket{0}}{2}+\frac{1}{8T^2}M^2_{0}\bra{0}\bar{q}q\ket{0}.
\end{eqnarray}
T is the Borel parameter. The currents employed for the calculation of the sum rule have an additional $\frac{1}{\sqrt{2}}$ factor (see \cite{Dai:1996yw}). Hence, in the heavy quark limit the masses of the doublets are degenerated. \\
The result of the sum rules evaluation in the channel with isospin is $\overline{\Lambda}=(1.00)GeV$ for the positive parity doublet and $\overline{\Lambda}=(0.95)GeV$ for the negative parity doublet. Hence, the positive parity states are supposed to be heavier than the negative parity states. Thus, the prediction 
of the mass for the $D-mesons$ with positive parity is $m_{D}\approx\overline{\Lambda}+m_{c}=2.3~GeV$. In the case of the negative parity states $m_{D}\approx\overline{\Lambda}+m_{c}=2.25~GeV$. The Borel curves are given in figure \ref{hqetborel}, the curves for the positive and negative parity states stay nearly equidistant for $T>1$.
Dai et al. where the first ones to analyze such sum rules (\ref{hqet_positive}), but they just analyzed the positive parity states. However, their value for the threshold $\omega_{c}$ was used for the calculations shown above. In their publication they obtained $\overline{\Lambda}=1.05 GeV$ with $\omega_{c}=2.65GeV$. The difference between the calculations above and their calculations is probably caused by a difference in the condensates. Unfortunately they do not give the values they used for the condensates.\\
However, in this calculation the values for the quark condensate, as given in section \ref{condensates}, has been used. Hence, in the negative parity channel the results can be compared with the masses of the $D^{0}(1869),D^{\pm}(1864),D^{*}(2007)^{0}$ and $D^{*}(2010)^{\pm}$ particles. In the positive parity channel the data of the $D_{1}(2420)^{0}$ and $D_{1}(2430)^{0}$ particles can be compared with the results. Obviously the positive parity states lie much closer to the results than the negative parity states. In the $P=+$ case the error is about $0.150~GeV$ while it is about $0.400~GeV$. This may be due to the OPE, an improved OPE may give better results. 

\subsubsection{The mass formula for heavy flavor sum rules \label{hqetcorrections}}

The intradoublet mass splitting is another topic. It stems from the breaking of the heavy quark symmetry. This violation is in QFT formulated as $\frac{1}{m_{heavy}}$ corrections. The Lagrangian with the first corrections is given by
\begin{eqnarray}
\mathscr{L}_{eff}=\bar{h}_{v}iv_{\mu}D^{\mu}h_{v}+\frac{\mathcal{K}}{2m_{Q}}+\frac{\mathcal{S}}{2m_{Q}}+\mathcal{O}\left(\frac{1}{m_{Q}^2}\right)
\end{eqnarray}
where $\mathcal{K}$ is the nonrelativistic kinetic energy operator, defined as
\begin{eqnarray}
\mathcal{K}=\bar{h}_{v}\left(iD_{\perp}\right)^2 h_{v}
\end{eqnarray}
with $D^2_{\perp}=D_{\mu}D^{\mu}-\left(v_{\mu}D^{\mu}\right)^2$, and $\mathcal{S}$ is the chromomagnetic interaction term
\begin{eqnarray}
\mathcal{S}=C_{mag}\left(\frac{m_{Q}}{\mu}\right)\bar{h}_{v}\frac{g_{s}}{2}\sigma_{\mu\nu}G^{\mu\nu}h_{v}
\end{eqnarray}
where $C_{mag}=\left[\frac{\alpha_{s}(m_{Q})}{\alpha_{s}(\mu)}\right]^{\frac{3}{\beta_{0}}}$ and $\beta_{0}=11-\frac{2}{3}n_{f}$ is the first coefficient of the $\beta$ function.\\
Taking into account the $\frac{1}{m_{Q}}$ corrections in the Lagrangian, the meson mass formula in HQET is expressed as
\begin{eqnarray}
M=m_{Q}+\bar{\Lambda}-\frac{1}{2m_{Q}}\left(\lambda_{1}+d_{m}\lambda_{2}\right)
\end{eqnarray}
where the two additional parameters $\lambda_{1}$ and $\lambda_{2}$ at the $\frac{1}{m_{Q}}$ order are defined by two matrix elements 
\begin{eqnarray}
2M\lambda_{1}=\bra{M}\mathcal{K}\ket{M},~~2d_{M}C_{mag}M\lambda_{2}=\bra{M}\mathcal{S}\ket{M}
\end{eqnarray}
the constant $d_{M}$ is spin-related $d_{M}=d_{j,j_{l}}$ and $d_{j_{l}-\frac{1}{2},j_{l}}=2j_{l}+2,d_{j_{l}+\frac{1}{2},j_{l}}=-2j_{l}$ .\\
The evaluation of $\lambda_{1}$ and $\lambda_{2}$ needs the consideration of the three-point correlator
\begin{eqnarray}
\Sigma(\omega,\omega `)=i^2\int d^4 xd^4 y\exp(ikx-ik'y)\bra{0}TJ^{\dagger}(x)O(0)J(y)\ket{0}
\end{eqnarray}
where the operator O can be $\mathcal{K}$ or $\mathcal{S}$ and J is still the generic interpolating current. The derivation of the formulas can be found in \cite{Ball:1993xv} and references herein.

\subsubsection{Shortcomings of the heavy flavor sum rules for the $D$-mesons}

The preceding sections have shown the characteristics of heavy flavor sum rules and their shortcomings. The mass formula shows that a lot of additional work has to be done in order to get the corrections to the heavy quark limit. Moreover, the operators in the HQET have to be renormalized in order to get the right results (see \cite{Bagan:1991sg}). This is also not necessary in the full theory. In summary the heavy flavor sum rules can be excluded as the right tool for the calculation of the properties of the spectral function, because except for additional work they supply nothing new nor an improvement against the sum rules in the full theory.\\
Even if the first corrections are taken into account the effective theory does not produce a satisfying accuracy (see for example \cite{Huang:2005ke}). This is a second argument against heavy flavour sum rules for the calculation of properties of spectral functions. 

\subsection{The QSR for heavy light-mesons in full QCD}
\begin{figure}[htbp]
 \begin{center}
  \includegraphics{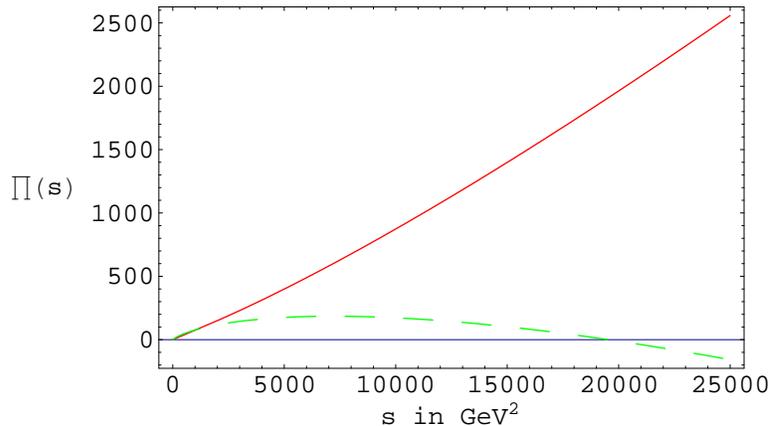}
  \caption{The correlator for the scalar current of a heavy-light system in the approximation $m_{light}=0$ 
  to first (dashed line) and second order in $\alpha_{S}$ (full line). The expressions are nearly identical for $q^2<1000~GeV^2$. Above $q^2=1000~GeV^2$ the curves differ strongly from each other until the point is reached where the first order approximation becomes negative. In the case of a vector current, the correlators in first and second order remain nearly equal to each other over the whole $q^2$ range.\label{hl_scalar_pert}}
  \end{center}
  \end{figure}

The first step is to set the boundary conditions for the OPE. In section \ref{importantunimportant} a criteria was formulated which allows a classification of the terms in the OPE. Important are all terms containing operators of dimension smaller or equal to the dimension of the operator product under consideration, $d(O_{j})\leq d(jj)$. All remaining terms are less important. The currents for the heavy-light mesons which are analyzed here are given by
\begin{eqnarray}
j_{S}=\left(m_{2}+m_{1}\right)\bar{q}_{2}q_{1}=\partial_{\mu}j_{V}^{\mu}~~~~\nonumber\\
j_{P}=\left(m_{2}-m_{1}\right)\bar{q}_{2}i\gamma_{5}q_{1}=\partial_{\mu}j_{A}^{\mu}\nonumber\\
j_{V}=\bar{q}_{2}\gamma_{\mu}q_{1}~~~~~~~~~~~~~~~~~~~~~~~~~~ \nonumber\\
j_{A}=\bar{q}_{2}\gamma_{5}\gamma_{\mu}q_{1}.~~~~~~~~~~~~~~~~~~~~~~~~\label{hlcurrents}
\end{eqnarray}
The scalar and pseudo scalar currents have dimension four while the vector and axialvector current have dimension three. Hence, all important operators in the scalar or pseudoscalar case have dimension less than or equal to eight, while in the vector or axialvector case the important operators have dimension of less than or equal to six. In the ideal case the OPE for the particels approximated by these currents would contain all of these terms. However, the situation in reality is far from that goal.\\ 
Historically the pioneers in the area of mass spectroscopy of heavy light systems with QSR were the authors Novikov, Voloshin, Shifman, Vainshtein and Zakharov (SVZ) in the late seventies \cite{Novikov:1978tn}. Moreover the authors Reinders,Rubinstein and Yazaki (RRY) continued their work in the early eighties. They computed the first QSR for this problem (see \cite{Reinders:1981ty} and references therein). They truncated the OPE on a very crude level, keeping only the quark and gluon condensates. However their work showed that the sum rules for such particels work and they could derive first estimates of properties for the B-meson family with moment sum rules. Properties for the D-meson family were first computed by Aliev in 1983. He used an OPE which contained more operators than RRY. This lead to predictions of leptonic decay constants for D-mesons (see \cite{Aliev:1983ra}). He included, in addition to the quark and gluon condensate, the mixed and four quark condensates in the OPE. Narison used an OPE which was similar to the one of Aliev for further investigations of heavy-light systems in the late eighties \cite{Narison:1987qc}. His work showed that it is possible to treat B- and D-mesons with such sum rules, but for D-mesons the use of Borel sum rules was necessary. Moment sum rules do not converge for the D-meson. Despite the differences in the approximations made by the groups they also have things in common. The continuum contribution ca not be neglected as it has been done for the charmonium system. This is nothing particular to heavy-light systems. The continuum is important in many cases. The $\rho$-meson sum rule is a classical example where the continuum has to be taken into account. During the nineties the heavy quark effective theory entered the realm of QSR but as shown in section \ref{heavyflavorsr} this approach has not brought much progress to mass spectroscopy with QSRs. Thus for many years Narison was the only author who pushed the field forward. He always used the sum rules close to the one constructed in \cite{Aliev:1983ra}. Aliev's sum rule was for D-mesons with $J^P=0^-$. Thus all analyzes based on this sum rule where done for $0^-$ or $0^+$ states, the sum rules for the two particels are connected by a transformation.  There have been many papers about QSR and heavy-light systems which simply used this sum rule in order to perform calculations, but they only applied Narison's work. The situation changed after the discovery of the positive parity doublet in the $D_{s}$-meson family. The interpretation of the nature of such states is still open and has been attacked in papers like \cite{Narison:2003td}, \cite{Hayashigaki:2004gq}, \cite{Kim:2005gt} and \cite{Nielsen:2005ia}. Where Narison's \cite{Narison:2003td} result implies that the dominant configuration of the $D_{S}(2317)$ is a quark anti-quark state, Hayashigaki \cite{Hayashigaki:2004gq} finds, using an unconventional approach that a more complicated state must dominate. In both publications two quark currents and the sum rule of the Aliev type are used for the $J=0$ particels. An important difference is that Hayashigaki uses a sum rule for the $J=1$ states, while Narison uses HQET arguments to derive the masses of the $J=1$ particels from the $J=0$ masses. Hence, Hayashigaki's calculation is superior in the $J=1$ channel. In the remaining publications a four quark current was used as the interpolating current for the $D_{S}(2317)$. However, there still remains much work to be done.\\
A short review of the sum rules used by Narison and Aliev exhibit the shortcomings of those sum rules.  The OPE for the sum rules is of course truncated and the Wilson coefficients are calculated to a certain order in the strong coupling constant $\alpha_{S}$. Moreover the limit of massless quarks is used. A curious point is that various authors do not agree on the form of the OPE. Some authors have even published several papers on heavy-light systems using the same Sum Rule but with different results. The difference enters the sum rule through the Wilson coefficients. During this work special attention was payed to the mixed condensate coefficient. A short review about the different results will now be given. In the first calculation of the $J=0$ channel the result was incorrect \cite{Novikov:1978tn} and was corrected in the publication \cite{Aliev:1983ra}.The sum Rule in the publication \cite{Aliev:1983ra} is today concerned as the correct one in the $J=0$ channel. Unfortunately Narison published an additional version of the mixed condensate coefficient in the $J=0$ which is incorrect \cite{Narison:1987qc}. He subsequently improves his results step by step until he arrives at the correct one in \cite{Narison:2003td}. There have been additional publications in which the Wilson coefficients do not agree with other authors. In most cases the quark and gluon condensate coefficients are correct, but for higher dimensions the results differ. In the $J=1$ channel the situation is similar. This problem was also recognized by K�pfer et al. (see \cite{Kaempfer:2005}). Narison's result in \cite{Narison:1989aq} differs from Hayashigaki's \cite{Hayashigaki:2004gq} in more than the mixed condensate terms. The correct result in the $J=0$ and $J=1$ channel are given in \cite{Jamin:1992se} and \cite{Generalis:1990id}. However, most authors who recognize the ambiguities state that the difference does not change the results significantly \cite{Narison:2003td}. The OPE used for further calculations is given by  
\begin{eqnarray}
\Pi_{5}\left(q^2\right)=\frac{1}{\pi}\int_{m^2}^{\infty}\frac{Im\left[ \Pi_{5,pert}(s)\right] }{s-q^2}ds\nonumber\\+\frac{m^2}{m^2-q^2}\left[\frac{\mele{\alpha_{s}G^2}}{12\pi}-m\mele{\bar{\psi}_{i}\psi_{i}}\right]+\frac{m^3q^2}{(m^2-q^2)^3}\frac{1}{2}\mele{g\bar{\psi_{i}}\sigma^{\mu\nu}\frac{\lambda_{a}}{2}\psi_{i}G^{a}_{\mu\nu}}\nonumber\\+\frac{m^2}{(m^2-q^2)^2}\left[2-\frac{m^2}{m^2-q^2}-\left(\frac{m^2}{m^2-q^2}\right)^2\right]\frac{\pi}{6}\alpha_{s} \mele{\bar{\psi}_{i}\gamma_{\mu}\lambda_{a}\psi_{i}\sum_{i=u,d,s}\left(\bar{\psi}_{i}\gamma_{\mu}\lambda_{a}\psi_{i}\right)}
\label{hlscalar}
\end{eqnarray}
for the pseudo-scalar current and by
\begin{eqnarray}
\Pi_{V}\left(q^2\right)=\frac{1}{\pi}\int_{m^2}^{\infty}\frac{Im\left[ \Pi_{V,pert}(s)\right] }{s-q^2}ds\nonumber\\+\frac{m^2}{m^2-q^2}\left[-\frac{\mele{\alpha_{s}G^2}}{12\pi}-m\mele{\bar{\psi}_{i}\psi_{i}}\right]+\left(\frac{m^2}{m^2-q^2}\right)^3\frac{1}{m} \frac{1}{2}\mele{g\bar{\psi_{i}}\sigma^{\mu\nu}\frac{\lambda_{a}}{2}\psi_{i}G^{a}_{\mu\nu}}\nonumber\\-\frac{m^2}{(m^2-q^2)^2}\left[4+8\frac{m^2}{m^2-q^2}-3\left(\frac{m^2}{m^2-q^2}\right)^2\right]\frac{\pi}{18}\alpha_{s} \mele{\bar{\psi}_{i}\gamma_{\mu}\lambda_{a}\psi_{i}\sum_{i=u,d,s}\left(\bar{\psi}_{i}\gamma_{\mu}\lambda_{a}\psi_{i}\right)}
\label{hlvector}
\end{eqnarray}
for the vector current (see \cite{Jamin:1992se,Generalis:1990id}). The OPE for the axial partners can be derived from the OPEs above by exchanging m to -m. Thus the OPE of the scalar particels can be derived from the one of the pseudo-scalar particels and the same holds for the axial-vector and vector particles. The terms that are changed by this procedure are the ones which contain odd powers in the mass. An inspection of at the OPEs above exhibits that these are terms which contain quarks in the condensates. Thus chiral symmetry breaking manifests itself through the quark condensates. For vanishing quark condensates the OPEs for the axial partners would be identical. Hence the interdoublet splitting is due to these terms. Another important aspect is the mass due to breaking of the heavy quark symmetry. Here the intradoublet mass splitting is mainly due to the mixed condensate term. In comparison with the finite mass corrections of section \ref{hqetcorrections} this is quite obvious because the operator of the mixed condensate is nearly the chromomagnetic interaction term which is responsible the biggest contribution to the finite mass corrections and these corrections give the mass splittings between the $J=0$ and $J=1$ states.\\
The Wilson coefficient of the unit operator $\mathds{1}$ is written down as a dispersion integral. Though this notation seems to be quite long winded it is very handy. In many cases the imaginary part of the perturbative expression of the current correlator is much simpler than the full expression of the current correlator. Moreover if the OPE is applied to a QSR calculation and the continuum contribution has to be taken into account it turns out that the notation quoted above implements the continuum effects very naturally. The reason is that on the phenomenological side of the OPE the continuum of the spectral function is approximated by the imaginary part of the perturbative piece of the 2-point correlator, which is included in a dispersion relation. Hence, the imaginary part of the perturbative piece of the two point correlator appears both on the left and right hand side of a sum rule and is on both sides embedded in a dispersion integral. Thus it can be eliminated from the phenomenological side by a simple subtraction and be incorporated by the perturbative Wilson coefficient on the theoretical side. In conclusion only the limits of integration in the perturbative Wilson coefficient change as the continuum contribution is accounted for.\\
The perturbative part of the current correlator for the currents given in (\ref{hlcurrents}), with the mass of light quarks set to zero, is given in \cite{Chetyrkin:2000mq}. In that paper a Mathematica package is presented that contains the perturbative part of the correlators. The use of the second order corrections is necessary for the scalar and pseudo-scalar channel if large $q^2$ are considered because the imaginary part of the correlator becomes negative for large $q^2$ if only the first order corrections are used. This problem is remedied when the second order corrections are included (see figure  \ref{hl_scalar_pert}).
The requirement of positivity for the imaginary part of the correlator stems from the connection between the spectral function and the imaginary part of the correlator. The spectral function has to be positive and is up to a constant and a power of $q^2$ equal to the imaginary part of the correlator. Hence, a sign switch is forbidden. In fact for small $q^2$ the expression for the correlator in first and second order of $\alpha_{S}$ are nearly equal, while they differ strongly with growing $q^2$. This is a beautiful example for a calculation where higher order corrections in the coupling constant are necessary in order to get a valid approximation.\\ 
In the OPEs (\ref{hlscalar}) and (\ref{hlvector}) the quark mass effects, as outlined in section \ref{gluonmix}, are taken into account. It is a remarkable coincidence that in the limit of vanishing light quark masses the coefficient of the gluon and the quark condensate nearly coincide. 

\subsubsection{Analysis with moment sum rules}

\begin{figure}[htbp]
 \begin{center}
  \includegraphics{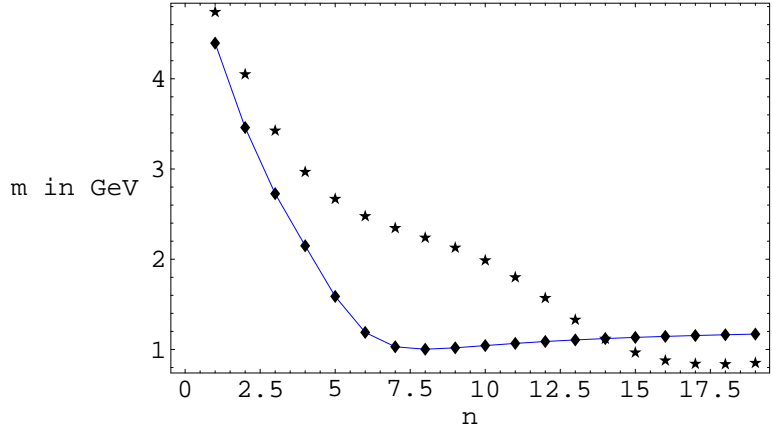}
  \caption{Results of the moment sum rules for the $J^P=0^+$ D-mesons. The full line with diamonds corresponds to the set $q^2=-1GeV^2,m=1.3GeV,t_{c}=50GeV^2$ while the other curve corresponds to the set $q^2=-5GeV^2,m=1.3GeV,t_{c}=50GeV^2$. The condensate values are those from section \ref{condensates}. No convergence is found for theses sum rules.\label{0+moments}}
  \end{center}
  \end{figure}

\begin{figure}[htbp]
 \begin{center}
  \includegraphics{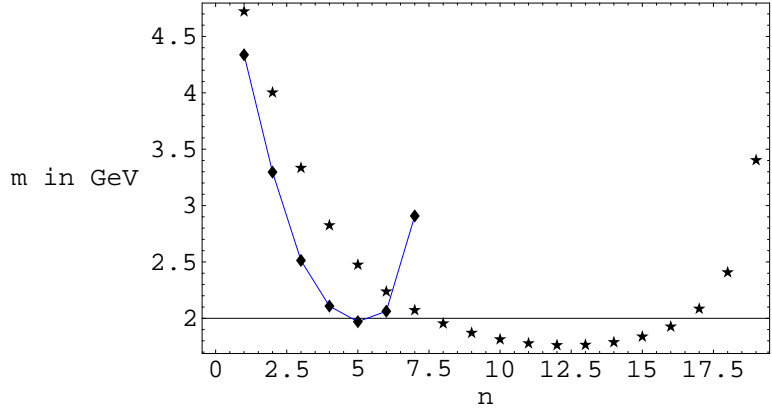}
  \caption{Results from the moment sum rules for the $J^P=0^-$ D-meson with isospin. The full line with diamonds corresponds to the set $q^2=-1GeV^2,m=1.3GeV,t_{c}=50GeV^2$ while the other curve corresponds to the set $q^2=-5GeV^2,m=1.3GeV,t_{c}=50GeV^2$. The condensate values are those from section \ref{condensates}. The horizontal line gives the experimental mass of $0^-$ heavy-light systems with isospin.\label{0-moments}}
  \end{center}
  \end{figure}

As the starting point for the analysis of such sum rules, moment sum rules and the narrow resonance plus continuum ansatz are used to evaluate the QSR for heavy-light systems. The qualitative result of such an analysis concerns the convergence of the sum rule. For D-mesons in the $J^P=0^-,1^-$ the sum rule converges, although the stability of the curves with respect to a shift in the parameters is weak. On the other hand in the $J^P=0^+,1^+$ channels no convergence is seen (see figure \ref{0+moments}). 
Hence, there is something wrong, with the Sum Rules in the positive parity channels. The moment sum rules in the channel with negative parity is also excluded, but why ? In the example at hand the OPEs (\ref{hlscalar}) and (\ref{hlvector}) with the condensate values of section \ref{condensates} have been used. The mass of the c-quark and the threshold value are the only free parameters. The result should be independent from $q^2$. The fit of the free parameters should yield nearly equal values independent of the isospin or strangeness of the system. Unfortunately this is not the case, the difference in the parameters are too large. The plots in figure \ref{0-moments} show the masses for the negative parity states which are fitted to reproduce the right result. The criteria for the right result was to hit a value which lies around the right value for D-mesons with isospin and $D_{s}-mesons$ with strangeness. This should lead to nearly equal c-quark masses, a threshold value $t_{c}$ which is close for the two cases and a independence from $q^2$ in a small errorbar window. The  threshold requirement is satisfied, but all others are not. In addition the plateau for the $q^2=-1$ sum rule is to small to be reliable. This does not exclude the sum rule but it diminishes the trust in the reliability of the sum rule drastically, without further improvements the sum rule ca not be used. In the $J=1$ channels a similar behavior is found.\\ 
In the case of B-mesons an analogous behavior is seen, but the convergence in the $J^P=0^-,1^-$ channel is much better. This is manifested in a wider plateau than in the D-meson case and in a better correlation of the parameters. Therefore the moment sum rules for the B-mesons can and have been used. At this stage a warning is necessary. In his paper \cite{Narison:1988ep} from 1988 Narison calculated the mass splittings of the B-meson hypermultiplett with $J^P=0^+,0^-,1^+,1^-$ of B-mesons with the sum rules (\ref{hlscalar}) and (\ref{hlvector}) and concluded from that the existence of B-mesons with $J^P=0^+,1^+$. From todays view point that conclusion has to be contradicted. The masses of the positive parity states have not been calculated directly, but by Narison's double moment sum rules which calculate the quotient of the masses which belong to the states $J^P=0^+,0^-$ or $J^P=1^+,1^-$. Hence the mas splittings of the states $J^P=0^+,0^-$ and the splitting of the states $J^P=1^+,1^-$ can be calculated by his method, but only the masses of the states $J^P=0^-,1^-$ can be calculated directly by moment sum rules. However no moment sum rule calculation of the D-meson masses has ever been published. The only published moment sum rule calculation concerning D-mesons was found in the book \cite{Narison:1989aq}, where an estimate for the decay constants of D-mesons based on moment sum rules is given. On the other hand, the calculation of the B-meson masses and decay constants by moment sum rules was published separately in a paper (see \cite{Narison:1989aq} and references herein). The reason for this is probably the bad convergence of the D-meson moment sum rules which have been used.\\ 
There exist many methods to fix the parameters which occur in the moment sum rules. The one chosen in this thesis is based on measurements of masses. The decay constants are no directly observable quantities, but the masses are. Hence, it can be tested if the sum rule reproduces the experimental mass spectroscopy. If it reproduces the mass spectroscopy and fullfills the stability criteria it can be used for the calculation of other quantities like the decay constants. In the case of the D-meson moment sum rule it did not prove to be reliable. Thus it can not be used for the calculation of the decay constants.\\
The moment sum rule in the $J^P=0^-$ channel is given by:
\begin{eqnarray}
M_{n}(q^2)=\frac{1}{n!}\left(\frac{d}{dq^2}\right)^n\Pi_{5}(q^2)\nonumber\\=\frac{1}{\pi}\int_{m^2}^{t_{c}}\frac{Im\left[\Pi_{5,pert}(s)\right]}{(s-q^2)^{n+1}}ds
+\frac{m^2}{(m^2-q^2)^{n+1}}\left[\frac{\mele{\alpha_{s}G^2}}{12\pi}-m\mele{\bar{\psi}_{i}\psi_{i}}\right]\nonumber\\ +\frac{m^3}{2}\left(\frac{(n+2)(n+1)}{(m^2-q^2)^{n+3}}q^2+\frac{(n+1)n}{(m^2-q^2)^{n+2}}\right)\frac{1}{4}\mele{g\bar{\psi_{i}}\sigma^{\mu\nu}\lambda_{a}\psi_{i}G^{a}_{\mu\nu}}\nonumber\\-\frac{m^2}{6}(n+1)\dfrac{n(n+8)m^4-3(n-6)q^2m^2-12q^4}{\left( m^2-q^2\right)^{n+4}}\frac{\pi}{6}\alpha_{s}\mele{\bar{\psi}_{i}\gamma_{\mu}\lambda_{a}\psi_{i}\sum_{i=u,d,s}\left(\bar{\psi}_{i}\gamma_{\mu}\lambda_{a}\psi_{i}\right)}.
\end{eqnarray}
Calculations of D-meson masses based on the OPE given in (\ref{hlscalar}) and (\ref{hlvector}) using moment sum rules are plagued by too many shortcomings, which prohibit high precision analyzes. Thus only rough estimates can be obtained by moment sum rules. The problem can be remedied by using Borel sum rules. The discussion of the results will follow in the next section.

\subsubsection{Analysis with Borel sum rules \label{hlborelope}}

The first step to obtain the Borel transformed sum rule is to transform the phenomenological part of the sum rule. This leads to a change of the integral kernel from a fraction to exponential function as shown in section \ref{dispersionrelations}. The imaginary part of the correlator stays untouched by transformation. This is the simplest part of the transformation. The transformation of the OPE is in many cases much more work. Here every Wilson coefficient has to be transformed. If the perturbative coefficient is given in the form of a dispersion integral at least this one is simple to transform. All other coefficients need explicit treatment. The results for the OPEs (\ref{hlscalar}) and (\ref{hlvector}) are shown below.\\
Here are the Borel transformed Wilson coefficients for the pseudoscalar correlator.
\begin{enumerate}
\item{Borel transformation of $C_{\bar{q}q}$:}
     \begin{eqnarray}
      C_{\bar{q}q}=\frac{m^2}{m^2-q^2}\longrightarrow\frac{m^2}{M^2}e^{-\frac{m^2}{M^2}}.
     \end{eqnarray}
 \item{Borel transformation of $C_{mixed}$:}
     \begin{eqnarray}
      C_{mixed}=\frac{m^3}{2}\frac{q^2}{(m^2-q^2)^3}\longrightarrow\frac{m^3}{2M^4}\left(1-\frac{1}{2}\frac{m^2}{M^2}\right)e^{-\frac{m^2}{M^2}} \label{borelcmixed}.
     \end{eqnarray}
 \item{Borel transformation of $C_{4q}$:}
     \begin{eqnarray}
      C_{4q}=\frac{m^2}{(m^2-q^2)^2}\left[2-\frac{m^2}{m^2-q^2}-\left(\frac{m^2}{m^2-q^2}\right)^2\right]\frac{\pi}{6}\nonumber\\  \longrightarrow\frac{\pi}{6}\frac{m^2}{M^4}\left[ 2-\frac{1}{2}\frac{m^2}{M^2}-\frac{1}{6}\frac{m^4}{M^4}\right] e^{-\frac{m^2}{M^2}}.
     \end{eqnarray}        
\end{enumerate}
Here are the Borel transformed Wilson coefficients for the vector correlator.
\begin{enumerate}
\item{Borel transformation of $C_{\bar{q}q}$:}
     \begin{eqnarray}
      C_{\bar{q}q}=\frac{1}{m^2-q^2}\longrightarrow\frac{e^{-\frac{m^2}{M^2}}}{M^2}.
     \end{eqnarray}
 \item{Borel transformation of $C_{mixed}$:}
     \begin{eqnarray}
      C_{mixed}=\frac{m^3}{2}\frac{1}{(m^2-q^2)^3}\longrightarrow\frac{m^3}{4}\frac{e^{-\frac{m^2}{M^2}}}{M^6}.
     \end{eqnarray}
 \item{Borel transformation of $C_{4q}$:}
     \begin{eqnarray}
      C_{4q}=-\frac{1}{(m^2-q^2)^2}\frac{\pi}{18}\left[4+8\frac{m^2}{m^2-q^2}-3\left(\frac{m^2}{m^2-q^2}\right)^2\right]\nonumber\\\longrightarrow-\frac{\pi}{18}\left[ 4+4\frac{m^2}{M^2}-\frac{m^4}{2M^4}\right]\frac{e^{-\frac{m^2}{M^2}}}{M^4} .
     \end{eqnarray}        
\end{enumerate}
This leads to the OPE 
\begin{eqnarray}
\widehat{\mathcal{B}}\Pi_{5}(Q^2)=
\frac{1}{M^2\pi}\int_{m^2}^{\infty}Im\left[ \Pi_{5,pert}(s)\right] e^{-\frac{s}{M^2}}ds+\left(\left[\frac{\mele{\alpha_{s}G^2}}{12\pi}-m\mele{\bar{\psi}_{i}\psi_{i}}\right]\right.\nonumber\\+\frac{m}{2M^2}\left(1-\frac{1}{2}\frac{m^2}{M^2}\right)\mele{g\bar{\psi_{i}}\sigma^{\mu\nu}\frac{\lambda_{a}}{2}\psi_{i}G^{a}_{\mu\nu}}\nonumber\\\left.+\frac{\pi}{6}\frac{1}{M^2}\left(2-\frac{1}{2}\frac{m^2}{M^2}-\frac{1}{6}\frac{m^4}{M^4}\right)\alpha_{s}\mele{\bar{\psi}_{i}\gamma_{\mu}\lambda_{a}\psi_{i}\sum_{i=u,d,s}\left(\bar{\psi}_{i}\gamma_{\mu}\lambda_{a}\psi_{i}\right)}\right)\frac{m^2}{M^2}e^{-\frac{m^2}{M^2}}=\nonumber\\
\frac{1}{M^2\pi}\int_{m^2}^{\infty}Im\left[ \Pi_{5,pert}(s)\right] e^{-\frac{s}{M^2}}ds+\left(\left[\frac{1.2*10^{-2}GeV^4}{12}-m\mele{-1.5625*10^{-2}GeV^3}\right]\right.\nonumber\\+\frac{m}{2M^2}\left(1-\frac{1}{2}\frac{m^2}{M^2}\right)\mele{-0.0125GeV^5}\nonumber\\\left.+\frac{\pi}{6}\frac{1}{M^2}\left(2-\frac{1}{2}\frac{m^2}{M^2}-\frac{1}{6}\frac{m^4}{M^4}\right)\left(-3.11\cdot 10^{-4}GeV^6 \right) \right)\frac{m^2}{M^2}e^{-\frac{m^2}{M^2}} \label{hlpseudoborelope}
\end{eqnarray}
for the pseudo scalar correlator and 
\begin{eqnarray}
\widehat{\mathcal{B}}\Pi_{V}(Q^2)=\frac{1}{M^2\pi}\int_{m^2}^{\infty}Im\left[ 
\Pi_{V,pert}(s)\right]e^{-\frac{s}{M^2}}ds+\left(
\frac{-\mele{\alpha_{s}G^2}}{12\pi}-m\mele{\bar{q}q}+\frac{m^3}{4M^4}\mele{g\bar{q}\sigma^{\mu\nu}\frac{\lambda_{a}}{2}qG^{a}_{\mu\nu}}\right.\nonumber\\\left.-\frac{\pi}{18}\frac{1}{M^2}\left(4+4\frac{m^2}{M^2}-\frac{1}{2}\frac{m^4}{M^4}\right)
\alpha_{s}\mele{\bar{q}\gamma_{\mu}\lambda_{a}q\sum_{i=u,d,s}\left(\bar{n}\gamma_{\mu}\lambda_{a}n\right)}\right)\frac{1}{M^2}e^{-\frac{m^2}{M^2}}\nonumber\\
=\frac{1}{M^2\pi}\int_{m^2}^{\infty}Im\left[ \Pi_{V,pert}(s)\right] e^{-\frac{s}{M^2}}ds+\left(-\frac{1.2*10^{-2}GeV^4}{12}-m\mele{-1.5625*10^{-2}GeV^3}\right.\nonumber\\\left.+\frac{m^3}{4M^4}\mele{-0.0125GeV^5}-\frac{\pi}{18}\frac{1}{M^2}\left(4+4\frac{m^2}{M^2}-\frac{1}{2}\frac{m^4}{M^4}\right)\left(-3.11\cdot 10^{-4}GeV^6 \right)\right)\frac{m^2}{M^2}e^{-\frac{1}{M^2}} \label{hlvectorborelope}
\end{eqnarray}
for the OPE of the vector correlator. After the transformation the situation for the positive parity states changed drastically. The sum rules now converge. Moreover the parameters of all four sum rules are now correlated in an acceptable way and the continuum threshold is much smaller. This is a welcome effect because the threshold should lie close to the first radial excitation which was not the case in the moment sum rule computations. The original analysis of the $J=0$ states in the strange D-meson channel was done by Narison (see \cite{Narison:2004th} and references herein). Hayashigaki performed an analysis of all four states \cite{Hayashigaki:2004gq}.\\ 
The results of these papers will now be discussed in order to serve as a basis for a further analysis. Narison's analysis can be classified as a conservative one. He uses the relative freedom in the choice of the continuum threshold $t_{c}$ and the c-quark pole mass $m_{c}$ to obtain agreement with the measurements in the $D_{s}$ channel for the $J=0$ states. The parameters of his choice are given by $t_{c}=(7.5\pm1.5)GeV^2=(2.725\pm0.275)GeV$ and $m_{c}=1.46GeV$. Although the measurements and the analysis agreed, the analysis had a huge problem. The Borel curves for the D-meson states are expected to have a plateau, as classical mesons like the $\rho$ had \cite{Shifman:1978bx}, but this was not the case. Narison's curves have all been hyperbola shape like.\\
Hayashigaki performed an unconservative analysis. He uses a larger value for the charm quark mass than Narison and uses another interval for the continuum threshold $t_{c}$. His analysis reproduces the masses of all particles from the hypermultiplet except the $0^+$ states. For the $0^+$ states he obtained masses larger than the experimental ones. This holds both for the  channel with isospin and strangeness. He concludes that the $D_{s}(2317)$ is therefore a four quark state, while he claims no clear conclusion for the $0^+$ state in the channel with isospin. Narison made a comparison of his and Hayashigaki's work in his paper \cite{Narison:2003td}, and argues against Hayashigaki. The difference between the two publications is probably due to a difference in the perturbative charm quark pole mass. Hayashigaki use a perturbative charm quark pole mass of $1.46GeV$, while Narison used $1.3GeV$. Another problem recognized during the work on this thesis considers the sign of the mixed condensate coefficient in the $J=1$ channel. In Narison's book \cite{Narison:1989aq} he claims a negative sign in the case of a vector current, while Hayashigaki has a positive one \cite{Hayashigaki:2004gq}. Hence, Hayashigaki's calculations in the vector channel are in question, while Narison is out of the game because he never performed calculations for D-mesons in the vector channel. He did it for B-mesons, which are now also in question. 
One thing holds for Hayashigaki's calculations in all cases, except of the $1^+$ case, the curves are hyperbolas and no plateau is visible. Narison just calculated the $J=0$ channel and there the same phenomenon occurred. The reason for this phenomenon is unclear. There are several possibilities
\begin{enumerate}
\item The OPE could be to inexact. The terms could yield too crude approximations or there may be terms missing.
\item The ansatz for the spectral function could be wrong.
\end{enumerate}
However, there must be something in the sum rules that is right. The likelyhood to write down a sum rule which by chance produces results that agree with the measurements is to small to seriously consider that possibility. Anyhow it is reasonable to search for possibilities to improve the convergence of the sum rule. 

\subsubsection{Testing new possibilities to improve the Borel sum rules for the D-meson hypermultiplet}

\begin{figure}[htbp]
 \begin{center}
  \includegraphics{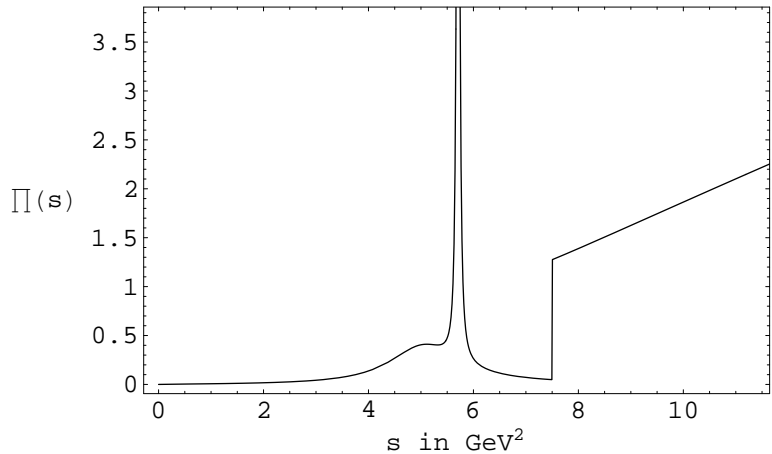}
  \caption{Sketch of the imaginary part of the scalar correlator as predicted by \cite{Hofmann:2003je} (see the data in table \ref{juliandata}). The spectral function contains two resonances a broad one and a narrow one which lie close to each other. The continuum is approximated by the imaginary part of the correlator. In the case of the axialvector D-mesons the sketch is similar. \label{julianspectrum02}}
  \end{center}
  \end{figure}

\begin{figure}[htbp]
 \begin{center}
  \includegraphics[scale=1.5]{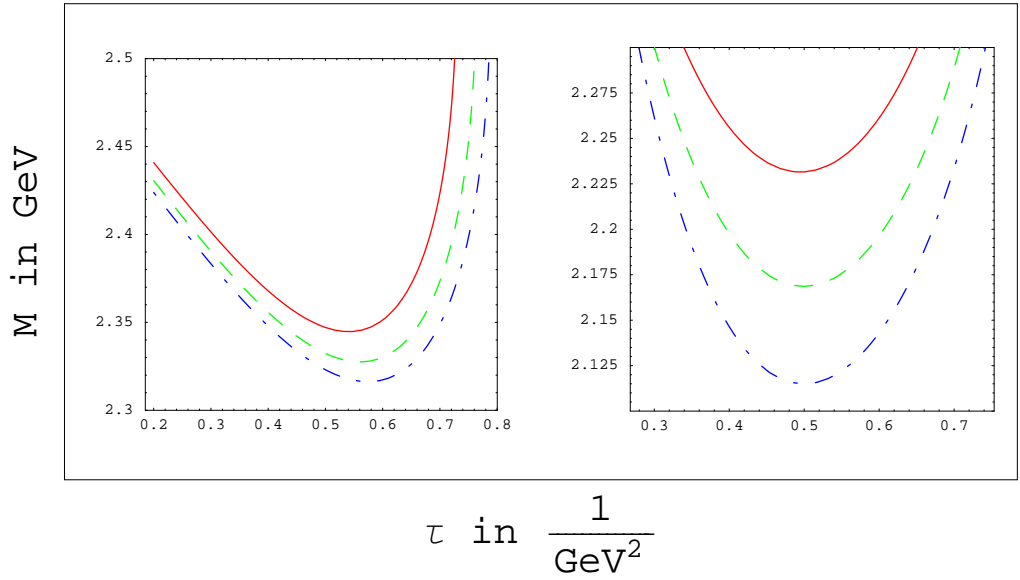}
  \caption{Example for the Borelcurves in the $J^P=0^+$ with isospin. On the left the state with m=2.255 GeV and on the right the state with m=2.389 GeV is found. The lines correspond to the following parameters: $t_{c}=8.25$ for all lines, full line  r=4, dashed line r=5 and dotted dashed line r=6. \label{rep1}}
  \end{center}
  \end{figure}

\begin{figure}[htbp]
 \begin{center}
  \includegraphics[scale=0.6]{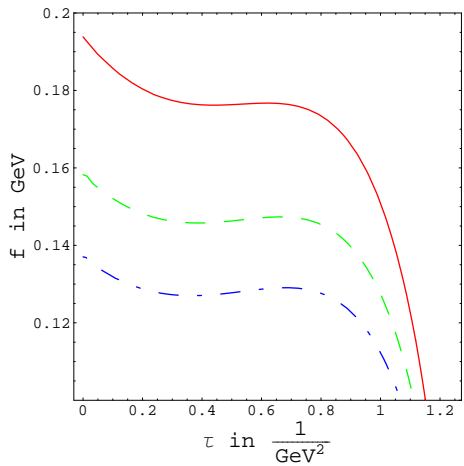}
  \caption{Example for the Borelcurves in the $J^P=1^+$ with strangeness for the state with m=2.459 GeV. The lines correspond to the following parameters: $t_{c}=9.75$ for all lines, full line  r=1, dashed line r=2 and dotted dashed line r=3.\label{rep6}}
  \end{center}
  \end{figure}

Recent measurements changed the picture of the spectral function owing to heavy-light systems. The narrow resonance ansatz is given by a single $\delta$-resonance and a continuum which is a step function. The distance between the resonance and the threshold of the continuum lies in the region of 0.5 to 1 GeV in the calculations performed by Narison and Hayashigaki. This distribution is justified from the $J/\psi$ and $\rho$ spectral function where the spectral functions are measured and the assumption holds. During this work no reason for this approximation in other channels was found. However, in the axial vector channel of D-mesons there are two cases where the narrow resonance plus continuum approximation does not hold. In those spectral functions two states lie very close to each other the ground state and a second one, which may be a radial excitation. The channels are the s- and u-quark channel of $1^+$ D-mesons.\\
In the s-quark case the resonances are of nearly equal width and mass, while the u-quark case consits of a broad and a narrow state (see table \ref{spectralfunc}).
\begin{table}[htbp]
\begin{tabular}{||c|l|l|l|l||}
\hline
$J^P_{(I,S)}$&$1^+_{(\frac{1}{2},0)}$&$1^+_{(\frac{1}{2},0)}$&$1^+_{(0,-1)}$&$1^+_{(0,-1)}$\\
\hline 
\hline
m (in GeV)&2.425  &2.427&2.459& 2.535\\
\hline  
$\Gamma$ (in MeV)&58 &384&5.5 &2.3\\ 
\hline                      
\end{tabular}
\caption{Data of the spectra as they are given by the particle data group (2005). The quantum numbers of the heavy particles in the spectral function are not yet sure, the errors are omitted.\label{spectralfunc}}
\end{table} 
The mass splitting is below $0.1GeV$. Therefore the narrow resonance plus continuum ansatz does not seem to be reasonable. The continuum has to start close to the second resonance and if the calculations are performed with Aliev type sum rules no convergence is seen with such a small threshold value. Hence, it is reasonable to introduce a second resonance and to work with two resonances and a continuum. Moreover, the second resonance in the u-quark channel is not narrow but very broad. Therefore the narrow resonance approximation is not reasonable any more. As an alternative a Breit-Wigner function is suitable.\\ 
Forced by such measurements the question arises if there may be a similar spectrum in the remaining channels $0^+$,$0^-$ and $1^-$. Unfortunately no measurements were found, but a theoretical prediction. Recently a new method for the calculation of exited states of heavy-light systems has been developed by M.F.M.Lutz. Based on this method J.Hofmann calculated the spectral function for the $J^P=0^+,1^+$ states with isospin or strangeness. The result exhibit something in common with the measurered spectral functions. Hofmann calculated for six spectral functions the part of the spectral function around the ground state. He did it for three channels, one with isospin, one with strangeness and one with anti-strangeness. In the channel with isospin he published results which can be compared with the measurements. In his publication \cite{Hofmann:2003je} the axial-vector channel with isospin is found which corresponds to the u-quark channel. The calculations which where published before the measurements where done agreed qualitatively with the data. Hence, the scheme seems to be valid in the positive parity doublet of the D-mesons with isospin.  Moreover, he calculated further spectral functions. In the $0^+$ channel with isospin he predicts a spectral function which again has two resonance close to each other where one is broad and the other is narrow.  Therefore a check of the spectral functions with QSR is reasonable for several reasons. The quantum number assignment for the measurered spectralfunction can be checked. Moreover, the model with which Lutz and Hofmann work does not have quarks and gluons but hadrons as basic degrees of freedom. QSR offer a method with which can be checked what QCD says to the predictions which have been made. Hence, a test of the method can be performed with such a calculation. Therefore the $0^+$ channel with isospin is checked (see the data in table \ref{juliandata}).
\begin{table}
\begin{tabular}{||c|l|l|l|l||}
\hline
$J^P_{(I,S)}$&$0^+_{(\frac{1}{2},0)}$&$0^+_{(\frac{1}{2},0)}$\\
\hline 
\hline
m (in GeV)&2.255  &2.389\\
\hline  
$\Gamma$ (in MeV)&360 &10\\ 
\hline                      
\end{tabular}
\caption{Data of the spectra as they were given in \cite{Hofmann:2003je} for the $0^+$ channel with isospin \label{juliandata}}
\end{table}  
In order to perform the calculations the usual method has to be modified, the parameterization of the spectral function must be modified. In the case of the s-quark channel two narrow resonances can be used and in the u-quark channel a narrow resonance and a Breit-Wigner resonance can be used
\begin{eqnarray}
S(s,m,\Gamma)=\frac{1}{\pi} \frac{\sqrt{s}~\Gamma}{(s-m^2)^2+s(\Gamma)^2}
\end{eqnarray}
where $\Gamma$ is the width of the resonance. The continuum is still approximated by the imaginary part of the correlator. Thus, the Breit-Wigner curves for the resonances are cut at the threshold $t_{c}$ and from there on the imaginary part of the correlator is used. The spectral function for the scalar D-meson is than given by 
\begin{eqnarray}
\Pi\left(s\right)= S(s,2.255GeV,0.36GeV)+S(s,2.389GeV,0.01GeV)+Im\Pi_{0^+,pert}(s) \label{julianspectrum01}
\end{eqnarray}
and plotted in figure\ref{julianspectrum02}.
Due to the Breit-Wigner function the QSR can not be solved for the mass corresponding to the broad state. Fortunately, with tools like Mathematica this problem is easily manageable. The width of the states will be treated as given by measurements or predictions. Thus the narrow resonances can also be approximated by a Breit-Wigner curve without much additional work. Figure \ref{julianspectrum02} and equation (\ref{julianspectrum01}) are not fully correct. The resonances have to carry factors which are given by squared matrix elements $\bra{0}j\ket{n}^2$. Those matrix have to be inserted into the spectral function
\begin{eqnarray}
\Pi\left(s\right)= \bra{0}j\ket{2.255}^2~S(s,2.255GeV,0.36GeV)+\bra{0}j\ket{2.389}^2~S(s,2.389GeV,0.01GeV)\nonumber\\+Im\Pi_{0^+,pert}(s) \label{julianspectrum03}.
\end{eqnarray}
The coefficients can be fitted to reproduce the mass spectroscopy, more exactly the ratio of the coefficient. This is due to the way the Borel sum rules are evaluated. As always the spectral function enters the Borel transformed dispersion integral. In order to eliminate as much as possible from the matrix elements in front of the resonances the logarithmic derivative of the dispersion integral is taken 
\begin{eqnarray}
\mathcal{R}(\tau)=-\frac{d}{d\tau}\log\left(\frac{\tau}{\pi}\int Im\Pi(s)e^{-s\cdot\tau}ds\right).
\end{eqnarray}
Hence a ratio between the original expressions and the derivative of this expressions is calculated. Thus one of the matrix elements can be factored out leaving a ratio between the matrix elements. This reduces the number of a priory unknown quantities from three to two. However, there are three new parameters in the QSR two widths, one for each particle and the ratio of the matrix elements. In QSR calculations with narrow resonances the widths have of course been absent and if only one resonance is considered only one matrix element enters the spectral function and this is cancelled from the evaluation of the sum rule due to the ratio following from the logarithmic derivative.\\
Thus the phenomenological part of the sum rules is determined. As the theoretical part the OPEs of (\ref{hlscalar}) and (\ref{hlvector}) are used. Therefore a two quark dominance of the states is assumed. The expression for the dispersion integral and the OPE are equal. Hence, the logarithmic derivative of the dispersion integral and the OPE are equal. This equation can be used to determine properties of the spectral function, assuming that the condensates are all known sufficiently exact. Overall six free parameters enter the sum rule.
\begin{enumerate}
\item The mass of the first resonance $M_{1}$.
\item The mass of the second resonance $M_{2}$.
\item The width of the first resonance $\Gamma_{1}$.
\item The width of the second resonance $\Gamma_{2}$.
\item The ratio of the matrix elements $r=\frac{\bra{0}j\ket{2.389}}{\bra{0}j\ket{2.255}}$.
\item The values of the continuum threshold $t_{c}$.
\end{enumerate}
Generally it is possible to measure everything except of $t_{c}$ and the matrix elements $r$. These quantities can be measured only in special cases. $t_{c}$ can be measured only for vector currents. A particle for which the matrix elements can be measured is the $\pi$ meson, where it can be determined from decays to a high accuracy. The goal is to check if its possible to fit $r$ and $t_{c}$ so that the sum rule reproduces the spectral function. Depending on whether this is possible or not statements concerning the physics can be made.\\
Before the results are presented a short review of the evaluation of Borel sum rules is given. The dispersion relation where on the left side the phenomenological and on the right side the theoretical part of the sum rules are found is evaluated in order to give the plot of the meson mass in dependency of the Borel parameter $M$. There are two borders for the Borel parameter. The first is given by the theoretical part, for small $M$ the OPE is not valid. The second is given by the phenomenological part, for large $M$ the continuum contribution gets dominant, but then the resonance properties can no longer be determined from the sum rule. The Borel window is determined to be a domain where the OPE is valid and the resonance dominates the phenomenological part of the sum rules. Unfortunately, it is not sure that a Borel window exists. In the case of small $M$ the Borel window is fixed, by the requirement that the $d=6$ terms contribute less than $10-20\%$ to the OPE for big $M$ the Borel window is fixed by the requirement that the contribution of the resonance to the dispersion integral is bigger than the contribution of the continuum
\begin{eqnarray}
\int_{0}^{t_{c}}dsIm\Pi(s)e^{-\frac{s}{M^2}}\leq\int_{t_{c}}^{\infty}ds~Im\Pi(s)e^{-\frac{s}{M^2}}.     
\end{eqnarray}
Hence the upper border of the Borel window depends on $t_{c}$, while the lower does not. The Borel window determines the regime where the sum rules are reliable in order to determine the meson mass. The larger the Borel Window is the more reliable is the corresponding determination of resonance properties.\\
\begin{table}[htbp]
\begin{tabular}{||c|l|l|l|l||}
\hline
$J^P_{(I,S)}$&$0^+_{(\frac{1}{2},0)}$&$0^+_{(\frac{1}{2},0)}$&$1^+_{(0,-1)}$&$1^+_{(0,-1)}$\\
\hline 
\hline
m (in GeV)&$2.255\pm0.01$&$2.389\pm0.01$&$2.459\pm0.06$&$2.535\pm0.1$\\
\hline  
$\Gamma$ (in MeV)&360&10&5.5 &2.3\\ 
\hline 
 $f$ (in MeV)&$148\pm 16$&$264\pm32$&$155\pm27$&$202\pm21$\\ 
\hline  
 $t_{c}$ (in GeV)&$8.25\pm0.75$&$8.25\pm0.75$&$9.875\pm1.125$&$9.875\pm1.125$\\ 
\hline
$r$ (in GeV)&$6\pm3$&$6\pm3$&$3\pm2$&$3\pm2$\\ 
\hline      
 $m_{c}$ (in GeV)&$1.15\pm0.01$&$1.15\pm0.01$&$1.2\pm0.02$&$1.2\pm0.02$\\ 
\hline
\end{tabular}
\begin{tabular}{||c|l|l||}
\hline
$J^P_{(I,S)}$&$1^+_{(\frac{1}{2},0)}$&$1^+_{(\frac{1}{2},0)}$\\
\hline 
\hline
m (in GeV)&$2.425\pm0.080$ &$2.427\pm0.050$\\
\hline  
$\Gamma$ (in MeV)&58 &384\\ 
\hline 
 $f$ (in MeV)&&\\ 
\hline  
 $t_{c}$ (in GeV)&$7.5\pm0.3$&$7.5\pm0.3$\\ 
\hline
 $r$ (in GeV)&$0.3\pm0.25$&$0.3\pm0.25$\\ 
\hline      
 $m_{c}$ (in GeV)&$1.2\pm0.01$&$1.2\pm0.01$\\ 
\hline                                            
\end{tabular}
\caption{Results from the Sum Rule analysis. Everything except of the widths has been determined from the Sum Rules.\label{results}}
\end{table}
As it was already discussed above the parameters of the sum rules have to be fixed. Therefore, experimental data on the particles are used. The parameters of the Sum Rule are adjusted to reproduce the experimental data. Moreover, the Sum Rule has to fulfill additional criteria in order to be reliable. The data which are used are the masses of the particles. The parameters are adjusted in order to shift the plateau of the corresponding Borel curve to the mass of the corresponding particle. In addition, the Borel curve is required to be stable against shifts in the parameters. This corresponds to an interval in which the parameters can be varied without changing the position of the plateau significantly. Finally the plateau has to be located in the Borel window. A large Borel window is of course much more reliable than a small Borel window. As the mass of the charm quark the running mass was used. The width of the particles was held fixed.\\
The intervals for the parameters are given in table \ref{results}. In these intervals it is possible to reproduce the masses of the corresponding particles. Hence, they can be used to determine the corresponding decay constants. After that the Borel window can be determined. Unfortunately, the Borel curve in the channel with strangeness shows no plateau in the plots for the decay constants. Due to this only one interpretation is possible, the spectral function can not be described by the QSR which was used. The other channels have a plateau in the plot for the decay constant, although they do not have a plateau in the plots for the masses. The plots for the masses always have the shape of a hyperbola. The decay constants agree with earlier calculations. The minimal range for the Borel window for every plot is $\tau=0.3-1.0$. Thus, all plateaus lie inside the Borel window. However, also the Sum Rules for these states show inconsistencies. The intervals for the parameters are small, far from the ideal case.
The curves for the masses look like a parabola and not like a plateau, some representive curves are shown in figure \ref{rep1} to \ref{rep6}. Due to these inconsistencies the sum rules are also regarded as unreliable. To be more exact, the Sum Rule in the channel with strangeness is less reliable than the other Sum Rules.\\
The results raise many questions. Even the improved spectral function did not improve the reliability of the Sum Rules. Moreover, their reliability even got worser. Without the improved spectral function the Sum Rule in the channel with strangeness is more reliable than it is here. If the values of the decay constants are correct or not can only be proven by experiments. Unfortunately, no data from which the decay constants can be extracted are available. The source which is supposed to produce the correct result is lattice QCD. The results in these calculations are slightly lower than the lattice results, but in agreement with earlier Sum Rule calculations. At this point further discussions are not reasonable only further investigations could help to solve the problems. The conclusions will offer a possibility to find such a solution.

  \begin{figure}[htbp]
 \begin{center}
  \includegraphics[scale=0.6]{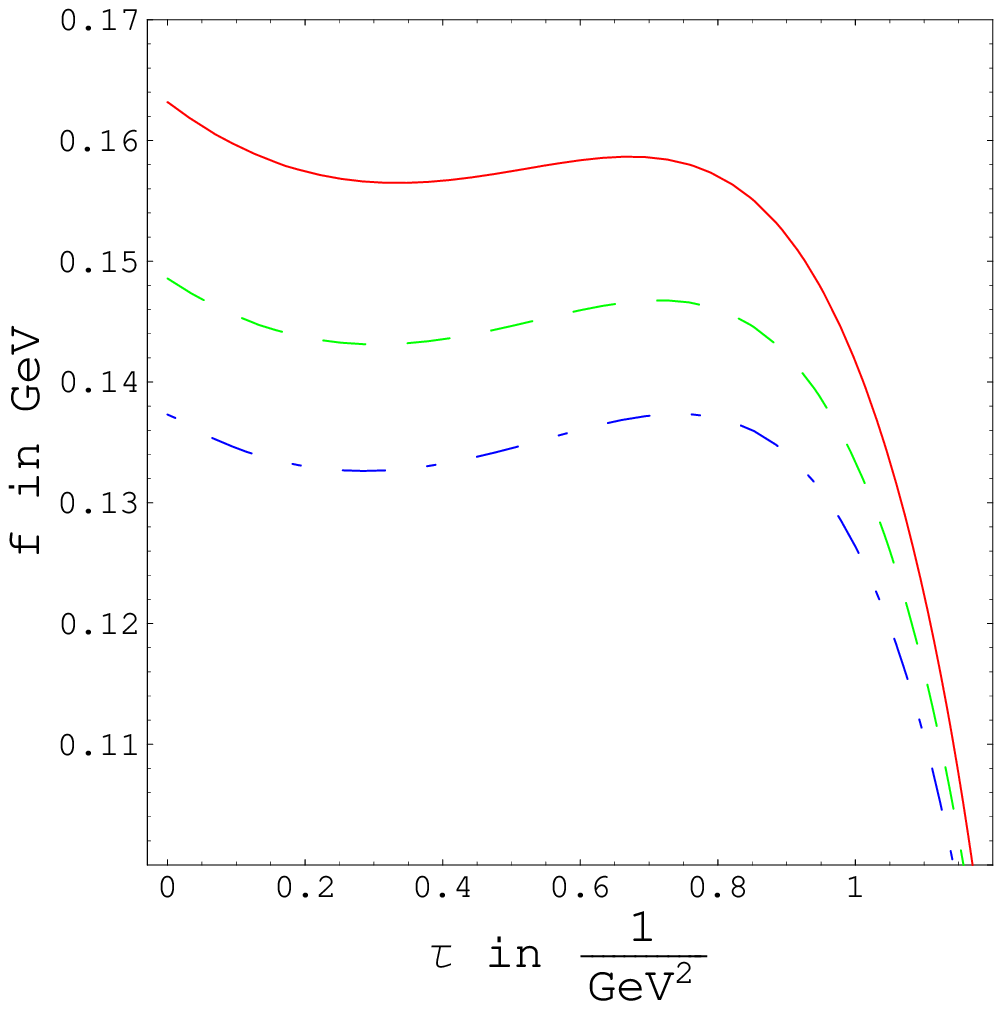}
  \caption{Example for the Borelcurves in the $J^P=0^+$ with isospin for the state with m=2.255 GeV. The lines correspond to the following parameters: $t_{c}=8.25$ for all lines, full line  r=4, dashed line r=5 and dotted dashed line r=6.}
  \end{center}
  \end{figure}
  
 \begin{figure}[htbp]
 \begin{center}
  \includegraphics[scale=1.5]{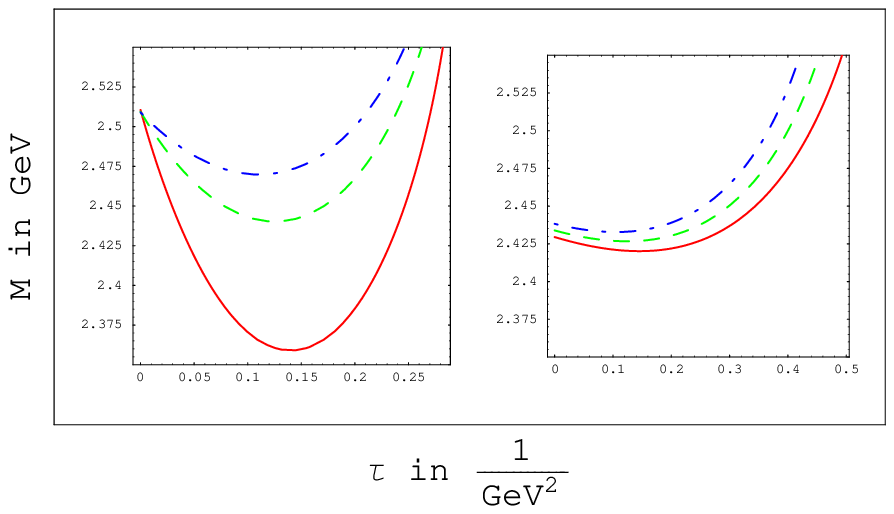}
  \caption{Example for the Borelcurves in the $J^P=1^+$ with isospin. On the left the state with m=2.425 GeV and on the right the state with m=2.427 GeV is found. The lines correspond to the following parameters: $t_{c}=7.5$ for all lines, full line  r=0.1, dashed line r=0.2 and dotted dashed line r=0.3.}
  \end{center}
  \end{figure} 
  
  \begin{figure}[htbp]
 \begin{center}
  \includegraphics[scale=0.6]{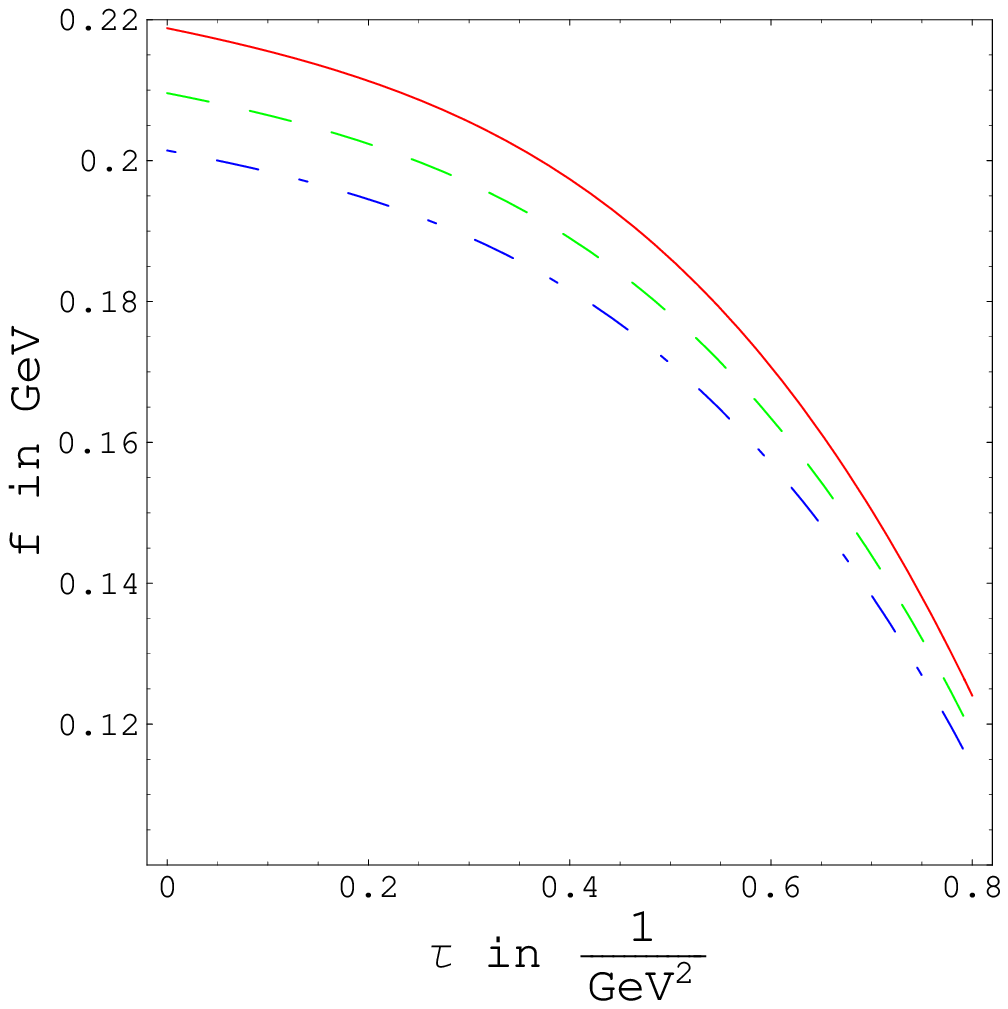}
  \caption{Example for the Borelcurves in the $J^P=1^+$ with isospin for the state with m=2.425 GeV. The lines correspond to the following parameters: $t_{c}=7.5$ for all lines, full line  r=0.1, dashed line r=0.2 and dotted dashed line r=0.3.}
  \end{center}
  \end{figure}
  
  \begin{figure}[htbp]
 \begin{center}
  \includegraphics[scale=1.5]{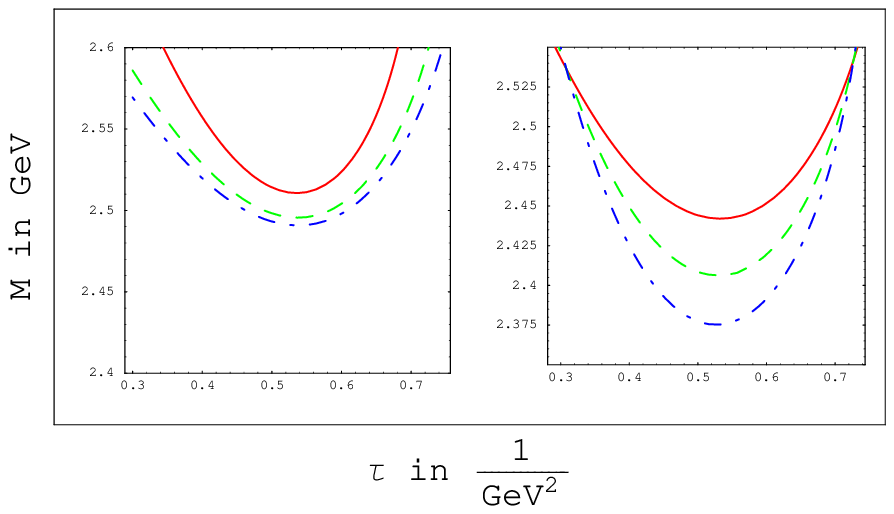}
  \caption{Example for the Borelcurves in the $J^P=1^+$ with strangeness. On the left the state with m=2.459 GeV and on the right the state with m=2.535 GeV is found. The lines correspond to the following parameters: $t_{c}=9.75$ for all lines, full line  r=1, dashed line r=2 and dotted dashed line r=3.}
  \end{center}
  \end{figure} 
  
\newpage

\section{Conclusions} 
The calculations made in this thesis use an improved ansatz for the spectral function. In most of the Sum Rule applications the spectral function is approximated by a single resonance and a continuum. In this work a spectral function with two resonances and a continuum is used. According to the QCD Sum Rule frame work the calculations are improved when a more realistic spectral function is used. Due to the second resonance an additional parameter enters the theory. The properties of the resonances are known from experiment or other theoretical approaches. Hence, they are used to fix the parameters of the theory. On that basis the decay constants of the corresponding hadrons are calculated (see table \ref{results}). The results of these calculations and the shape of the corresponding Borel curves are used to extract statements concerning the analyzed hadrons.\\
The hadrons which are addressed are D-mesons. D-mesons are believed to have two valence quarks, a charm quark together with a up, down, or strange quark. Therefore, the Sum Rules used here are based on a two quark structure. In particular three systems are analyzed. A system with $J^P=0^+$ with charm and isospin two systems with $J^P=1^+$ one with charm and isospin and the other with charm and strangeness. The results are of quantitative and qualitative nature and can be divided into two groups.\\ 
In the first group the D-mesons with  $J^P=0^+$ and isospin and the D-meson with $J^P=1^+$ and strangeness are contained. Although, the Sum Rules for these particles can reproduce the data, they are not satisfactory. The OPE does not seem to reproduce the spectral function reliably. Although it can not be excluded that an OPE which contains more terms would do it. The quantitative results for the D-mesons in the first group are the decay constants in the two quark picture. The results for the decay constants of the resonances have again to be split into two groups. The first one is the group of the resonances which are the lightest ones. The ground state so to say. Their values agree within error bars with earlier calculations. The second group is the group of resonances which are higher in mass. No earlier calculations or estimates on their value has been found during this work. These properties have most probably never been calculated before. For these states the decay constants are always higher than for the ground state particles.\\
In the second group the D-mesons with $J^P=1^+$ and isospin is contained. In that channel the OPE can not reproduce the spectral function. In consideration of the experience gained during the work on this thesis even an OPE with more terms should not change this situation. The calculation of decay constants was impossible. The sum rule in that channel did not show any sign of a plateau for the decay constant. This point was discussed in section \ref{sumrules}.\\
However, not long ago some papers on D-mesons and QCD Sum Rules appeared which implement a four quark structure of D-mesons \cite{Kim:2005gt,Nielsen:2005ia}. In these publications many problems with which the two quark versions are plagued with are absent. Hence, from the viewpoint of the current investigation four quark structures have to be taken into account in QSR calculations in the hope of maintaining Sum Rules which behave more reasonable. The minimal recommendation which can be extracted from this thesis is to consider Sum Rules for D-meson systems which implement a two and a four quark structure of D-Mesons in order to extract the mixing angle between those structures. This means that it is necessary to consider four-quark structures in addition to the standard two-quark ones.

\newpage
\vspace{1cm}
Erkl\"arung zur Diplomarbeit gem\"ass 19 Abs. 6 DPO/AT.\\
\\
Hiermit versichere ich, die vorliegende Diplomarbeit ohne Hilfe Dritter nur mit den angegebenen Quellen und Hilfsmitteln angefertigt zu haben. Alle Stellen, die aus den Quellen entnommen wurden, sind als solche kenntlich gemacht worden. Diese Arbeit hat in gleicher Form noch keiner Prfungsbeh\"orde vorgelegen.
\\
\\
Darmstadt, den 31.Januar 2006\\ 
\\ 
\\
---------------------------------------\\
(Unterschrift)

\end{document}